\documentclass[10pt,onecolumn]{aastex631}
\usepackage{graphicx}
\usepackage{amsmath}
\usepackage{amsfonts}
\usepackage{amssymb}%
\providecommand{\U}[1]{\protect\rule{.1in}{.1in}}

\usepackage{array}
\usepackage[utf8]{inputenc}

\usepackage[most]{tcolorbox}

\usepackage{tikz}
\usetikzlibrary{shapes,shadows,calc}
\usetikzlibrary{positioning,calc}

\usepackage{yfonts}
\usepackage[normalem]{ulem}
\usepackage[T1]{fontenc}

\usepackage[yyyymmdd,hhmmss]{datetime}

\newcommand{\beq}{\begin{equation}}
\newcommand{\eeq}{\end{equation}}
\newcommand{\ba}{\begin{array}}
\newcommand{\ea}{\end{array}}
\renewcommand{\L}{L_{52}}
\newcommand{\E}{E_{53}}

\newcommand{\M}{\dot{M}_{-6}}
\newcommand{\n}{n_{0,3}}

\newcommand{\ee}{\epsilon_{e,0}}

\newcommand{\D}{{\mathcal {D}}}
\newcommand{\R}{{\mathcal {R}}}
\renewcommand{\O}{{\mathcal {O}}}

\def\be{\begin{equation}}
\def\ee{\end{equation}}
\def\lsim{\raisebox{-0.3ex}{\mbox{$\stackrel{<}{_\sim} \,$}}}
\def\gsim{\raisebox{-0.3ex}{\mbox{$\stackrel{>}{_\sim} \,$}}}

\newcommand{\oh}{\frac{1}{2}}

\usepackage{color}
\usepackage{tabularx}
\usepackage{xcolor,colortbl}
\definecolor{LightCyan}{rgb}{0.88,1,1}
\definecolor{Chartreuse}{rgb}{0.5, 1, 0}

\newcommand{\I}{{\mathcal {I}}}
\newcommand{\J}{{\mathcal {J}}}

\begin{document}

\title{cuHARM : a new GPU accelerated GR-MHD code and its application to ADAF disks}

\shorttitle{cuHARM}
\shortauthors{B\'egu\'e et al.}

\author[0000-0003-4477-1846]{D. B\'egu\'e}
\affiliation{Bar Ilan University, Ramat Gan, Israel}
\correspondingauthor{D. B\'egu\'e : cayley38@gmail.com}

\author[0000-0001-8667-0889]{A. Pe{'}er}
\affiliation{Bar Ilan University, Ramat Gan, Israel}

\author[0000-0001-6545-4802]{G.-Q.  Zhang}
\affiliation{School of Astronomy and Space Science, Nanjing University, Nanjing 210093, China}
\affiliation{Bar Ilan University, Ramat Gan, Israel}

\author[0000-0003-4111-5958]{B.-B. Zhang}
\affiliation{School of Astronomy and Space Science, Nanjing University, Nanjing 210093, China}
\affiliation{Key Laboratory of Modern Astronomy and Astrophysics (Nanjing University), Ministry of Education, China}

\author{B. Pevzner}
\affiliation{Bar Ilan University, Ramat Gan, Israel}

\begin{abstract}
We introduce a new GPU-accelerated general-relativistic magneto-hydrodynamic (GR-MHD) code based on HARM which we call cuHARM. The code is written in CUDA-C and uses OpenMP to parallelize multi-GPU setups.  Our code allows us to run high resolution simulations of accretion disks and the formation and structure of jets without the need of multi-node supercomputer infrastructure. A $256^3$ simulation is well within the reach of an Nvidia DGX-V100 server, with the computation being a factor about 10 times faster if only the CPU was used.
We use this code to examine several disk structures all in the SANE state. We find that: (i) increasing the magnetic field, while in the SANE state does not affect the mass accretion rate; (ii) Simultaneous increase of the disk size and the magnetic field, while keeping the ratio of energies fixed, lead to the destruction of the jet once the magnetic flux through the horizon decrease below a certain limit. This demonstrates that the existence of the jet is not a linear function of the initial magnetic field strength; (iii) the structure of the jet is a weak function of the adiabatic index of the gas, with relativistic gas tend to have a wider jet. 
\end{abstract}


\section{Introduction}


Accretion disks into compact objects are central to many astronomical objects of interest, including, among others, active galactic nuclei (AGNs), X-ray emitting binaries (XRbs), and even gamma-ray bursts (GRBs). Theoretically, several types of disks have been identified, whose structure mainly depend on (i) the mass accretion rate and the resulting optical depth; and (ii) the configuration of magnetic field inside the disk  \citep[for reviews, see, e.g.,][]{NM08, AF13}.

At high accretion rate (close to the Eddington limit, $\dot M_{acc} \lesssim \dot M_{Edd}$), the gas is radiatively efficient, and the disk radiates approximately $10\%$ of the rest mass energy during the accretion process, resulting in a geometrically thin disk \citep{SS73,NT73}, with height to radius ratio $H/r \ll 0.1$. Such disks are found around all types of black-holes (stellar and supermassive).  
At lower accretion rate,  $\dot M \ll \dot M_{Edd}$, the gas is radiatively inefficient, and the accretion flow is underluminous, $L_{disk} \ll 0.1 \dot M_{acc} c^2$. The resulting disks, termed ADAFs \citep[standing for Advection Dominated Accretion Flow, ][]{NY94, ACK95,NM08}, are geometrically thick, $H/r \lesssim 1$, and are characterized by large radial velocity, implying relatively short accretion time.  In ADAF disks,  the accreted gas is tenuous and have long radiative cooling time relative to the accretion time, $t_{cool} \gg t_{acc}$\footnote{In fact, a second type of ADAF disk occurs when  particle do cool fast,  $t_{cool} \ll t_{acc}$, but the optical depth for scattering is large enough such that most photons do not escape. This can occur in very high accretion rates, $\dot M_{acc} > \dot M_{Edd}$ and results in a "slim disk". \cite{ACL88} We will not discuss this further in this work.}. Such disks are termed "Radiatively Inefficient Accretion Flow" (RIAF) \citep{NY94}.

A second defining property of accretion disks is the magnetic field configuration. Seed magnetic fields are amplified by magneto-rotation instability (MRI) as the flow accretes, resulting in a steady configuration that affects the inner radii of the disk structure. Two main types of distinct, steady state configurations were identified in the literature. The first is Magnetically Arrested Disks \citep[MAD;][]{BR74,MIA03}. In the MAD accretion state, magnetic field lines accumulate near the horizon, and the resulting magnetic pressure regulates the accretion by delaying or even stopping the in-falling
matter \footnote{In fact, a large magnetic flux threading the horizon is the defining property of a MAD state}. As a result, in the inner parts of the disk, the infalling gas accretes in filaments or streams \citep{FCG17, WDP21}. This configuration is found to be accompanied by ejection of matter into strong relativistic jets, where the jet power only weakly depends on
the initial disk setup, and are limited by the magnetic flux threading the horizon \citep{TMG14}. 

The second steady state configuration of the evolving disk is coined "Standard And Normal Evolution" \citep[SANE;][]{NSP12,SNP13}. In a SANE disk, the magnetic flux threading the horizon is not large, thereby enabling smooth accretion. Magnetic fields mainly contribute to the
transport angular momentum to larger radii. In these states as well, relativistic jets were (numerically) observed, although typically these are less powerful than jets created during MAD states.



Indeed, relativistic jets are known to exist in many astrophysical systems, including AGNs, XRBs and GRBs \citep[e.g.,][]{BMR19, Rom21}. The connection between accretion disks and jets had been thoroughly studied by many authors, and is well established both
observationally and theoretically \citep{MSA12,FB04, FHB09_1, SFT10}. Yet, many
important unsolved questions remain, such as the mechanisms for powering,
launching and collimating jets which are not yet fully specified and understood
\citep[e.g. ][]{MRN11, DMM14}.

While in early days, analytical models of steady state accretion configurations \citep[e.g., ][]{SS73,NT73,NY94} and energy
extraction \citep{BZ77, BP82} were built, in recent years, global numerical simulations became feasible. Such simulations are used for time dependent study
of these complex accreting (and ejecting) dynamical systems in full general relativity (e.g., Kerr metric), shading light on the magneto-hydrodynamical processes in the close vicinity of
a black-hole \citep[e.g.][for a recent review]{Miz22}. 
Over the years, several
codes solving the general relativistic magneto-hydrodynamic (hereinafter GRMHD)
equations have been developed \citep[see \textit{e.g.}][to name a few]{DHK03, GMT03,AFS05,
DZB07, SGT08,EPH15,POM17}.

These codes have been used to numerically study
different types of disks. Thanks to the advances of computational
facilities over the two last decades, important numerical results were obtained. 
%
%
%
Thick disks, with $H/r \sim 1$, were the first types of disk to be studied numerically \citep[e.g. ][]{DHK03, MG04}. In particular, GRMHD simulations were used to establish the structure of an accreting system, comprising of a disk, a jet or funnel, and a corona \citep{MG04}. In addition, simulations were performed to determine the sensitivity of the initial assumptions on the accretion rate and the morphology of the accreting system. For instance, \citet{BHK08} studied the influence of different assumption for the initial magnetic field, finding that the morphology of the disk is not strongly affected by any choice, but that the presence of a powerful jet is. Another example is the study of tilted disk by \citet{FBA07,FLA09}. Moreover, using 3D numerical simulations, \cite{TNM11}, followed by \cite{MTB12}, showed that the jet power is larger than the accretion power, thereby demonstrating that part of the jet energy should be extracted from the
rotational energy of the spinning black-hole, therefore numerically demonstrating the existence of the Blandford and Znajek (hereinafter BZ)
mechanism \citep{BZ77}. Recently, with the addition of external radiative transfer codes, these simulations were used in extracting physical information on the properties of the BH in M87, as detected by the Event Horizon Telescope (EHT) \citep{EHT19, EHT_21b, YWY22}.
 
Thin disks disks were also numerically studied. They are challenging to be
numerically resolved and they require cooling functions and specific numerical
grids \citep[see \textit{e.g.}][]{SMN08,NKH10, PMN10, KPS11, AMR16, LTI19}.
Numerical simulations in the context of thin disks were used by \citet{SMN08}
to contrast the specific angular momentum against the analytical solution of
\citet{NT73}. Another example is the study of the radiative efficiency of thin
disk by \citet{AMR16}. Finally, more recent simulation of thin disks were aimed
at studying the Bardeen-Petterson alignment of tilted accretion disks
\citep{LHT20}.


Clearly, such simulations, despite their great success, are very time consuming and computationally expensive, necessitating access to High-Performance Computing facilities (HPC). Naturally, this provides a physical limitation on the ability to perform such calculations. However, in the past ten years, HPC has seen the development of accelerators
\citep{GR14,KWB10} with the generalisation of Graphical Process Units
(hereinafter GPU) on computing nodes. Indeed, in the latest TOP 500 fastest computers list published
in November 2021, 7 out of the 10 first supercomputers have GPU accelerators.
Developing applications for GPU accelerators is a challenge for many research
fields as it requires a large investment of resources and time \citep[see
\textit{e.g.}][]{LRB18}. Yet, time-explicit grid based MHD simulations can
be efficiently run on GPU-accelerators \citep{VSS11,NS14}, with a large time
gain (usually $> \times 10$), compared to their CPU counterparts. Although there
are a plethora of hydro-dynamical codes and even relativistic codes that have the capability of running
on GPU-based systems \citep[see \textit{e.g.}][]{SR15, SZG18}, to our knowledge there exist
only two general relativistic codes aimed at studying accretion disks with GPU capabilities: grim \citep{CFG17}, which uses the
library ArrowFire \citep{YAM15}, and H-AMR \citep{LHT18, CLT19} which is based
on the hierarchical use of MPI, OpenMP and CUDA. Both codes are not publicly available.

In this work, we present
a new GPU-based numerical solver for the GR-MHD equations in Kerr metric. Our code is developed based on the publicly available code HARM \citep{GMT03,NGM06}, albeit, of course, with many major modifications and improvements, which we therefore simply call cuHARM (cuda-HARM). In its current
form, cuHARM is designed to run a single multi-GPU workstation by splitting
the volume in equal shares between each GPU. This already allows to run
3D simulations with resolution around $\sim 192^3$, taking about 72 hours to reach
$t = 10^4$M on a Nvidia DGX-8xV100 server. 


In this paper, we first review the equations of
GRMHD and then explain their numerical discretization, emphasizing the GPU implementation in cuHARM.
We then present several runs, where we study ADAF disks in the SANE accretion regime. The simulations are  designed
to compare the influence of different initial conditions on the steady outcome.
In particular, we aim at studying the effects of 
different initial magnetization, different disk size and 
different gas adiabatic index on the accretion rate
and structure of the accretion disk as well as its accompanying magnetized jet. In addition, we 
present two simulations with initial conditions out of equilibrium to study the disk evolution under these conditions.


The paper is structured as follow. In Sections \ref{sec:GRMHD}, we review the
ideal GRMHD equations and their numerical discretization following the seminal
work of \cite{GMT03}. Section \ref{sec:cuHARM} introduce our GPU solver cuHARM.
Sections \ref{sec:numerical_setup} and \ref{sec:physical_model} respectively present our numerical setup and physical models. Our results are presented in Section \ref{sec:results}. The conclusion follows in section \ref{sec:discussion}.

\section{GR-MHD equations and their numerical solution}

\label{sec:GRMHD}

\subsection{Review of GRMHD formulation}

In this section, we review the ideal non-resistive GRMHD equations, their
numerical discretized version and the numerical techniques applied in cuHARM
to evolve them in time. Further details on the GRMHD equations and on their discretization
can be found in \citet{Ani90, MM03_MHD, Fon08, RZ13} and \citet{MM15}.
This system of equations describes the motion of ideal magnetized plasma in arbitrary fixed space time.
They characterise the time and space evolution of the gas properties: density $\rho$, internal energy density $u$, 4-velocity
$u^\mu$, gas pressure $p_g$, entropy $S = p_g/\rho^{\hat \gamma-1}$ and magnetic field. Here,
$\hat \gamma$ is the adiabatic index of the gas, whose value we take here to be either 4/3 or 5/3.

Let $T^{\mu \nu}$ be the stress energy tensor and $F^{\mu \nu}$ the Faraday tensor. The conservation
of mass, energy and momentum conservation as well as the homogeneous Maxwell's equation are written as
\begin{align}
    \nabla_\mu \left ( \rho u^\mu \right ) &= 0   \label{eq:mass_conservation}\\
    \nabla_\mu \left ( T^{\mu\nu}\right ) &= 0  \label{eq:energy_momentum_conservation}\\
    \nabla_\mu \left ( ^* F^{\mu \nu}\right ) &= 0  \label{eq:maxwell} 
\end{align}
where $^*F^{\mu\nu}$ is the dual of the Faraday tensor $F^{\mu\nu}$. The magnetic field
4-vector is defined as $b^\mu \equiv ~^*F^{\mu\nu} u_\nu$ (note that the anti-symmetry of the Faraday tensor ensures that the 4-vector magnetic field $b^\mu$ is orthogonal to the 4 velocity $u^\mu$). Under the assumption of ideal MHD, namely
$u_\mu F^{\mu \nu} = 0$, the dual of the Faraday tensor can be expressed as \citep{Lic67}
\beq
    ~^* F^{\mu \nu} = b^\mu u^\nu - b^\nu u^\mu,
    \label{eq:4}
\eeq
and the stress energy tensor is given by
\beq
    T^{\mu\nu} = (h + b^2) u^\mu u^\nu + \left (p_g + \frac{b^2}{2} \right )g^{\mu \nu} - b^\mu b^\nu.
\label{eq:5}
\eeq
Here, $ h = \rho + u + p_g$ is the enthalpy, $ b^2 = b^\mu b_\mu$ and $g^{\mu \nu}$
is the metric tensor. The system of equations is closed by the equation of state
of an ideal gas, $p_g = (\hat \gamma -1) u$. 

In order to solve the conservation equations \eqref{eq:mass_conservation}-\eqref{eq:maxwell} we express the
GRMHD  equations as follows. Following \cite{Kom99}, we write the components of the 4-vector magnetic field $b^\mu$ using
the 3-vector field, $B^{i} = {^*F}^{it}$. Using the orthogonality of $b^\mu$
and $u^\mu$, as well as $B^i = b^i u^0 - b^0 u^i$ (Equation \ref{eq:4}) one obtains 
$b^t = B^i u^\mu g_{i\mu}$ and $b^i = (B^i + b^t u^i)/u^t$. The conservation
equations \eqref{eq:mass_conservation}-\eqref{eq:maxwell} are expressed in
conservative form :
\begin{align}
     \partial_t \left ( \sqrt{-g} \rho u^t \right ) &= - \partial_i \left ( \sqrt{-g}  \rho u^i\right ), \label{eq:mass_conservation_cb}\\ 
     \partial_t \left ( \sqrt{-g} T^t_{~\nu} \right ) &= -\partial_i \left ( \sqrt{-g} T^i_{~\nu} \right ) + \sqrt{-g} T^\kappa{}_{\lambda} \Gamma^{\lambda}_{\nu \kappa}, \label{eq:energy_momentum_conservation_cb}\\
     \partial_t \left ( \sqrt{-g} B^i \right ) &= -\partial_j \left ( \sqrt{-g} \left [ b^j u^i - b^i u^j\right ]\right ). \label{eq:maxwell_cb}
\end{align}
where $g$ is the determinant of the metric tensor $g^{\mu \nu}$.
The (locally) divergence free condition of the magnetic field, $\vec \nabla \cdot \vec B = 0$ is written in the form 
\beq
    \partial_i\left ( \sqrt{-g} B^i\right ) = 0. 
    \label{eq:solenoid_constraint}
\eeq
Following \cite{GMT03}, we define the following vector of conserved quantities: $U = \sqrt{-g}(\rho u^t,
T^t_{~t}, T^{t}_{~i}, B^i)$, as well as a vector of primitive variables, $P  = (\rho,
u, \tilde u^i, B^i)$.  Following \citet{MG04} and \citet{NGM06}, in order to
enhance computational stability we use modified velocities $\tilde u^i$. These
are obtained from the four velocities $u^\mu$ and the shift $\beta^i = g^{ti} \alpha^2$ where $\alpha^2 = -1/g^{tt}$ is the square of the lapse, via 
\begin{align}
\tilde u^i \equiv u^i + \frac{\Gamma \beta^i}{\alpha},    
\label{eq:u_tilde}
\end{align}
with $\tilde u^0 = 0$. Here, the Lorentz factor is $\Gamma = \sqrt{1 + g_{ij} \tilde u ^i \tilde u^j}$.
This change of variable was found to increase numerical stability, since the corresponding 4-velocity $u^\mu$ is always a time-like vector. It has therefore now become a usual consideration in GRMHD code.

While equations \eqref{eq:mass_conservation_cb}-\eqref{eq:maxwell_cb} augmented with the equation
of state form a closed system, we follow \citet{NKH09,SNT13}, and we evolve the
equation of entropy conservation
\begin{align}
    \nabla_\mu \left ( S u^\mu\right ) &= 0 \label{eq:entropy_conservation},
\end{align}
in addition to the mass, energy and momentum conservation equations. 
We use this conservation law instead of the energy conservation equation in two cases. First,
it is used in highly magnetized regions as the energy equation is prone to numerical errors and failures in
case of large magnetic field. And second, it is used as a backup when the numerical scheme fails;
see Section \ref{sec:inversion_U_to_P} below for further details.

\subsection{Numerical discretization}

\label{sec:numerical_discretization}


In order to numerically solve Equations
\eqref{eq:mass_conservation_cb}-\eqref{eq:maxwell_cb}, we follow the finite
volume methodology. We first introduce a structured partition of space
in some coordinate system $(X^1, X^2, X^3)$, the full details of which will be described below.
In three spacial dimensions, each element of the partition is assumed to be a cube,
with 6 faces. For each cell, we label the center by integer numbers  and the faces by integers
plus half, such that $(X^{1}_I, X^{2}_J, X^{3}_K)$ is the center of
the cell indexed by $(I,J,K)$, while $(X^{1}_{I+1/2}, X^{2}_J, X^{3}_{K})$
is the center of the surface separating cells $(I,J,K)$ and $(I+1,J,K)$.\footnote{We use capital Latin letters to describe the cell coordinates, in order to avoid confusion with the [3-d] vector / tensor indices, which are marked with small Latin letters.} Hereinafter, indices $I$, $J$ and $K$ have their usual meaning, namely index $I$ is always associated to coordinate $X^1$, $J$ to $X^2$ and $K$ to $X^3$.

We integrate Equations \ref{eq:mass_conservation_cb}-\ref{eq:maxwell_cb} over the volume of each cell.
As an example for the continuity equation (\ref{eq:mass_conservation_cb}), after integrating over the volume one obtains 
\beq
    \int_{V}  \partial_t \left (\sqrt{-g} \rho u^t \right ) dX^1 dX^2 dX^3 = \int_{V}  \partial_i \left (\sqrt{-g} \rho u^i \right ) dX^1 dX^2 dX^3
    \label{eq:cont}
\eeq
As typical for Godunov type method, it is further assumed that the value of each primitive quantity (a component of the vector $P$) and conserved quantity (a component of the vector $U$) are uniform within the entire volume of each cell. This approximation is denoted here as a bared
quantity: for instance, the density is written as $\bar \rho$. 
Using this assumption for the four velocity $u^\mu = \bar {u}^\mu$ and the density
$\rho = \bar {\rho}$ in Equation \ref{eq:cont} and applying the divergence theorem, one finds 
\begin{align}
    \partial_t \left ( \sqrt{-g} \bar \rho \bar{u}^t \right ) \int_{V} dX^1 dX^2 dX^3 =& \bar F_{I-\frac{1}{2}} \int_{\partial V_{I-\frac{1}{2}}}  dX^2 dX^3 - \bar F_{I+\frac{1}{2}} \int_{\partial V_{I+\frac{1}{2}}} dX^2 dX^3
      \nonumber \\ 
    + & \bar F_{J-\frac{1}{2}} \int_{\partial V_{J-\frac{1}{2}}}   dX^1 dX^3 - \bar F_{J+\frac{1}{2}} \int_{\partial V_{J+\frac{1}{2}}} dX^1 dX^3 \label{eq:discretizes_eqaution_fluxes} \\
    + & \bar F_{K-\frac{1}{2}} \int_{\partial V_{K-\frac{1}{2}}}   dX^1 dX^2 - \bar F_{K+\frac{1}{2}} \int_{\partial V_{K+\frac{1}{2}}} dX^1 dX^2 \nonumber
\end{align}
where $\bar F_{M-(1/2)}$ $(M \in {I,J,K})$ is an approximation of the flux $\sqrt{-g} \rho u^M$ in the $M^{\rm th}$ direction (see below), calculated at the center of the surface $\partial V_{M-(1/2)}$ between cells of indices $M$
and $M-1$. Here and below, for keeping light notations, we absorbed
the complete position reference of a cell surface to a single changing index: e.g.,
$\bar F_{I-(1/2)}$ actually represents $\bar F_{I-(1/2), J, K}$. Furthermore, in the remaining of the manuscript, $\bar F$ represents the fluxes of all conserved variables indistinctivly; e.g., when calculating the fluxes of the conserved variables $U^i$, $\bar F(U^i)$, we omit to write the conserved variable $U^i$.

The same technique is used for the other conservation equations. It only remains to specify the
treatment of the source term due to the appearance of the metric connections in the energy
and momentum conservation equation. We use the following approximation
\begin{align}
    \int_{V} \sqrt{-g} T^\kappa_{~\lambda} \Gamma^{\lambda}_{~\nu \kappa} dX^1 dX^2 dX^3 \sim  \left ( \sqrt{-\bar g} \bar T^\kappa_{~\lambda}  \bar \Gamma^{\lambda}_{~\nu \kappa} \right )_{I,J,K} \int_{V}  dX^1 dX^2 dX^3 
\end{align}
where the metric connection and determinant assume their value at the center of
each cell. Note that this discretization is somehow different from that presented in \cite{POM17}
who keeps the metric terms inside the volume and surface integrals.

The fluxes are obtained as follow. We adopt the original method used in HARM \citep{GMT03} and
use a MUSCL scheme to define the numerical fluxes \citep{vLe79}. As a first step, the
values of the primitive variables are reconstructed at the cell boundaries by the limited
piecewise linear interpolation method (hereinafter PLM). This method requires the calculation of the gradient of each variable. 
The gradient $dq_I$  in cell $I$ is
computed via the monotized central (MC) limiter, using the values of the primitive variables $q$ in cells $I-1$, $I$ and $I+1$ :
\begin{align}
    dq_I = \left \{
    \begin{aligned}
    &0 & ~~~~~ & {\rm if~~ } (q_{I} - q_{I-1})(q_{I+1} - q_{I}) < 0 \\
    &2(q_{I} - q_{I-1}) & & {\rm else~if~~ } |q_{I} - q_{I-1}| < |q_{I+1} - q_{I}| {\rm ~~ and ~~ } 2|q_{I} - q_{I-1}| < \frac{1}{2}|q_{I+1} - q_{I-1}| \\
    &2(q_{I+1} - q_{I}) & & {\rm else~if~~ } 2|q_{I+1} - q_{I}| < \frac{1}{2}|q_{I+1} - q_{I-1}| \\
    &\frac{1}{2}|q_{I+1} - q_{I-1}| & & {\rm otherwise.}
    \end{aligned}
    \right.
\end{align}
This gradient is then used to compute the primitives on the left and right
of the cell boundaries as
\begin{align} 
\begin{aligned}
    q_{I + \frac{1}{2}}^l &= q_I + \frac{1}{2} dq_I  \\
    q_{I + \frac{1}{2}}^r &= q_{I+1} - \frac{1}{2} dq_{I+1}
\end{aligned}
\end{align}
where from hereon, the $l$ and $r$ superscripts refer to the left and 
right sides of the boundary $I+(1/2)$ respectively. It is well-known that piecewise parabolic
method \citep[hereinafter PPM, ][]{CW84,MM96, MPB05} gives more accurate results and better
convergence order than PLM for smooth flows, and we are planing to include it in the next
version of our code. Currently, we use the PLM reconstruction for its implementation simplicity, and we 
discuss below our results in the light of this approximation.

As a second step, the right and left values of the primitive variables at the cell
boundaries are used to compute the corresponding conserved variables $U^r$ and $U^l$,  as well as
the fluxes $F^r$ and $F^l$ (all these are directly obtained - see their definitions above).
Then, each flux through a given boundary is expressed using the Lax-Friedrichs method: 
\begin{align}
    \bar F_{I + \frac{1}{2}} = \frac{1}{2} \left [ F^l_{I+1/2} + F^r_{I+1/2} - c_w \left ( U^r_{I + 1/2 } - U^l_{I + 1/2}  \right )\right ].
\label{eq:fluxes_riemann}
\end{align}
Here, $c_w$ is a characteristic wave speed, which is computed following the prescription of \cite{GMT03} by solving the dispersion relation. In a relativistic magnetized plasma, the dispersion relation is a quartic equation \citep{Ani90},
which is numerically expensive and cumbersome to solve. We therefore resort to the simpler
dispersion relation given by Equation (28) of \citet{GMT03}, which is a second order polynomial
equation \citep[see further discussion in][]{POM17}:
\begin{align}
    \omega^2 = \left [ v_a^2 +c_s ^2 \left( 1 - {v_a^2}\right )\right ]k^2
    \label{eq:simplified_wave_equation}
\end{align}
Here, $v_a^2 = b^2/(b^2 +
\rho + u +p_g) $ is the Alfven speed (normalized to the speed of light, $c$) and the sound speed is given by
\begin{align}
    c_s^2 = \left ( \frac{\partial (\rho + u)}{\partial p}\right )^{-1} = \frac{\hat \gamma p_g}{\rho+ u + p_g}.
\end{align}
The dispersion relation, given by Equation \eqref{eq:simplified_wave_equation} is solved for waves propagating in
each direction separately. As an example, in the $X^1$ direction the wave-vector takes the form
$k_\mu = (-\omega, k_1, 0, 0)$, where $\omega$ is the wave frequency and $k_1$ is the
wave-number of a wave propagating in the $X^1$ direction. The wave-speeds in the direction $X^1$
are then given by $\omega / k_1$, where $k_1$ admits up to two values (for the simplified dispersion relation in Equation \ref{eq:simplified_wave_equation}).
Writing $\omega = k_\mu u^\mu$, $k^2 = K_\mu K^\mu$, where $K_\mu = (g_{\mu \nu} + u_\mu u_\nu) k^\nu $,
Equation \eqref{eq:simplified_wave_equation} can be solved as a quadratic equation in $\omega/k_1$,
which then posses two solutions: $c_1$ and $c_2 < c_1$, associated with two wave-speeds.
To obtain $c_w$ in Equation \eqref{eq:fluxes_riemann}, namely, at the boundary between the neighbouring cells, the dispersion relation is solved
two times, on the right and left of each boundary, using the values of the primitive variables $q^l$ and $q^r$ in the neighbouring cells. Finally, introducing,
$c_{\rm max} = |\max(0, c_1^r, c_1^l)|$ and $c_{\rm min} = |\max(0, -c_2^r, -c_2^l)|$, we define
$c_w \equiv \max(c_{\rm max}, c_{\rm min})$ \citep[see][for further details]{vLe79}. Using the simplified dispersion
relation in Equation \ref{eq:simplified_wave_equation} makes the scheme more diffusive at most by
a factor 2 than if the accurate dispersion relation was considered \citep{GMT03}.


In the current version of the code, we have also implemented the HLL fluxes \citep{HLL83}
\begin{align}
    \bar F_{I + \frac{1}{2}} =  \frac{ c_{\rm max} F^l_{I+1/2} + c_{\rm min} F^r_{I+1/2} - c_{\rm max} c_{\rm min} \left ( U^r_{I + 1/2 } - U^l_{I + 1/2}  \right ) } {c_{\rm max} + c_{\rm min} }, \label{eq:fluxes_HLL}
\end{align}
but it is not used in this study, as we found that the Lax-Friedrichs method provide less numerical failures for the setups we tested (see Section \ref{sec:physical_model} below). The computation of the fluxes is repeated in all three directions, such that all terms
in Equation \ref{eq:discretizes_eqaution_fluxes} are calculated. 

A critical aspect of (GR-)MHD codes is their ability to satisfy the solenoid
condition given by Equation \eqref{eq:solenoid_constraint}. The two
main methods discussed in the literature are the divergence cleaning \citep{DKK02} and the constrained
transport \citep{EH88, Tot00}. The former involves modifying the Faraday
equation such that the divergence errors are transported to the boundaries
and eventually dumped. The later relies on modification of the fluxes computed
as in Equation \eqref{eq:fluxes_riemann}, such that the divergence free condition
is exactly satisfied locally. A detailed comparison between the two methods can
be found in \cite{BK04,ZF16}, which conclude in the superiority of the constrained
transport scheme. Here, we implemented the constrained transport (CT) algorithm using the 'flux-CT' method of \citet{Tot00} and
\citet{GMT03} since it requires only cell centered values of the magnetic field,
although we are aware of its limitation \citep[see, e.g., ][]{GS05,DZB07,OPD19}.

The implementation is done by modifying the fluxes in the  following way. We introduce the variables $E^1$, $E^2$ and $E^3$ representing
the components of the electric field at the 12 edges of each cell, e.g., at $(I, J+1/2, K+1/2)$. The values of these variables are computed using (i)  the definition of the 
Faraday tensor, and (ii) the conservation Equation \eqref{eq:maxwell_cb} integrated over each cell surface. As a first step, we compute the electric fields at each face center, expressed in terms of the fluxes $\bar F = \bar F(B^i)$ obtained using Equation \ref{eq:fluxes_riemann} above.  The face centered values are averaged for all
four faces associated to the considered edge.
For instance, the electric field $E^1$ on the edge $(I, J+1/2, K+1/2)$ is given by 
\begin{align}
    E^1_{I,J+1/2,K+1/2}  &= \frac{1}{4} \left [  \bar F^2_{I, J+\frac{1}{2}, K+1}(B^3) + \bar F^2_{I,J+\frac{1}{2},K}(B^3)  - \bar F^3_{I,J+1,K+\frac{1}{2}}(B^2) - \bar F^3_{I,J,K+\frac{1}{2}}(B^2)\right ].
\end{align}
Then, the fluxes of the magnetic field components $B^i$ calculated in Equation \ref{eq:fluxes_riemann} above are replaced by 
\begin{align}
    \left \{  \begin{aligned}
    \tilde{F}_{ I+ \frac{1}{2}, J, K }^1(B^1) &= 0 \\
    \tilde{F}_{I+ \frac{1}{2}, J, K }^1(B^2) &= \frac{1}{2} \left [  E^3_{I+\oh,J+\oh,K} + E^3_{I+\oh,J+\frac{3}{2},K} \right ] \\
    \tilde{F}_{ I+ \frac{1}{2}, J, K }^1(B^3) &= -\frac{1}{2} \left [  E^2_{I+\oh,J,K+\oh} + E^2_{I+\oh,J,K+\frac{3}{2}} \right ] 
    \end{aligned} \label{eq:fluxCT_F1} \right. 
\end{align}
where  it was assumed that the grid has uniform spacing in each direction. Similar formulae are used
in calculating the fluxes in the other two directions. This expression of the flux preserves the following numerical
approximation of the divergence, to machine accuracy :
\begin{align}
    \left(\nabla \cdot B\right)_{I+\oh, J+\oh, K+\oh} \propto \sum_{l_1, l_2 l_3 = 0, 1 } \frac{(-1)^{1+l_1} \bar B^1_{I+l_1, J+l_2, K+l_3}}{2\Delta x^1}  + \frac{(-1)^{1+l_2} \bar B^2_{I+l_1, J+l_2, K+l_3}}{2 \Delta x^2}  + \frac{(-1)^{1+l_3} \bar B^3_{I+l_1, J+l_2, K+l_3}}{2\Delta x^3}, 
\end{align}
where the $\Delta x^i$ is the grid spacing in direction $i$. Note that this formula implicitly assume a uniform grid. A detailed derivation of these expressions and a discussion on non-uniform grid can be found in Appendix C of \citet{POM17}.

It is well-known that the flux-CT approach used here has several drawbacks, mainly that the stencil is large
and that this technique does not reduce to the proper limit in 1D flow calculations. It is also known to
lack upwinding. Although several methods have been developed to overcome those limitations \citep[see, e.g.,][]{GS05,DZB07, OPD19}, we have elected to use this method for its simplicity.
In a future upgrade, we will consider staggered magnetic field and more advance formulation
of the constrained transport algorithm.

Once all the fluxes are computed from the primitive variable, the time evolution can
be performed. Introducing the surface area
\begin{align}
    \mathcal{S}^1_{I-\frac{1}{2}} = \int_{\partial V_{I-\frac{1}{2}}}  dX^2 dX^3 
\end{align}
($\mathcal{S}^2~,~\mathcal{S}^3$ are obtained by permutation of the indices), the conserved quantities $\bar U$ are evolved in time by second order
time stepping method 
\begin{align}
    \hat U_{I,J,K}^{t + \frac{1}{2} \Delta t} &=  U_{I,J,K}^{t} + \frac{\Delta t}{2 V_{I,J,K}} \left [
        \mathcal{S}^1_{I-\frac{1}{2}} \bar F^1_{I - \frac{1}{2} } - \mathcal{S}^1_{I + \frac{1}{2}} \bar F^1_{I + \frac{1}{2} }
     +  \mathcal{S}^2_{J-\frac{1}{2}} \bar F^2_{J - \frac{1}{2} } - \mathcal{S}^2_{J+\frac{1}{2}} \bar     F^2_{J + \frac{1}{2} }
     +  \mathcal{S}^3_{K-\frac{1}{2}} \bar F^3_{K - \frac{1}{2} } - \mathcal{S}^3_{K+\frac{1}{2}} \bar     F^3_{K + \frac{1}{2} }\right ], \nonumber \\
    U_{I,J,K}^{t+ \Delta t} &=  U_{I,J,K}^{t} + \frac{\Delta t}{V_{I,J,K}} \left [
        \mathcal{S}^1_{I-\frac{1}{2}} \hat F^1_{I - \frac{1}{2} } - \mathcal{S}^1_{I + \frac{1}{2}} \hat F^1_{I + \frac{1}{2} }
     +  \mathcal{S}^2_{J-\frac{1}{2}} \hat F^2_{J - \frac{1}{2} } - \mathcal{S}^2_{J+\frac{1}{2}} \hat     F^2_{J + \frac{1}{2} }
     +  \mathcal{S}^3_{K-\frac{1}{2}} \hat F^3_{K - \frac{1}{2} } - \mathcal{S}^3_{K+\frac{1}{2}} \hat     F^3_{K + \frac{1}{2} }\right ]
\end{align}
where the terms $\hat F$ are computed similarly to $\bar F$, using the values at half time step,
$\hat U^{t + \Delta t/2}$.
In addition, the fluxes of the magnetic fields are replaced by their
modified expression for the flux-CT method given by Equations
\eqref{eq:fluxCT_F1}. The surface variable $\mathcal{S}^1_{I-\frac{1}{2}}$ should
not be confused with the entropy $S$. 

In order to maintain the stability of the numerical scheme, the time step $\Delta t$ is varied at each
iterations. We consider the time it takes for a wave propagating at speed $c_w$ to cross a
cell of size dx. Therefore, for each cells and a given direction $l$, we consider the minimum of
$dt^l = \min(\xi \Delta x^l /c_w)$ where $\xi$ is a numerical coefficient which we set to $\xi = 0.45$. We found that this value provides good stability for our current setups.
The next time step is then obtained as 
\begin{align}
   ( \Delta t )^{-1} = \sum {(dt^l)}^{-1}.
\end{align}

\subsection{Conserved to primitive inversion techniques and fixing strategy}

\label{sec:inversion_U_to_P}

In this section only, we simplify the notations by removing the barred symbol representing the constant approximation of any variable inside a given cell.
Once the conserved quantities are advanced in time, the primitive variables need to be recovered. This stage requires the numerical solution of a non-linear system and is central to any numerical code solving the magneto-hydrodynamics equations. In current literature, there exist many methods to perform this inversion \citep[e.g.,][for a partial list]{NGM06, MM07, SMD18, KKC21}. Recently, \citet{DCB21} used machine learning to speed up the recovery. Here, we use three inversion methods: (i) the two dimensional (2D) method solving for $v^2$ and the modified enthalpy $W \equiv \Gamma^2 h = \Gamma^2 (\rho + u + p_{\rm g})$; (ii) the one dimensional (1D) method solving for $W$ only \citep{NGM06};  and (iii) a 1D method using the entropy equation \eqref{eq:entropy_conservation} in place of the energy equation \citep{NKH09,SNT13}. The 2D numerical system is solved by Newton-Raphson, while the 1D equations are solved by the Brent method.

In order to establish the link between primitive and conserved variables we follow \citet{NGM06}. Considering the normal observer's 4-velocity $n_\mu = (-\alpha, 0, 0, 0)$, it is convenient to define the following two 4-vectors:
\begin{align}
\begin{aligned}
    \mathcal{Q}_{\mu} &= -n_\nu T^{\nu}_{~\mu}, \\
    \tilde {\mathcal{Q}}^{\mu} &= j^\mu_{~\nu} \mathcal{Q}^{\nu}. 
\end{aligned}
\end{align}
Here, $j_{\mu \nu} = h_{\mu \nu} + n_\mu n_\nu$ is projection normal to the normal
observer and $h_{\mu \nu} = g_{\mu\nu} + u_\mu u_\nu$ is a projection normal to
the fluid velocity. Since $T^\mu_{~\nu}$ are conserved variables, $\mathcal{Q}^{\mu}$ and $\tilde{\mathcal{Q}}^{\mu}$ can be directly computed. Note that $(\tilde{\mathcal{Q}})^{2}= \tilde{\mathcal{Q}}^\mu \tilde{\mathcal{Q}}_\mu$ is independent on $\tilde{\mathcal{Q}}^{0}$, and therefore computing its value does not require the use of the energy conservation equation (which is part of Equation \ref{eq:energy_momentum_conservation}). Further introducing the 4-vector 
\begin{align}
    \mathfrak{B}^\mu = -n_\nu {^*F}^{\mu \nu},
\end{align}
the system of equations linking conservative to primitive variables can be reduced to \citep{NGM06}:
\beq
    \tilde {\mathcal{Q}}^2 = v^2 \left ( \mathfrak{B}^2 + W \right )^2 - \frac{\left(\mathcal{Q}_\mu \mathfrak{B}^\mu\right)^2 \left ( \mathfrak{B}^2 + 2W \right )}{W^2}, \label{eq:eq_1_inversion_Utop}
\eeq
\beq
\mathcal{Q}_\mu n^\mu = \frac{\mathfrak{B}^2}{2} \left ( 1 + v^2 \right ) + \frac{ \left(\mathcal{Q}_\mu \mathfrak{B}^\mu\right )^2}{2 W^2} - W + p_g(u,\rho), \label{eq:eq_2_inversion_Utop}
\eeq
where $ v^2 \equiv 1 - 1/\Gamma^2$. In this system of equations, the only two unknowns are $v^2 < 1$ and $W$, as the pressure is calculated using the equation of state, 
\begin{align}
    p_g =\left(\frac{\hat \gamma -1}{\hat \gamma}\right) \left [ W \left (1-v^2 \right ) - D\sqrt{1-v^2}\right ] 
    \label{eq:p_of_W}
\end{align}
where $D \equiv \Gamma \rho = \rho u^t$ is one of the conserved variables. For the 1D and 2D methods, (i) and (ii), equations \ref{eq:eq_1_inversion_Utop} and \ref{eq:eq_2_inversion_Utop} are solved to calculate the values of $v^2$ and $W$.

To compute all the terms in Equation \ref{eq:eq_2_inversion_Utop}, one needs to know the evolution of $T^{00}$, namely, solve the energy conservation equation (which is part of Equation 2). In the case of inversion failure, or in highly magnetized regions, the energy conservation equation is replaced by the entropy conservation equation (11). In this case, using the definitions of the entropy, $S$, $W$ and $D$, Equation \ref{eq:eq_2_inversion_Utop} is replaced by    
\begin{align}
    W = \Gamma D + \Gamma \frac{\hat \gamma}{\hat \gamma -1 } \left ( \frac{D}{\Gamma} \right )^{\hat \gamma -1} S \label{eq:W_of_S}
\end{align}

To recover the primitive variables, we proceed as follows. Firstly, Equations
\eqref{eq:eq_1_inversion_Utop} and \eqref{eq:eq_2_inversion_Utop} are solved by a
Newton-Raphson method to an accuracy of $10^{-10}$. If the root solver succeeds
and if the recovered values are physical ($\rho,p_g > 0$, $v^2 < 1$), the results
are accepted. If the numerical solution results in non-physical values of the density, pressure or velocity, we implement the entropy fix described below.  
If the numerical method fails to produce results, 
we resort to analytically expressing $v^2$ from Equation \eqref{eq:eq_2_inversion_Utop},
use its expression in Equation \eqref{eq:eq_1_inversion_Utop} which is then numerically
solved using the Brent method. 
If this method succeed, the values are accepted. Otherwise,
we use the entropy fix method: the expression of $W$ given by Equation \eqref{eq:W_of_S} is
directly used in Equation \eqref{eq:eq_1_inversion_Utop}, which is solved for $v^2$ via
the Brent method. 

In highly magnetized regions, defined by 
\begin{align}
    \beta = \frac{p_g}{p_b} < 10^{-2}, 
\end{align}
we apply the entropy fixed method directly (as a first and only choice). 
Numerically, the gas and magnetic pressures, $p_g$ and $p_b = b^2 /2$ are computed from the primitive variables at the
previous time step. This is similar to the approach used in BHAC \citep{POM17}.

There are two main reasons for the inversion to fail in a given cell. First the conserved variables can
be in a non-physical state, with $D < 0$ or $S < 0$, which prevents the use of any of the aforementioned  inversion methods. Alternatively, the recovered
primitive variables can also obtain non-physical values, namely $\rho <0 $, $u < 0$, $v^2 > 1$ or
$v^2 < 0$. In all these cases, we update the failed cells by averaging all the primitive 
but the components of the  magnetic field from the neighbouring cells. We consider two types of averaging procedure: 
\begin{align}
\begin{aligned}
     p_{I,J,K} &= \frac{1}{4} \left [ p_{(I+1),J,K} + p_{(I-1),J,K} + p_{I,(J+1),K} + p_{I,(J-1),K}   \right ]  \\
    {\rm and ~~~~~} p_{I,J,K} &= \frac{1}{4} \left [ p_{(I+1),(J+1),K} + p_{(I+1),(J-1),K} + p_{(I-1),(J+1),K} + p_{(I-1),(J-1),K}   \right ]
\end{aligned}
\end{align}
where the neighbouring cells used in the averaging procedure are cells for which the conserved to primitive
inversion was successful or which were already updated by the averaging procedure. We iterate the
domain several times until all cells that need fixing are fixed, although in practice
the number of failed cells is small and very rarely involved neighboring cells.


\subsection{Flooring model}

It is well known that grid methods for the solution of GR MHD equations are
vulnerable to numerical errors when treating low density medium. Therefore,
numerical floors are used for the density and the internal energy density.
Many choices of floors have been proposed. For the results presented in this
paper, we utilise the density floor suggested in \cite{PCN19}, but use somewhat 
higher floors for the internal energy density $u$:
\begin{align}
\begin{aligned}
    \rho &= \max \left ( \rho, ~ 10^{-20}, ~10^{-5} {r}^{-\frac{1}{2}}   \right ), \\
    u &= \max \left ( u, ~10^{-20}, ~\frac{10^{-5}}{3}  r^{-\frac{5}{2}} \right ),
\end{aligned}
\end{align}
where the radius $r$ is normalized to the gravitational radius, $r_g$. 
In the highly magnetized region, we also limit the ratio of magnetic energy
density to internal energy density and to rest mass energy density by adding
mass in the frame of the zero angular momentum observer \citep[e.g.][]{MTB12,RTQ17}.
The considered limits are 
\begin{align}
    \frac{b^2}{\rho} &< 50 \\
    \frac{b^2}{u} &< 2.5 \times 10^3
\end{align}
Finally, we also limit the Lorentz factor to be smaller than 50. When the Lorentz
factor is larger than 50, the velocities are re-scaled  such that the maximum
Lorentz factor be equal to $\Gamma_{\rm max} = 50$. This is required for the stability of our numerical scheme, although for the problems studied here, this situation nearly does not occur. We note that some inversion strategies, e.g., \citet{MM07, KKC21} can overcome this problem but are not necessary for the current setup.

\section{cuHARM: numerical implementation}

\label{sec:cuHARM}


State of the art 3D GR-MHD simulations are numerically very expensive. A
natural solution to improve wall clock time is to use hardware accelerators
such as GPUs. There already exists numerical solvers for the GRMHD equations
running on GPUs. For example, \citet{CFG17} uses the library ArrowFire
to allow the code \textit{grim} to run on nodes with different architectures.
Another code using GPU is H-AMR \citep{LHT18}, which uses hierarchical cuda,
openMP and MPI to achieve computation on NVIDIA GPU accelerators on multiple
nodes.

We have designed and implemented a multi-GPU solver for the GRMHD equtions
based on the numerical scheme from the original HARM code \citep{GMT03, NGM06}. In our implementation
using CUDA-C, nearly all computations are done on the GPUs, with the CPU
only dealing with exporting the data when required and the data transfer
between GPU devices. In fact, our multi-GPU implementation is simple: each
GPU is assigned to a specific openMP thread and computes a pre-determined slice of
the total grid \citep{VSS11}. Currently, the split is made in the $\theta$
direction, such that each GPU deals with the same number of cells. 

In its current version, the code can only be ran on a single node with arbitrary number
of GPUs. The simulations presented in this paper were made on a Nvidia
DGX-V100 server with 8 V100 GPUs, each with dual Nvlink-2 setup, enabling
 efficient data transfers between neighbouring GPUs without utilizing the
host memory as a bridge. We are shortly planing for an extension to allow the code
to run on several nodes with accelerators.

The code consists of two main kernels: one computing the fluxes
and the second one updating the conserved variables and performing the inversion
between them and the primitive variables. A direct and naive implementation
of those kernels leads to straightforward compute ($\sim 70$\%) and memory
($\sim 60$\%) throughput on a RTX gaming GPU. However, the computation and
memory efficiencies substantially drop on HPC cards like the V100. Detailed
compute and memory profiles show global memory bottlenecks limiting the
computation speed and efficiency. This issue is partially solved by mostly using
three advance CUDA features. Firstly, large amount of shared memory are used
and reused to (i) reduce the number of global memory calls, and (ii) align the
memory calls such that shared memory access is coalesced. Secondly, warp-level primitives
are used to exchange data between threads in a warp and avoid multiple memory loading
when performing flux limiting reconstruction at the cell interfaces. Finally, 
computations requiring the metric tensor are always performed in the azimuthal
direction. Indeed, since the metric tensor does not depend on the azimuth angle $\phi$,
the memory can be loaded one time for all cells with the same $r$ and $\theta$ and the values shared by the whole
block, thereby reducing the required memory throughput. All in all, those techniques
allow to reduce the computation time by a factor $\sim 4$ compared to the naive
version. However, the computation and memory throughput on HPC cards still remain at the order of $\sim 10-15\%$.


With the modern GPU workstation we used, the memory was not a bottleneck:
the largest resolution runs (256x256x128) presented in this paper required
$\sim 35$Gb of ram memory, both on the CPU and GPU, while the GPU memory
available to us was $260$Gb. The runs presented in this paper takes from 2
days to 10 days to complete.

 







\section{Numerical setup}
\label{sec:numerical_setup}
In order to i) ensure the reliability of our code and ii) to study the structure of accretion disks in the SANE regime,

\subsection{Metric, numerical grid and boundary conditions}

The simulations are performed using the modified horizon penetrating Kerr-Schild (KS) coordinate system $(t,r,\theta,\phi)$, which describes the space around a rotating black-hole. Setting the BH mass as well  as $c=G=1$, its line element is
\begin{align}
    ds^2 = &  -\left ( 1 - \frac{2r}{\rho^2} \right )dt^2 + \left ( \frac{4r}{\rho^2} \right ) dr dt + \left ( 1 + \frac{2r}{\rho^2}\right ) dr^2 + \rho^2 d\theta^2 + \sin^2(\theta) \left [ \rho^2 + a^2 \left ( 1 + \frac{2r}{\rho^2} \right ) \sin^2(\theta)\right ]d\phi^2 \nonumber \\ 
    & - \left ( \frac{4ar \sin^2(\theta)} {\rho^2}\right ) d\phi dt - 2a\left ( 1 + \frac{2r}{\rho^2} \right ) \sin^2(\theta) dr d\phi,
\end{align}
where $-1 \leq a \leq 1$ is the normalized BH spin, and $\rho^2 = r^2 + a^2 \cos^2(\theta)$.

In our numerical experiments presented below, we have used the modified KS coordinate system with grid refinement towards the equator \citep{GMT03}. The GR-MHD equations are solved on the modified KS coordinate system $(t, X,Y,Z)$ whose spatial coordinates are linked to the spherical coordinates $(r, \theta, \phi)$ by \footnote{\url{https://github.com/atchekho/harmpi}}
\begin{align}
    r &= \left \{
    \begin{aligned} 
    &R_0 + \exp\left ( X\right ) &  &X < x_b \\
    &R_0 + \exp\left ( X +  \left (X-x_b \right )^{n_X} \right ) & ~~~~~~ &X \geq x_b,
    \end{aligned} \right. \nonumber \\
    \theta &= \frac{\pi}{2}(1 + Y ) + \frac{1-h}{2} \sin \left (\frac{\pi(1+Y)}{2}  \right ), \nonumber \\
    \phi &= Z, 
\end{align}
where the numerical values of the parameters for the simulations presented herein are $h = 0.3$, $n_X = 4$, $R_0 = 0$. The parameter $x_b$ is chosen such that the transition from exponential to hyper-exponential happens at radius $r = 400$M. For the size of the disk studied here, this is sufficient to ensure that the full disk is properly resolved. The outer grid boundary is at $r = 5 \times 10^3$M.

\begin{figure}
    \centering
    \includegraphics[width = 0.3\textwidth]{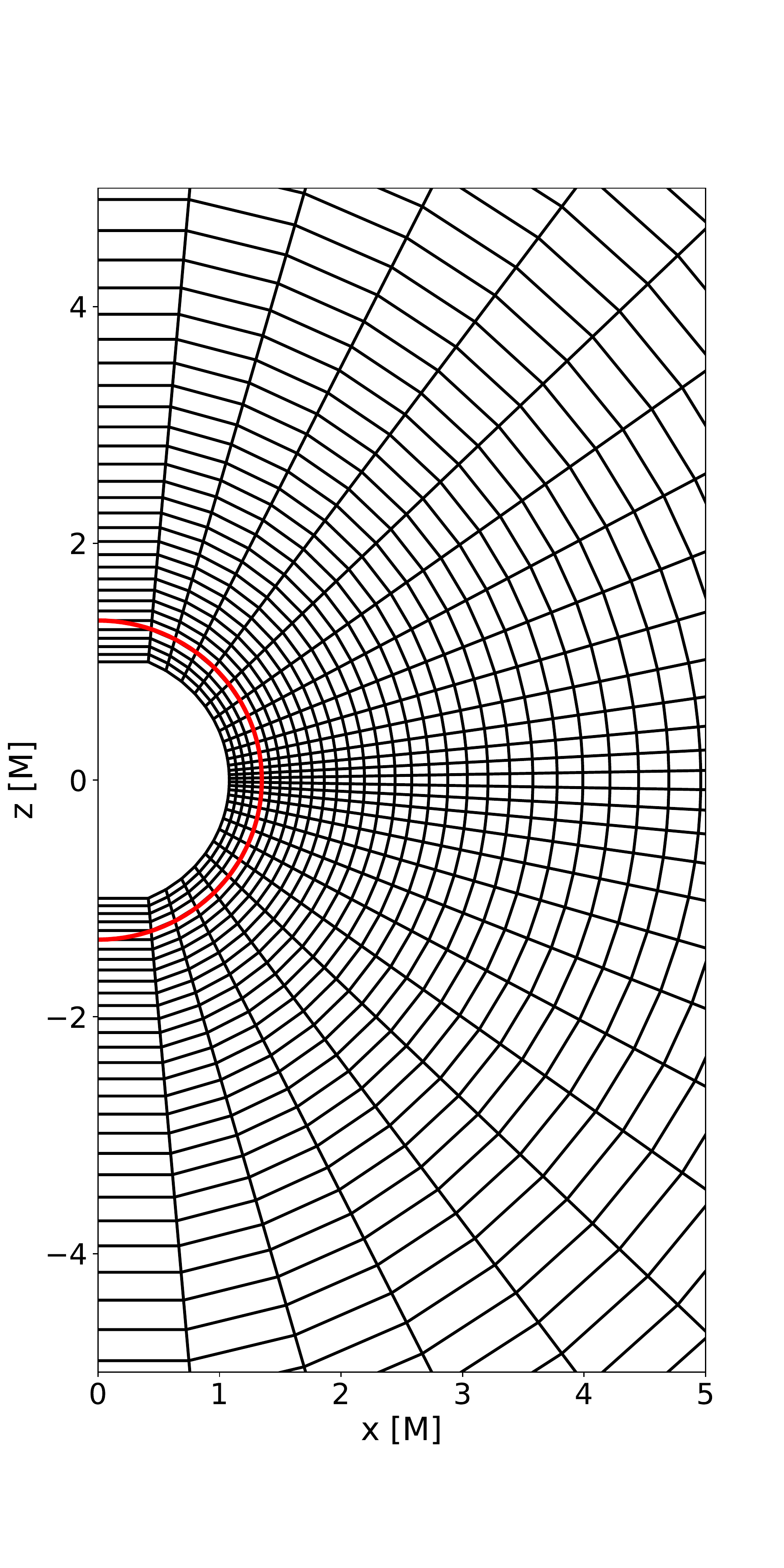}
    \caption{Numerical grid for the computation presented in this paper. Here the resolution has been reduced to $N_r = 128$ and $N_\theta = 32$ for clarity of the figure. The red line represents the outer horizon $r_h$. }
    \label{fig:grid}
\end{figure}

This coordinate system is tailored to resolve the disk by increasing the number of grid cells at the equator. This choice of the angular distribution of cells in $\theta$ increases the aspect ratio of cells close to the $\theta=(0, \pi)$ poles, therefore reducing the stability requirement on the time step. Yet, we find that in order to further increase the stability, a second transformation is required. Therefore, the cells closest to the polar boundary are ''cylindrified'' \citep{TNM11}. This transformation also allows for a faster computation by reducing the Courant condition on the time-step at the expense of de-resolving the pole closest to the black-hole. An example of grid used in this paper (albeit with a lower resolution, for demonstration purpose) is shown in Figure \ref{fig:grid}. The concentration of grid cells towards the equator and the "cylindrification" at both poles are clearly visible.

We use inflow boundary conditions in the radial direction, and periodic boundary conditions for the azymuthal boundary ($\phi$). For the polar boundary ($\theta$), reflective boundary conditions are used at the pole: the direction of the poloidal component of the velocity and the magnetic field is flipped at the pole ($u^2 \rightarrow -u^2$ and $B^2 \rightarrow -B^2$). Moreover to increase the numerical stability in the polar region, the polar component of the 4-velocity in the two cells closest to the pole is modified, by an interpolation to zero at the pole.

\subsection{Physical diagnostics}

To ensure that the MRI is properly resolved and that our numerical resolution is adequate, for each run, we first compute the MRI quality factor $Q^{(i)}$ in all three directions ($i = r, \theta, \phi$). These quality factors are an estimate of the number of cells available to resolve the fastest growing MRI mode. They are given by  
\begin{align}
Q^{(i)} &= \frac{\lambda^{(i)}}{\Delta x^{(i)}},
\end{align}
where
\begin{align}
    \lambda^{(i)} = \frac{2\pi}{\Omega \sqrt{\rho h + b^2}} b^\mu e_\mu^{(i)}, &~~~~&
\Delta x^{(i)} = [\Delta X(i)]^\mu e_\mu^{(i)}, &~~~~&
\Omega = \frac{u^\phi}{u^t}. 
\end{align}
Here, $\Delta X(i=1) = (0, \Delta r, 0,0 )$, and similarly for $i=2,3$. In the definitions of $\Delta X(i)$, $\Delta r$, $\Delta \theta$ and $\Delta \phi$ are the local grid spacing. 
The tetrads $e_\mu^{(i)}$ along the locally non-rotating reference frame are given by Equations (11)-(14) of \cite{Tak08}. Table \ref{tab:tab_run_caracteristics} gives the quality factor averaged over the disk area and over time, after the initial transition. Specifically, we define 
\begin{align}
    \langle Q^{(i)} \rangle &= \int_{t = 5\times 10^3 {\rm M}}^{t = 10^4 {\rm M}} dt^{'} \int_{\theta = \pi/3}^{\theta = 2\pi/3} d\theta \int_{\phi = 0}^{2\pi} d\phi \sqrt{-g} Q^{(i)}. \label{eq:volume_average_q}
\end{align}
The integration limits for the polar angle $\theta$ are chosen such to include only the disk region. We comment on the values of the MRI quality factor in the discussion.
For each runs, we compute the following diagnostics at the horizon : 
\begin{itemize}
    \item Mass accretion rate
    \begin{align}
        \dot M = \int_{\theta = 0}^{\pi} \int_{\phi = 0} ^{2\pi} \sqrt{-g} \rho u^r d\theta d\phi. \label{eq:diag_Mdot}
    \end{align}
    \item Magnetic flux threading the horizon 
    \begin{align}
        \phi_B &=  \frac{1}{2} \int_{\theta = 0}^{\pi} \int_0^{2\pi} \sqrt{-g} \left | ^*F^{rt} \right | d\theta d\phi. 
        \label{eq:diag_MAD}
    \end{align}
    \item Rate of accreted angular momentum 
    \begin{align}
        \dot L = \int_{\theta = 0}^{\pi} \int_{\phi = 0}^{2\pi} T_{~\phi}^r \sqrt{-g} d\theta d\phi. \label{eq:angular_momentum_horizon}
    \end{align}
    \item Energy flux through the horizon 
    \begin{align}
        \dot E = \int_{\theta = 0}^{\pi} \int_{\phi = 0}^{2\pi} -T_{~t}^r \sqrt{-g} d\theta d\phi. \label{eq:energy_flux_horizon}
    \end{align}
\end{itemize}
It is convenient to normalise the angular momentum and the energy flux given by Equations \eqref{eq:angular_momentum_horizon} and \eqref{eq:energy_flux_horizon} by the mass accretion rate at the horizon $\dot M$, and to further define the MAD parameter as
\begin{align}
    \Phi_B \equiv \frac{\phi_B}{\sqrt{ |\dot M|}}.
\end{align}


In order to study the disk structure and the accretion mode of our numerical models, we introduce several additional diagnostics. 
\begin{itemize}
    \item The barycentric radius
    \begin{align}
        \langle r \rangle(t) = \frac{\int_0^{2\pi} \int_0^\pi \int_{r_H}^{\bar r_{\rm max}}  r \rho \sqrt{-g} dr d\theta d\phi}{\int_0^{2\pi} \int_0^\pi \int_{r_H}^{\bar r_{\rm max}} \rho \sqrt{-g} dr d\theta d\phi},
        \label{eq:barycentric_radius}
    \end{align}
    gives the radius of the center of mass of the disk. Here $r_h = 1 + \sqrt{1 - a^2}$ is the outer horizon radius. The evolution of this radius characterises the spreading of the disk due to viscosity, as well as the contraction and bouncing in non-equilibrium models. The radius  $\bar r_{\rm max}$ limits the integration and is taken to be 80 in the figures below. Its exact value does not affect the conclusions, provided it is large enough.
    
    \item Disk-averaged quantities \citep{PCN19}
    \begin{align}
        \langle q \rangle(t,r) = \frac{\int_0^{2\pi} \int_{\theta = \pi/3}^{2\pi/3} q \sqrt{-g} d\theta d\phi}{ \int_0^{2\pi} \int_{\theta = \pi/3}^{2\pi/3} \sqrt{-g} d\theta d\phi},
    \end{align}
    where the limits on the integral over angle $\theta$ are chosen such that only disk material is considered. The parameter $q$ represents various physical quantities, such as  $\rho$, $\beta$, $p_g$ and $u^\phi$.
    \item The disk thickness defined as 
    \begin{align}
        H(t,r) = \frac{\int_0^{2\pi} \int_{0}^{\pi} \left | \frac{\pi}{2} - \theta \right | \rho \sqrt{-g} d\theta d\phi}{ \int_0^{2\pi} \int_{0}^{\pi} \rho \sqrt{-g} d\theta d\phi}.
    \end{align}
    \item The inward and outward angular momentum fluxes, which are used to characterise steady state \citep{SMN08}
    \begin{align}
        \dot l_{in}(r,t) &= \frac{\int_0^{2\pi}\int_0^{\pi}  (\rho + u_g + p_g + b^2 ) u^r u_\phi  \sqrt{-g} d\theta d\phi}{\int_0^{2\pi}\int_0^{\pi}  \rho  u^r  \sqrt{-g} d\theta d\phi}, \label{eq:lin}  \\ 
        \dot l_{out}(r,t) &= \frac{\int_0^{2\pi}\int_0^{\pi}  b^r b_\phi  \sqrt{-g} d\theta d\phi}{\int_0^{2\pi}\int_0^{\pi}  \rho  u^r  \sqrt{-g} d\theta d\phi}.
        \label{eq:lout}
    \end{align}
    Note that the outward flux of the angular momentum is dominated by contribution from the magnetic fields.
    In a steady state, both  $\dot l_{in}(r,t)$ and $ \dot l_{out}(r,t)$ are independent of $t$ and their difference $\dot l_{in}(r,t) - \dot l_{out}(r,t)$ is independent on radius.
    \item We also introduce the density-weighted Lorentz factor \citep{FWR12},
    \begin{align}
        \langle \Gamma \rangle = \frac{\int \sqrt{-g} \rho \sqrt{-g_{tt}} u^t d\theta d\phi}{\int \sqrt{-g} \rho d\theta d\phi}
        \label{eq:Lorentz_factor}
    \end{align}
\end{itemize}
In several sections of the analysis carried below, the angular integration is restricted to specific angles, in which the flow satisfies one or several specified conditions, for example $\sigma \equiv b^2/\rho> 1$.


\begin{figure}[t]
\centering
\begin{tabular}{cc}
\includegraphics[width = 0.48 \textwidth]{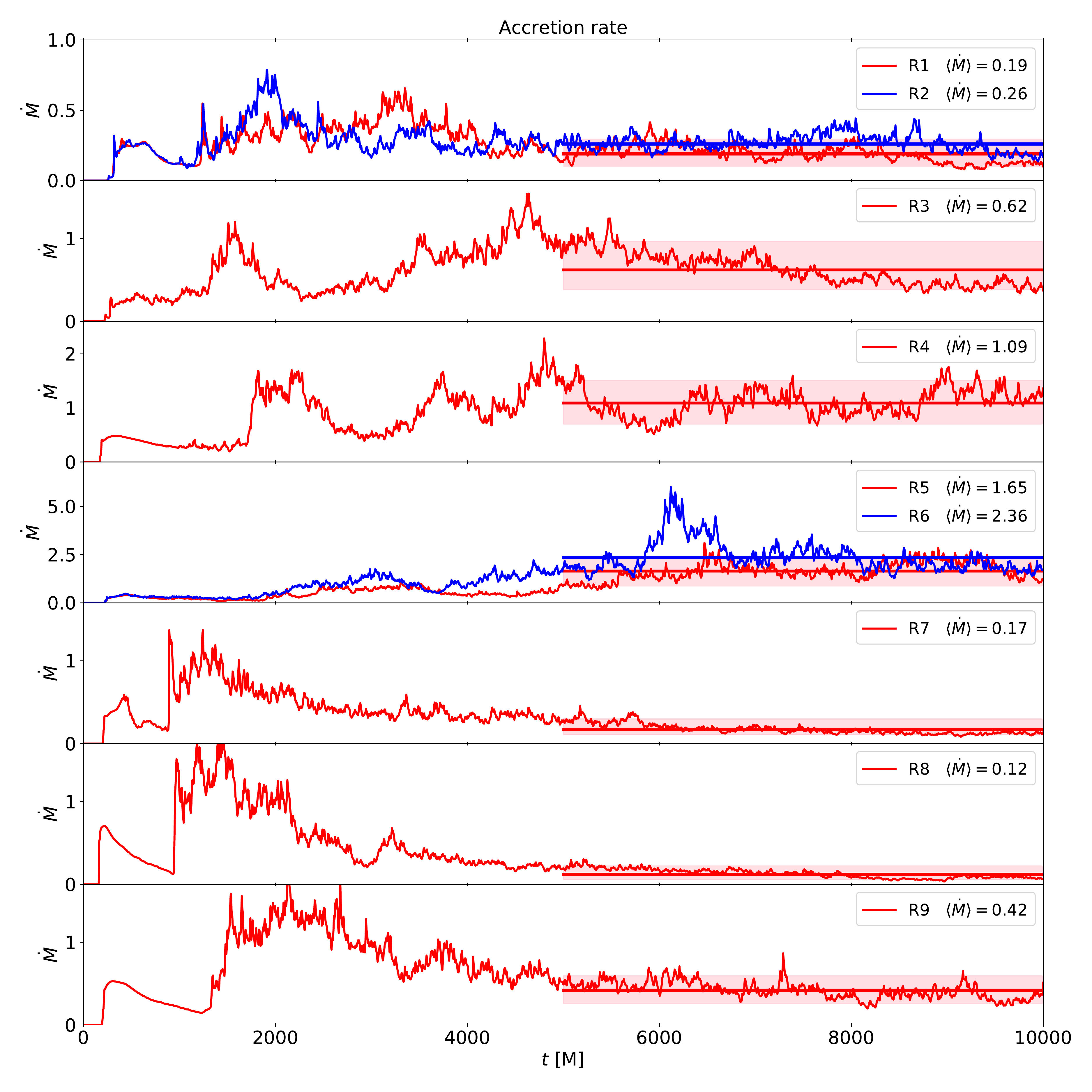}  &
\includegraphics[width = 0.48 \textwidth]{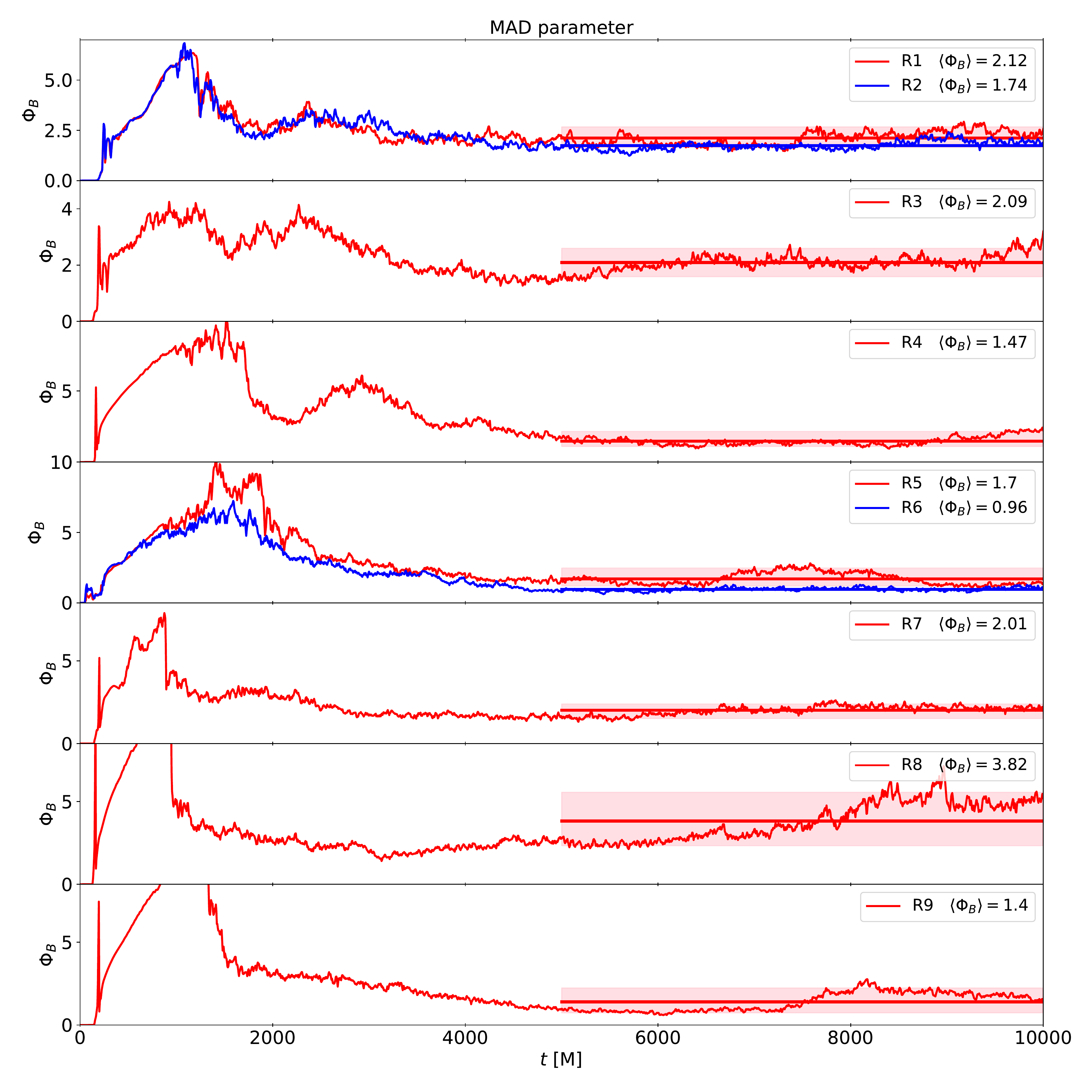} \\
\includegraphics[width = 0.48 \textwidth]{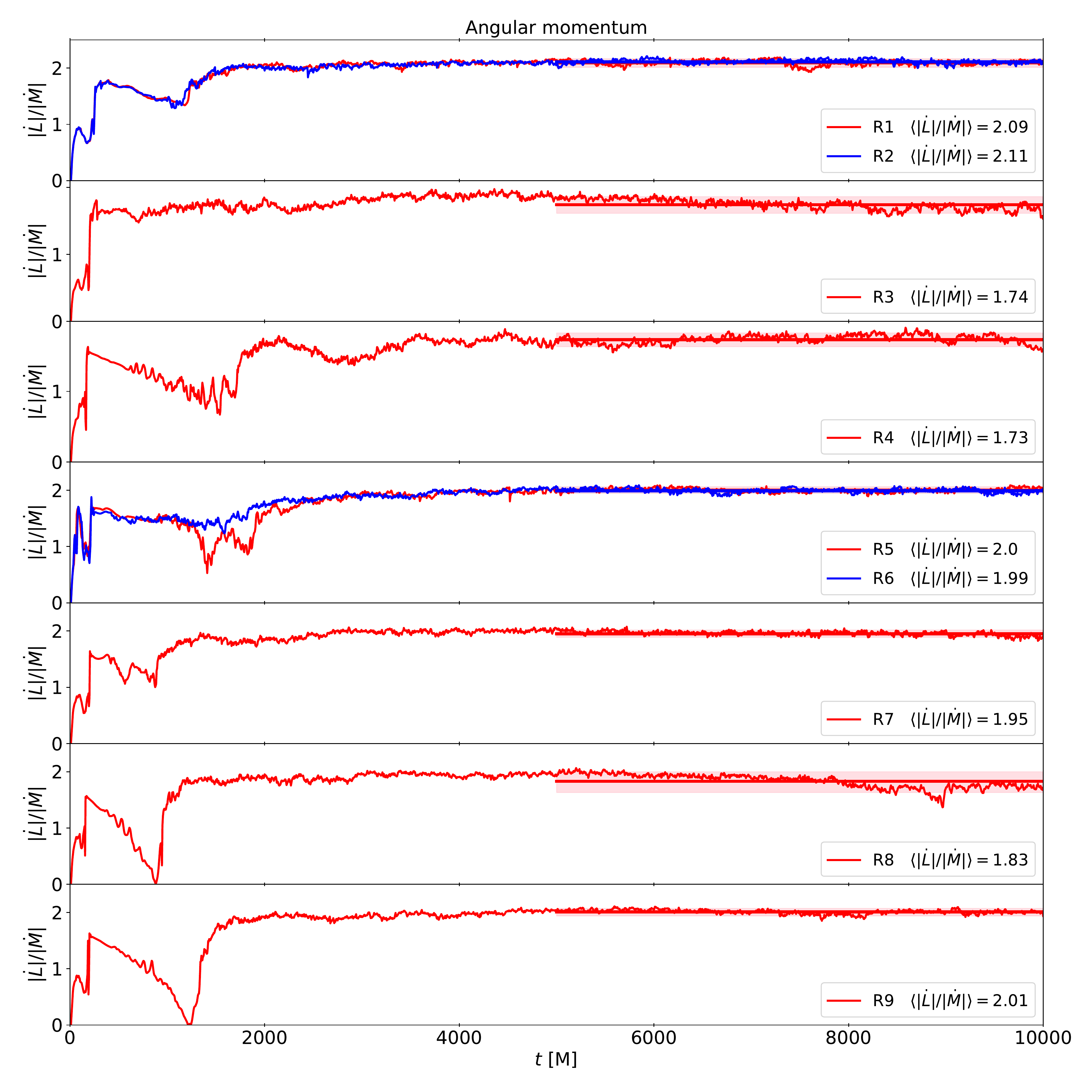}  & 
\includegraphics[width = 0.48 \textwidth]{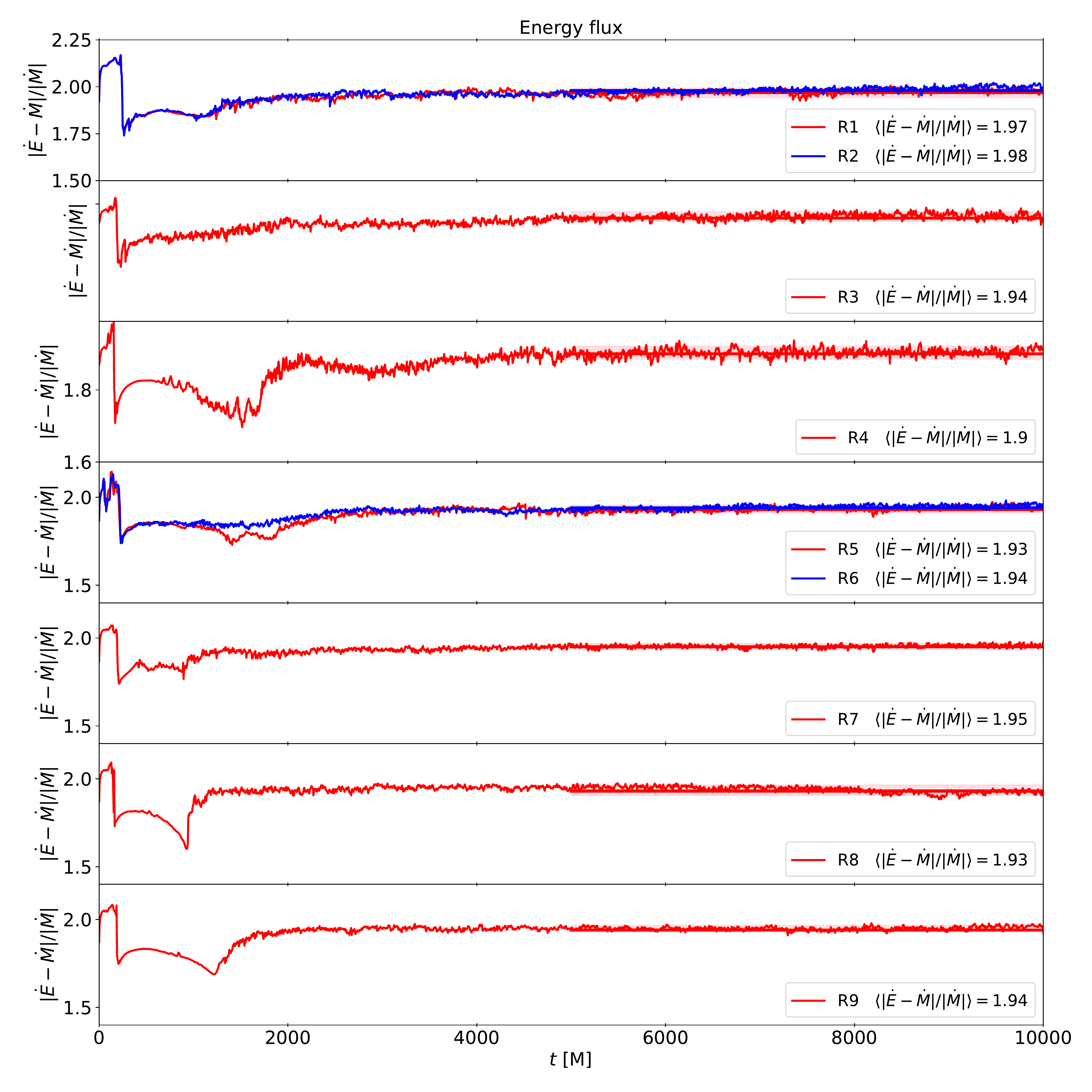}
\end{tabular}
\caption{Accretion diagnostics at the horizon. Top left - accretion rate $\dot M$, given by Equation \eqref{eq:diag_Mdot}. Top right - MAD parameter, given by Equation \eqref{eq:diag_MAD}. Bottom left - Angular momentum through the horizon, given by Equation \eqref{eq:angular_momentum_horizon}. Bottom right - energy flux, given by Equation \eqref{eq:energy_flux_horizon}. The horizontal line represent the time average for time between 5000M and $10^4$M, which value is given in the legend of each plot and with its $2\sigma$ variation in Table \ref{tab:diagnostic_horizon}. Runs with different resolutions only, that is to say R1-R2 and R5-R6, are shown on the same sub-figure. We note that the vertical scaling of each sub-figure is changing.} \label{fig:horizon_diagnostics}
\end{figure}






\section{Physical models}
\label{sec:physical_model}

\begin{table*}[t]
    \centering
    \begin{tabular}{|c|c|c|c|c||c|c|c||c|c|c|}
        \hline
\multicolumn{5}{|c||}{Physical setup} & \multicolumn{3}{c||}{Resolution} & \multicolumn{3}{c|}{Initial properties} \\\hline
Name &  $r_{\rm max}$ (M) & $\beta_0$ & $\hat \gamma$  & Equilibrium & $N_{\rm r}$ & $N_{\rm \theta}$ & $N_{\rm \phi}$ & $M$ & $E_B$ &  $E_B/M$    \\ 
(1) & (4) & (5) & (6) & (7) & (8) & (9) & (10) & (11) & (12) & (13) \\ \hline
R1  & 12 & 100 &  4/3 & Yes & 256 & 128 & 64  & $9.95 \times 10^3$ & $9.8 \times 10^{-2}$ & $9.8\times 10^{-6}$   \\  
R2  & 12 & 100 &  4/3 & Yes & 256 & 128 & 128 & $9.95 \times 10^3$ &  $9.8 \times 10^{-2}$ & $9.8\times 10^{-6}$  \\
R3  & 12 & 100 &  5/3   & Yes & 256 & 256 & 128 & $2.4 \times10^4$ & $1.2 \times 10^{-1}$ & $5 \times 10^{-6}$ \\
R4  &  13 & 44 & 5/3 &Yes & 256 & 128 & 128 & $9.4 \times 10^4$ & $4.6 \times 10^{-1}$ &  $5 \times 10^{-6}$  \\
R5  &  13 & 44 & 4/3 & No (5/3) & 256 & 128 & 128  & $9.4 \times 10^4$ & $2.3 \times 10^{-1}$  &  $2.5 \times 10^{-6}$  \\
R6  &  13 & 44 & 4/3 & No (5/3) & 256 & 256 & 128  & $9.4 \times 10^4$ & $2.3 \times 10^{-1}$  &  $2.5 \times 10^{-6}$    \\
R7  &  12 & 44 & 4/3 & Yes & 256 & 256 & 128  & $9.95 \times 10^3$ & $2.2 \times 10^{-1}$ & $2.3 \times 10^{-5}$  \\
R8  &  12 & 20 & 4/3 & Yes & 256 & 128 & 128  & $9.95 \times 10^3$ & $4.9 \times 10^{-1}$ & $4.9 \times 10^{-5}$  \\
R9  &  13 & 44 & 4/3 &Yes & 256 & 128 & 128 &  $2.8 \times 10^{4}$ & $4.0 \times 10^{-1}$  & $1.4\times 10^{-5}$    \\\hline
\end{tabular}
\caption{Initial setup of the simulations and their characteristics. Column (5) indicates if the initial disk is in equilibrium and if not which adiabatic index $\hat \gamma$ was used to initialise the disk. Column (9) contains the initial disk mass in arbitrary units. Column (10) gives the initial magnetic energy $E_B$, also in arbitrary units, and column (11) gives the ratio $E_B/M$. All simulations have spin $a = 0.9375$ and the inner radius of the torus is at $r_{in} = 6M$.}
\label{tab:tab_run_caracteristics}
\end{table*}

\begin{table*}[t]
    \centering
    \begin{tabular}{|c||c|c|c||c|c|c|c||c|}
    \hline
  & \multicolumn{3}{c||}{MRI quality factor} & \multicolumn{4}{c||}{Diagnostics at $r_h$} &  \\\hline
Name & $\langle Q^r \rangle_t$ &  $\langle Q^\theta \rangle_t$ & $\langle Q^\phi \rangle_t$ & $\langle \dot M \rangle_t$ & $\left \langle \frac{\dot \Phi_B}{\sqrt{-\dot M}} \right \rangle_t$ & $\left \langle \frac{\dot L}{{|\dot M|}} \right \rangle_t$ & $\left \langle \frac{\dot E - \dot M}{{|\dot M|}} \right \rangle_t$ & Form a jet ? \\ \hline
R1 &  6.2$^{+1.85}_{-1}$ & 6.5$^{+0.8}_{-0.9}$ & 7.6$^{+1.4}_{-0.9}$ & $0.2 \pm 0.1$ & $2.12 \pm 0.5$ & $2.09 \pm 0.07$ &  $1.97\pm 0.025$ & Yes  \\  
R2 &  6.9$\pm 1.3$ & 8.4$^{+2.1}_{-2}$ & 16$^{+2.4}_{-2.3}$ & $0.27 \pm 0.1$ & $1.74^{+0.4}_{-0.3}$ & $2.11 \pm 0.05$ & $1.98^{+0.03}_{0.02}$ & Yes  \\
R3 &  10.8$^{+1.6}_{-1.4}$ & 29.2$\pm 4.5$ & 21.4 $\pm 3$ & $0.63^{+0.33}_{-0.25}$ & $2.09 \pm 0.5$  & $1.74 \pm 0.12$  & $1.94^{+0.03}_{0.015}$  & Yes \\
R4 &  8.7$^{+1.1}_{-1}$ & 12.3$^{+1.9}_{-1.4}$ & 19$\pm 2$ & $1.1 \pm 0.4$  &  $1.47^{+0.7}_{-0.36}$  & $1.73 \pm 0.1 $& $1.9^{0.02}_{-0.015}$  & No \\
R5 &  6.4$^{+1.1}_{-0.9}$ & 7.5$^{+1.4}_{-1.1}$ & 15.6$^{+2.1}_{-1.5}$ & $1.65 \pm 0.8$ &  $1.7^{+0.77}_{-0.5}$ & $2.0 \pm 0.06 $& $1.93^{0.025}_{0.02}$  & Intermittent \\
R6 &  $8.4^{+0.9}_{-1.7}$  & $20.6^{+3}_{-4.7}$  & $17.0^{+1.9}_{-2.4}$ & 2.4$^{+1.65}_{-0.84}$ & $0.96 \pm 0.2 $  & $1.99 \pm 0.06 $ & $1.94 \pm 0.02$  & No\\
R7 &  $9.5^{+1.4}_{-0.9}$  & $23.5^{+3.1}_{-2.8}$  & $19.0^{+3.1}_{-1.7}$ & $0.17^{+0.12}_{-0.06}$  & $2.01^{+0.38}_{-0.5} $ & $1.95 \pm 0.06$ &  $1.95 \pm 0.015$  & Yes \\
R8 &  $9.8^{+1}_{-1.1}$  & $13^{+1.3}_{-1.5}$   & $20.7^{+1.6}_{-2.1}$  &  $0.12^{+0.1}_{-0.06}$ & $3.82^{+1.7}_{-1.5} $ & $1.83^{+0.17}_{-0.20}$ & $1.93^{+0.03}_{-0.02}$   & Yes \\
R9 &  $7.9^{+1.2}_{-1}$  & $10.3^{+1.2}_{-1.3}$  & $17^{+2.8}_{-2}$  &  $0.42 \pm 0.15$ &  $1.4^{+0.84}_{-0.67}$& $2.01 \pm 0.07$  &  $1.94^{+0.03}_{-0.01}$  &  Intermittent \\\hline
\end{tabular}
\caption{Time average properties of each runs. The errors represent the $2\sigma$ variance. The last column indicates if a jet is observed in the simulation between $t = 5\times 10^3$M and $t = 10^4$M, intermittent meaning a weak jet that forms and disappear during this period. }
\label{tab:diagnostic_horizon}
\end{table*}

\begin{figure}
    \centering
    \includegraphics[width=0.75\textwidth]{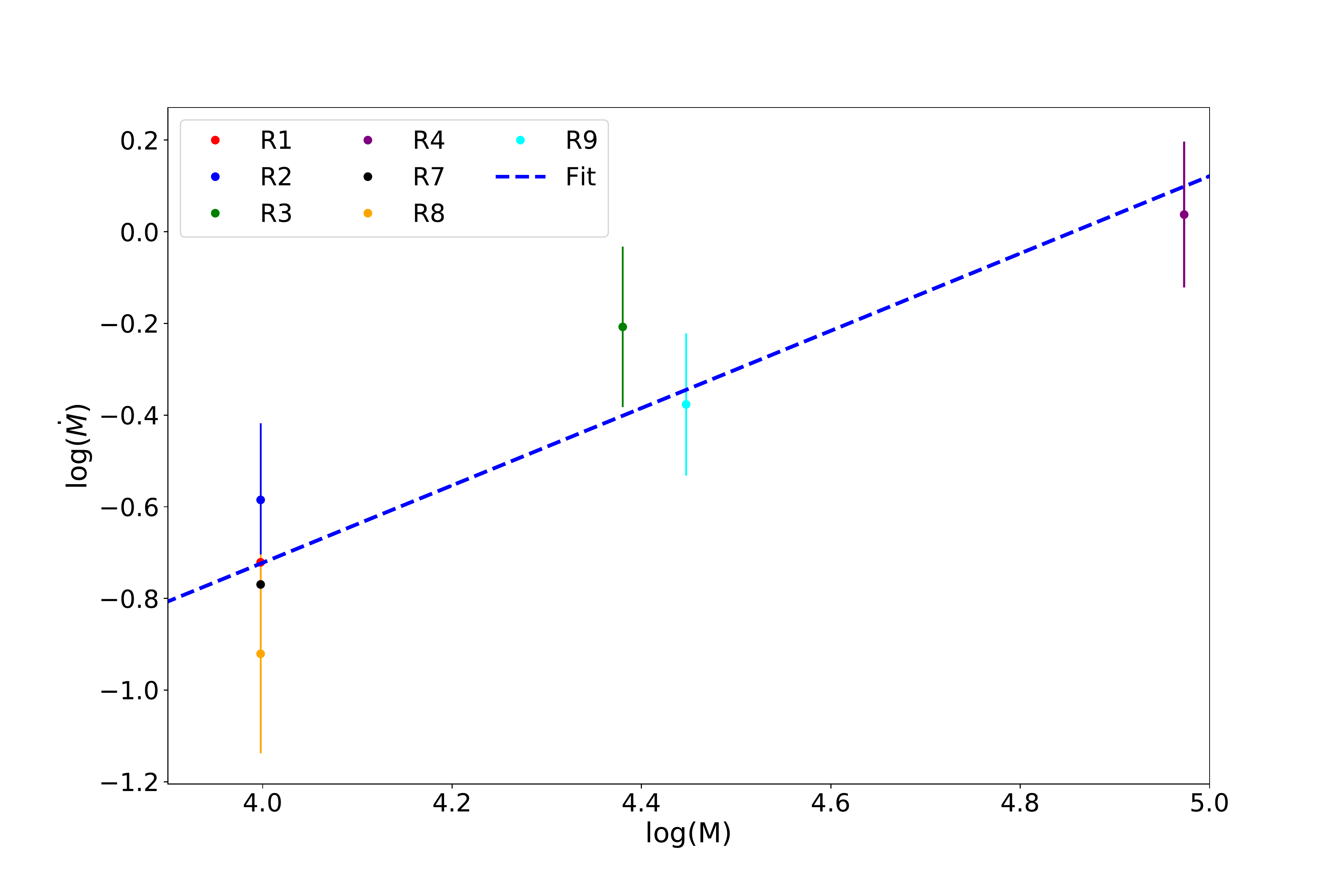}
    \caption{Correlation between the initial torus mass $M$ and the quasi-steady state mass accretion rate. The slope of the fit is 0.84.}
    \label{fig:correlation_M_dotM}
\end{figure}

\begin{figure}[t]
\centering
\includegraphics[width = 0.90 \textwidth]{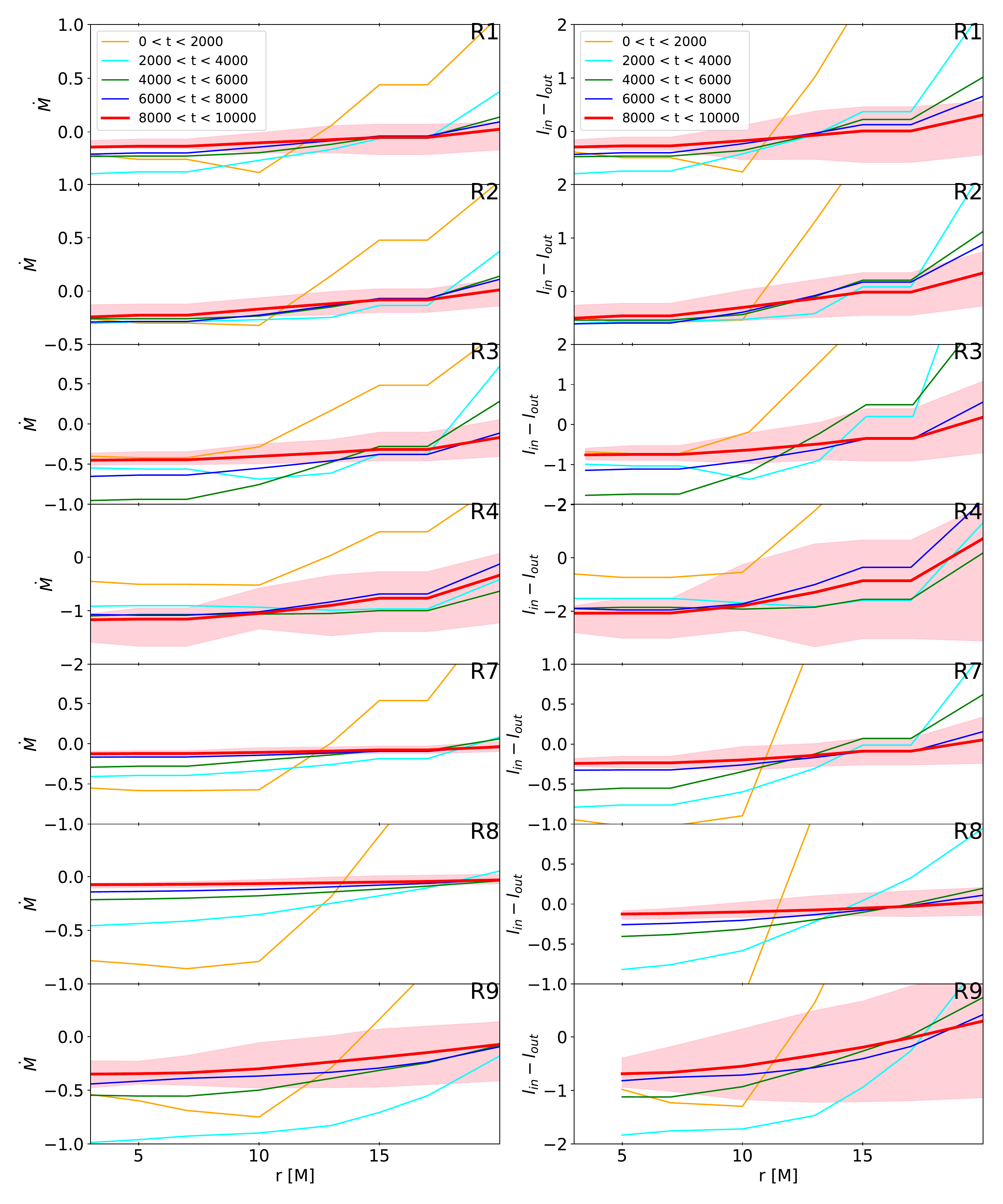} 
\caption{Left - Time-averaged mass accretion rate as a function of radius $r$ for different time intervals. Right - Time averaged angular momentum flux as a function of radius. The pink region corresponds to the 2-$\sigma$ variance during the last time interval, $8000 < t < 10^4$M. From this figures, it is clear that at the end of our run, quasi steady state up to radius $r = 10$M has been reached for R1, R2, R3, R7 and R8 and is only marginally achieved for the remaining runs. }
\label{fig:MdotLinlout_r}
\end{figure}


We use our numerical code to study accretion in the SANE regime around a
black-hole. In the current work, we only consider BH spin $a = 0.9375$. We
first test the reliability of our numerical results by using a similar setup
to the one used in the GR MHD code comparison paper \citep{PCN19}. Then we
run several simulations with different initial conditions and typical
resolution in the range  $(N_r, N_\theta, N_\phi) = (256,128,64)$ to
$(256,256,128)$ (see Table \ref{tab:tab_run_caracteristics}). We note
that $N_\theta = 128$ was dubbed minimum angular resolution for MRI
amplification of the magnetic field to efficiently develop \citep{PCN19}.
Therefore, in all our numerical experiments the number of cells in the
$\theta$ direction is equal or larger than this value. All simulations
are initialised with a torus in hydrostatic equilibrium, following
\cite{FM76}. This particular equilibrium solution is described by the
inner radius of the torus $r_{\rm in}$ and the radius of maximum pressure
$r_{\rm max}$. In this work, we set $r_{\rm in} = 6M$, while $r_{\rm max}$
can be either $12M$ or $13M$, which results in an outer disk boundary
location at $\sim 50M$ and $\sim 80M$, respectively. Typically, the
disks simulated have height $H/r \lesssim 1$ ("thick disks"). 

Since the scale of the disk density does not need to be specified
in the \citet{FM76} solution, the density is normalised such that its
maximum value in the initial disk is $\rho = 1 $ \citep{MG04}. The pressure
is correspondingly re-normalised. This initial setup is initially
axisymmetrically stable.  It is completed by the addition of a subdominant
poloidal magnetic field. Similar to previous GR-MHD simulations of magnetized
torus \citep[e.g.][]{GMT03,PCN19}, we consider a simple initial magnetic
field topology defined by the vector potential
\begin{align}
    A_\phi= {\rm max } \left ( \frac{\rho}{\rho_{\rm max}} - 0.2 , 0\right). \label{eq:potetial_vector}
\end{align}
We further normalise the intensity of the magnetic field such that the
maximum value of $ \beta_0 \equiv p_{\rm g, max} / p_{\rm B, max}  \gg 1 $.
This subdominant magnetic field serves as seed for the development of MRI,
thereby enabling transport of angular momentum through the disk and
accretion of matter to the black-hole. In order to trigger accretion,
the initial disk model is made unstable. To this end, a small random
perturbation of magnitude 4 percent is added to the gas pressure. For
each run, we present in Table \ref{tab:tab_run_caracteristics} the
properties of the initial torus and the value of $\beta_0$, which ranges
from a few tens to a hundred.

In the next section, we present the analysis of 9 GR-MHD simulations in
the SANE regime. Their initial properties, listed in Table
\ref{tab:tab_run_caracteristics}, are selected such that we can address
the following problems. (i) We first want to ensure that our code gives
reliable results. Therefore, runs R1 and R2 which only differ by their
azimuthal resolution, have the same initial conditions as the runs studied
in the code comparison paper \citep{PCN19}. The numerical results we obtained
are then confronted to those results presented in \citet{PCN19}. We
essentially find the same results. (ii) We want to study the effect of 
the value of the adiabatic index. For this, we run two sets of simulations
which differ only by their adiabatic index, being either relativistic
$\hat \gamma = 4/3$ or non-relativistic $\hat \gamma = 5/3$. Specifically, R2
is compared to R3, while R4 is compared to R9. The two simulation sets differ
by $r_{\rm max}$ and $\beta_0$. (iii) Next, the effect of changing diverse
initial conditions are studied. Specifically, R2, R7 and R8 differ only by
the initial normalisation of the magnetic field $\beta_0$, and R7 and R9
have a different $r_{\rm max}$. (iV) We also study an initial situation which
is out of equilibrium. This out-of-equilibrium setup is obtained by
initialising the torus with the non-relativistic adiabatic index
$\hat \gamma = 5/3$ and compute its evolution by using the relativistic
adiabatic index $\hat \gamma = 4/3$.

For each run and each direction, we compute the MRI quality factor $Q$,
given by Equation \eqref{eq:volume_average_q}. This computation is restricted
to the disk by limiting the angular integration to $\pi/3 < \theta < 2 \pi /3$
and a time average is performed with $t > 5\times 10^3 M$. The results are given
in Table \ref{tab:diagnostic_horizon} with 2$\sigma$ variance. These quality
factors $Q$ characterize the number of grid cells necessary to resolve the
growth of the fastest MRI mode. Different minimal values have been proposed.
The consensus seems to require $Q^z \geq 10$ and $Q^\phi \geq20-25$
\citep{HGK11,HRG13}. Lower limits for the quality factor where also given
by \citet{SRS12} ($Q^z \geq 10-15$ and $Q^\phi \geq10$) and \citet{SIT04}
(6 grid zones), while other metrics were proposed considering the coupling
between poloidal and toroidal components $Q^z Q^\phi \geq 250$ \citep{NSP12,DS19}.

The GR-MHD code comparison paper \citep{PCN19} underlines how important it
is to achieve a resolution sufficient to resolve the MRI growth rate. In
particular, it was noticed the appearance of ''mini-torus'' in low resolution
runs of BHAC, while the value of the MAD parameter of all code converges only
when the MRI quality  factors are large enough. Inspection of Table
\ref{tab:diagnostic_horizon} reveals that all our runs sufficiently resolves
the MRI according to the criteria of \citet{SIT04}, while only runs R3, R4,
R6, R7, R8 and R9 satisfy or at least marginally statisfy the criteria of
\citep{HGK11,HRG13}. Therefore this discussion prompts caution when discussing
the results of R1, R2 and R5.

For the need of the analysis, we also introduce the initial mass and the initial magnetic energy of the accretion disks after the normalisation of the density. They are respectively defined as 
\begin{align}
    M = \int_{\rm disk} \sqrt{-g} \rho dV
\end{align}
for the initial mass, while the initial magnetic energy is 
\begin{align}
    E_B = \int_{\rm disk} \sqrt{-g} b^2 dV
\end{align}
where the integral is performed only on the volume of the disk. The value of these two parameters for all simulations are given in Table \ref{tab:diagnostic_horizon}.


\section{Results}
\label{sec:results}

All our numerical models but R5 and R6 (which are not in equilibrium) present the same time evolution morphologhy.
After an initial highly active period, lasting about $5 \times 10^3$M, the
disk-jet\footnote{If the jet exists. When it does not exist, the disk is
surrounded by magnetized plasma with $\beta \sim 1$.} systems settle in a
quasi-steady state. Therefore all time average diagnostics are measured for a
time between 5000M and the end of the simulation at $10^4$M. Since R5 and R6
are initially out of equilibrium, we discuss their analysis independently in
Section \ref{sec:out_of_equilibrium_results}. 

\subsection{Horizon diagnostics and transport of mass and angular momentum with radius}

For all diagnostics, the average values and their standard deviation are given
in Table \ref{tab:diagnostic_horizon} for all runs, while their time variations
for the full simulations duration are presented in Figure
\ref{fig:horizon_diagnostics}. In details, the diagnostics of R1 and R2 for which only the
azimuthal resolution is changed by a factor of 2, are similar. The quasi
steady-state accretion rate and the MAD parameter values are consistent for
both runs within $2\sigma$ variation (e.g., $\dot M(R1) = 0.2\pm0.1$, $\dot M(R2) = 0.27\pm0.1$, see Table \ref{tab:diagnostic_horizon}). We find that these values, as well as the values of all other diagnostics we use (magnetic flux, energy flux and angular momentum flux through the horizon, radial disk structure and the existence of jets) are all consistent with the
results obtained by several independent numerical codes for the same initial setup, as presented in \citet{PCN19}. This provides us with a stronger confidence in the reliability of our code. 

The simulation R3 has the
non-relativistic adiabatic index $\hat \gamma = 5/3$, which is different than
the one used in R1 and R2 ($\hat \gamma = 4/3$, see Table \ref{tab:tab_run_caracteristics}). The initial burst of accretion due to  the
establishment of the steady state is followed by another one. Overall, the
average accretion rate is $\dot M(R3) = 0.62$ with a large variance. This
accretion rate is substantially larger than the accretion rate in R1 and R2 which have a relativistic
adiabatic index (see further discussion in Section \ref{sec:adiabatix_index} below). However, the difference
is mostly due to the fact that the disk is initially 2.5 times more massive - for higher adiabatic index, the disk is much thicker; see Table \ref{tab:tab_run_caracteristics}.
Indeed, the accretion rate in R4, with a larger disk and $\hat \gamma = 5/3$, is even larger, as would its
initial mass suggest, while the accretion rate of R7 and R8, which have similar disk sizes and adiabatic indices as R1, R2, and differ by the parameter $\beta_0$, are comparable to
that of R1 and R2 which have a similar initial mass. 
As the initial setups of these four runs (R1, R2, R7 and R8)
only differs by the initial $\beta_0$, we conclude that the initial magnetic field
only has a minor effect on the accretion rate, seemingly only impacting the
resolution with the highest $\beta_0$. Finally, R9 has an accretion rate of
$\dot M(R9) = 0.42$, about 2 times larger than that of R7. The two simulations 
differ by their initial radius of maximum disk pressure, $r_{\rm max}$, changing the mass of the disk by a
factor $\sim 3$.

We find that the accretion rate nearly only depends on the
initial mass of the disk. We show in Figure \ref{fig:correlation_M_dotM} the
correlation between the initial disk mass and mass accretion rate. An empirical fit gives $\log(\dot
M) = 0.84 \log(M) - 4.1$, demonstrating that the initial magnetic field or the adiabatic
index do not substantially influence the mass accretion rate for the SANE/RIAF disks considered here.

Figure \ref{fig:MdotLinlout_r} shows the mass accretion rate and the rate of angular
momentum change respectively, given by Equations \eqref{eq:diag_Mdot}, \eqref{eq:lin}
and \eqref{eq:lout} as functions of radius (for $r<20$~M), averaged over time intervals of
duration $\Delta t = 2\times 10^3$M. The pink region corresponds to the
$2\sigma$ variance of the last time interval $ 8\times 10^3 {\rm M} < t < 10^4 {\rm M}$.
In the quasi steady-state, it is expected that both $\dot M$ and $l_{\rm in}
- l_{\rm out}$ be independent of the radius and of time \citep{SMN08}. 
Indeed, we observe that quasi steady state inflow and outflow is achieved at least up to
$r \lsim 15$~M at $t = 10^4$M for R1, R2, R3, R7 and R8. This is supported by
the diagnostics at the horizon given in Figure \ref{fig:horizon_diagnostics},
which shows that the various diagnostics reach their quasi steady state values after $t \lsim 5\times 10^3$M.

However, the two larger disks R4 and R9 did not reach outflow equilibrium, even
after $9\times 10^3$M, prompting caution when concluding about those runs.
In fact several authors investigated the quasi-state in very long
time simulations with $t > 10^5$M; see \textit{e.g.} \citet{NSP12,SNP13, WQG20}.
This is required for the large disks they used and for studying the quasi-steady
shape of the disk until large radii $r \gsim 10^2$M. Yet, since our disks are
smaller, our simulations end time are sufficient for reaching inflow equilibrium
until past the initial radius of maximum pressure $r_{\rm max}$.





\subsection{Structure of the disk-jet system: temporal and azymuthal average}

\begin{figure}
    \centering
    \begin{tabular}{ccc}
    \includegraphics[width=0.25\textwidth]{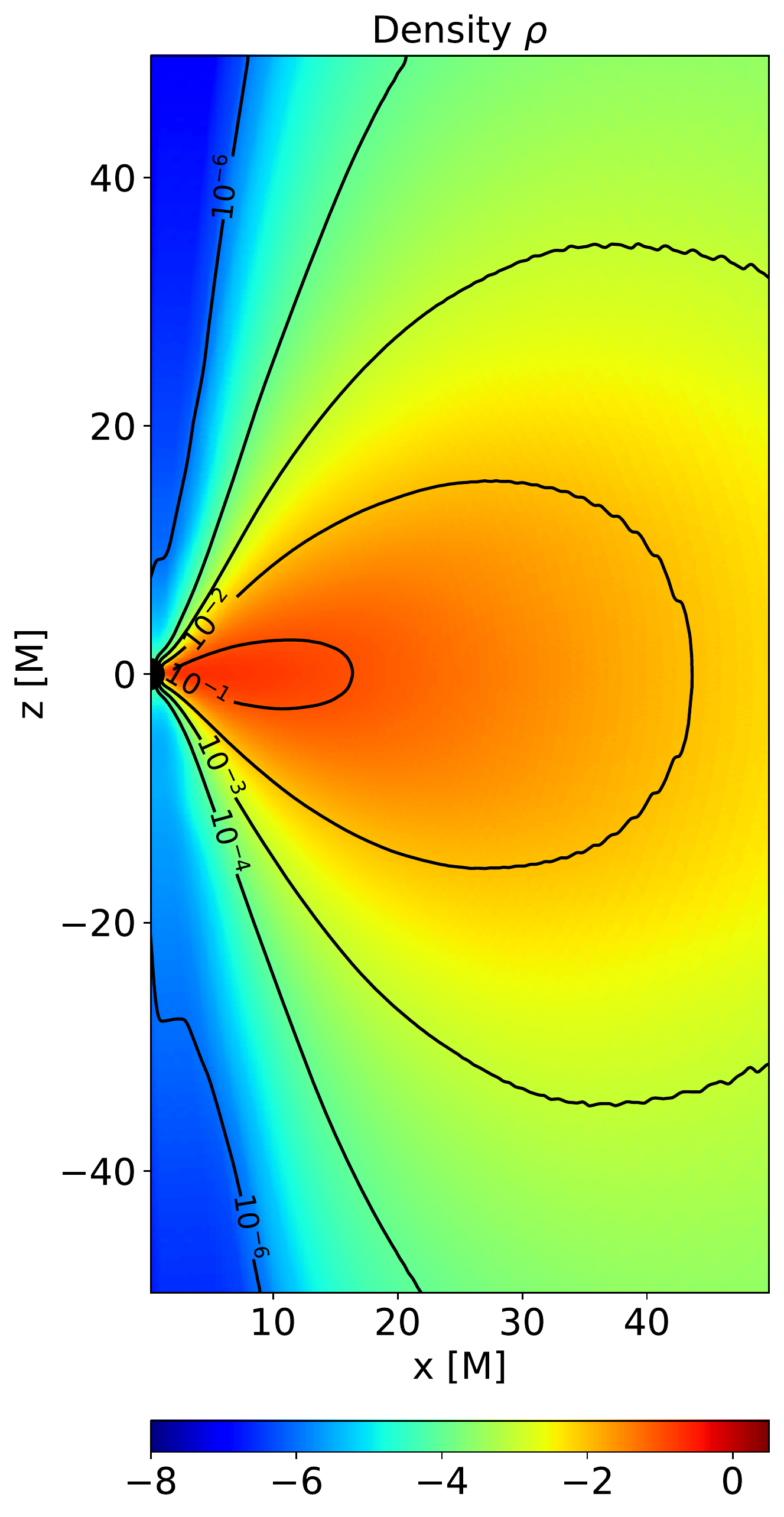}  &
    \includegraphics[width=0.25\textwidth]{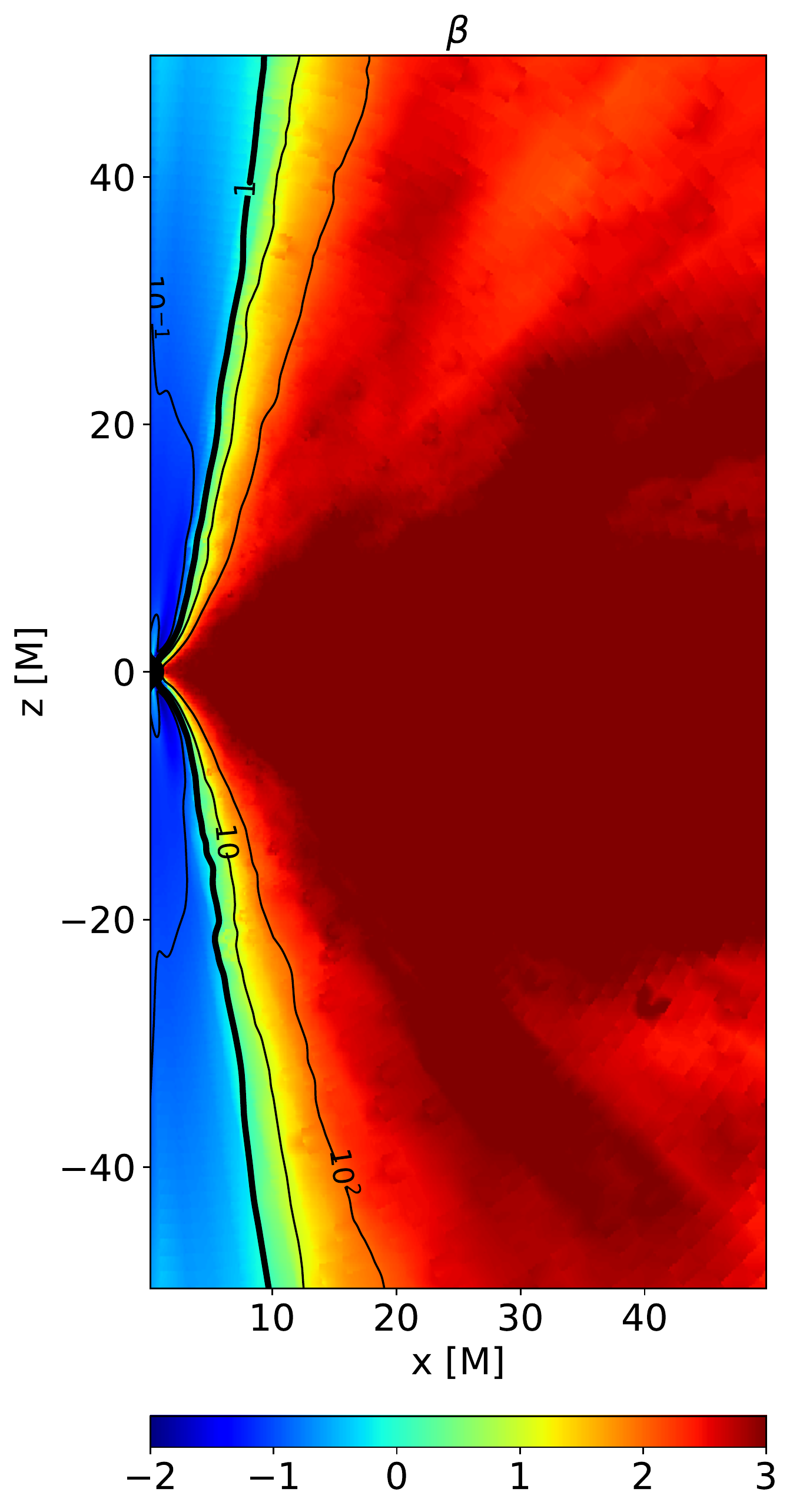} &
    \includegraphics[width=0.25\textwidth]{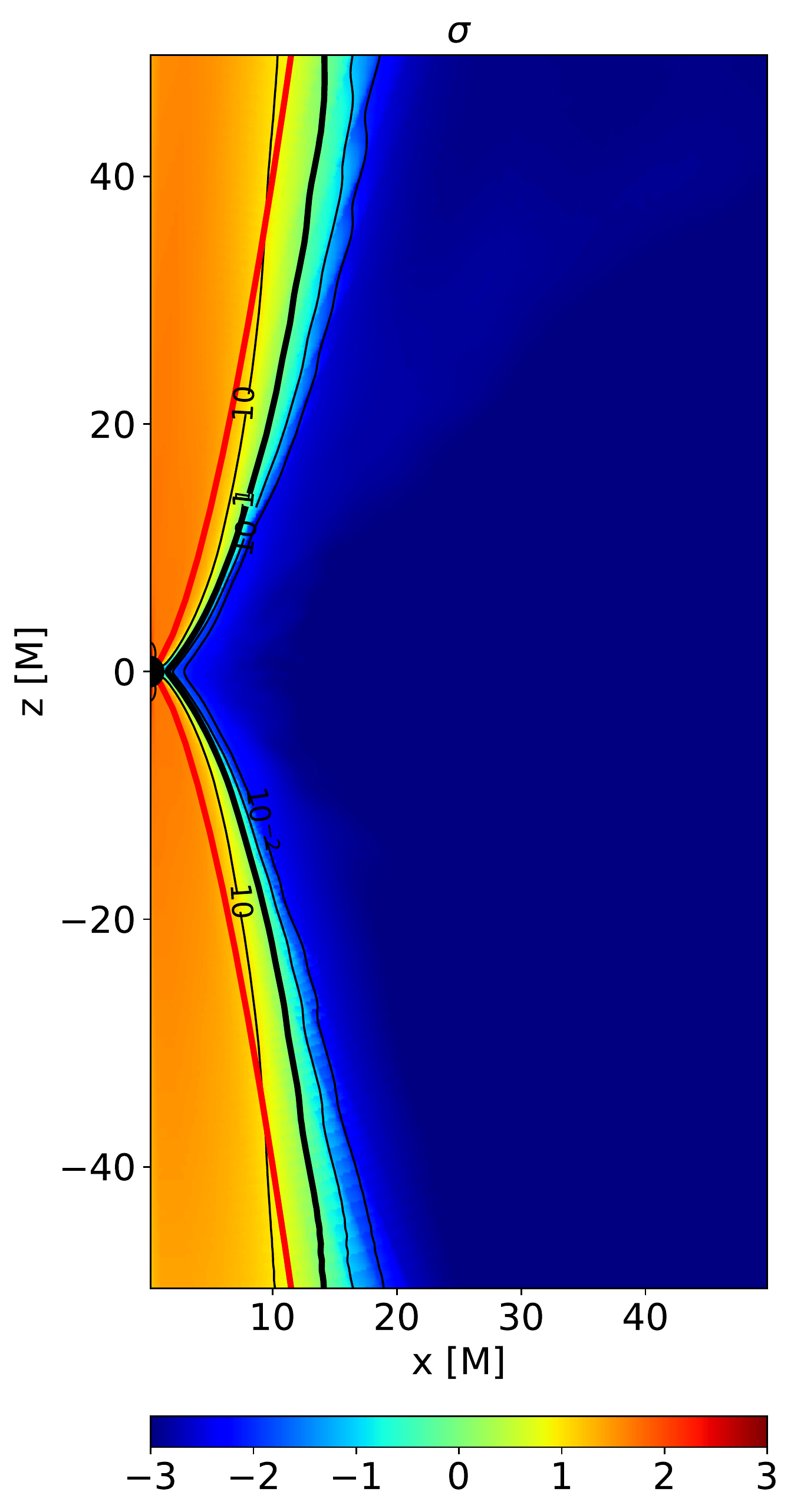} \\ 
    \includegraphics[width=0.25\textwidth]{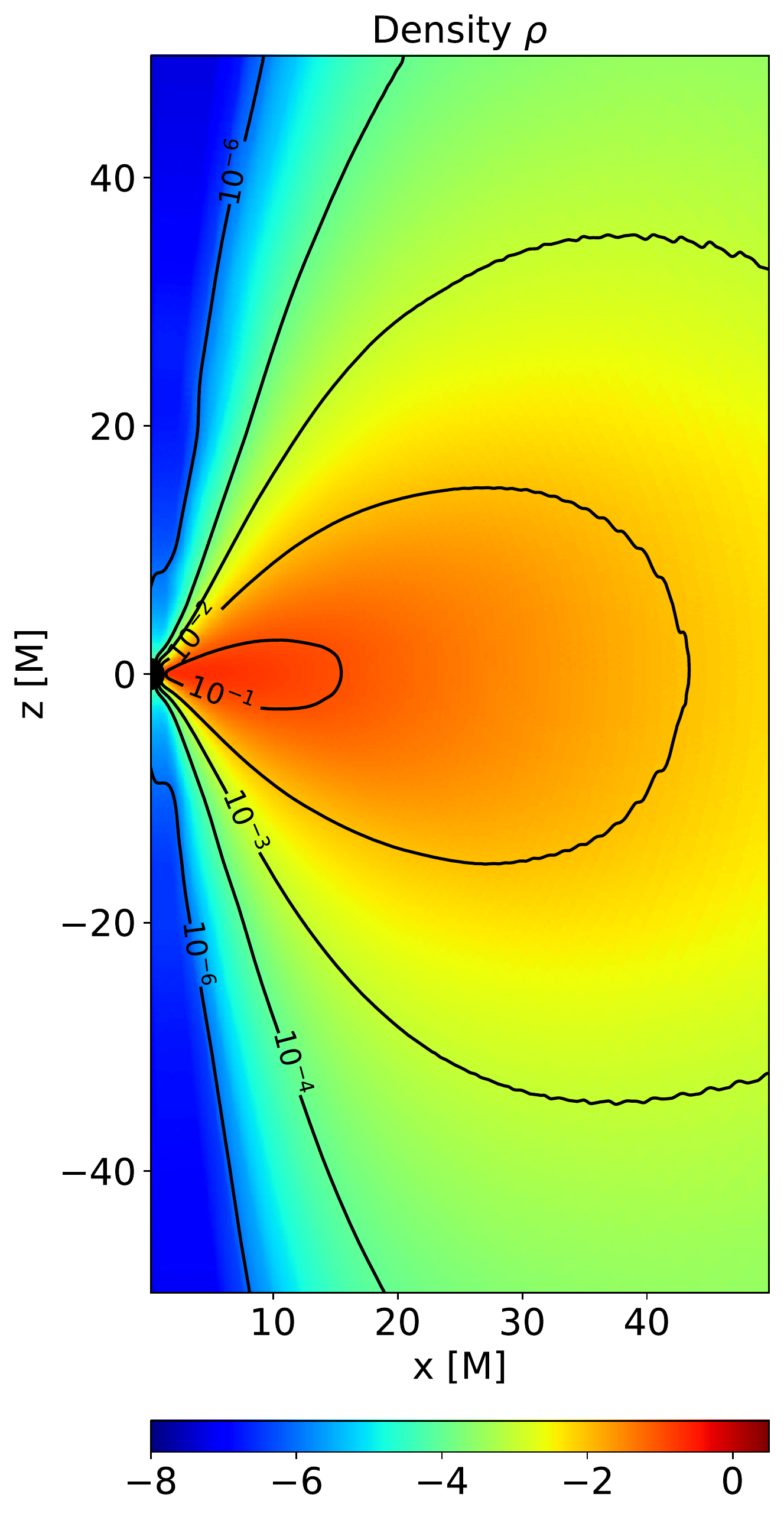}  &
    \includegraphics[width=0.25\textwidth]{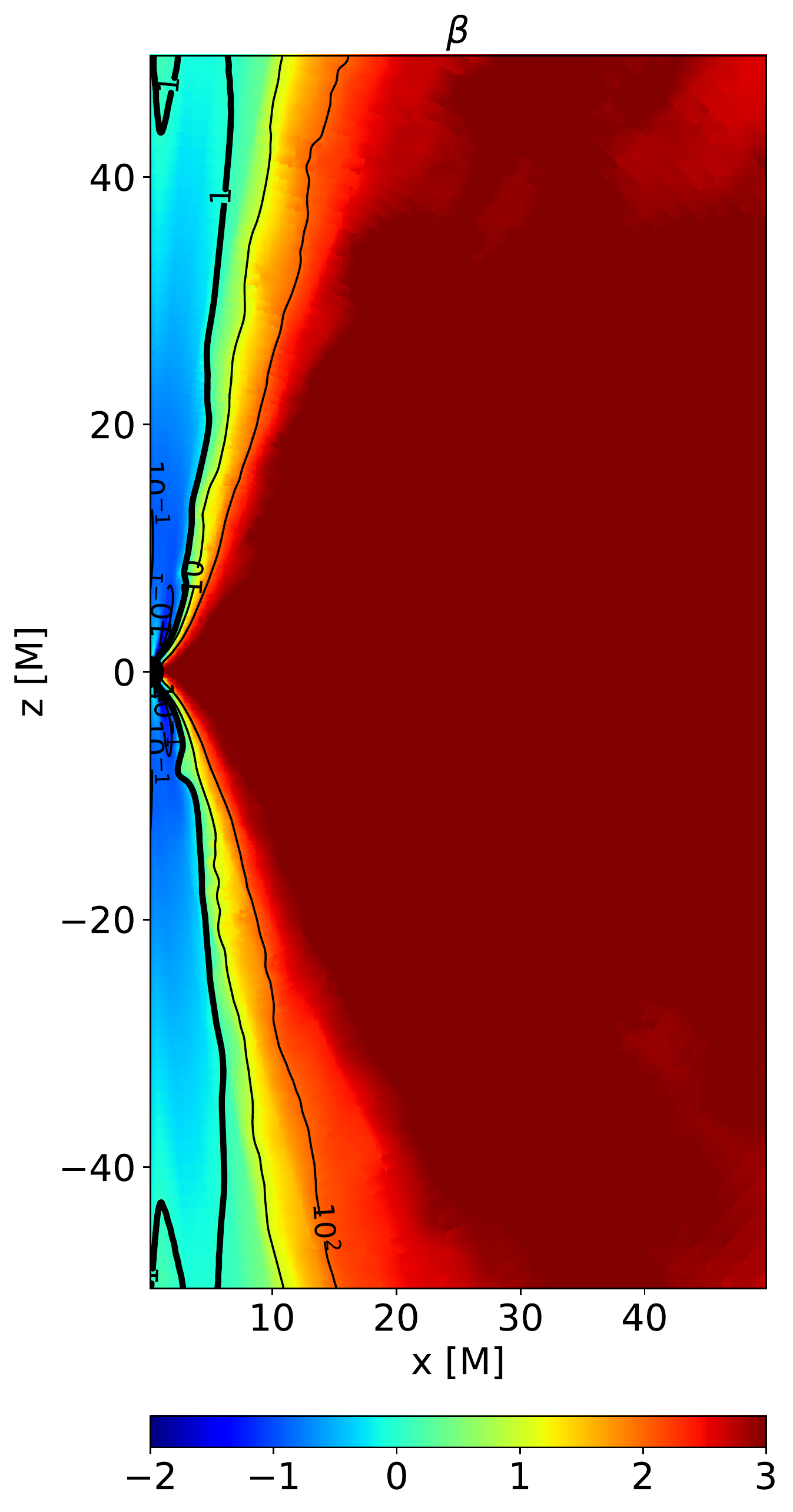} &
    \includegraphics[width=0.25\textwidth]{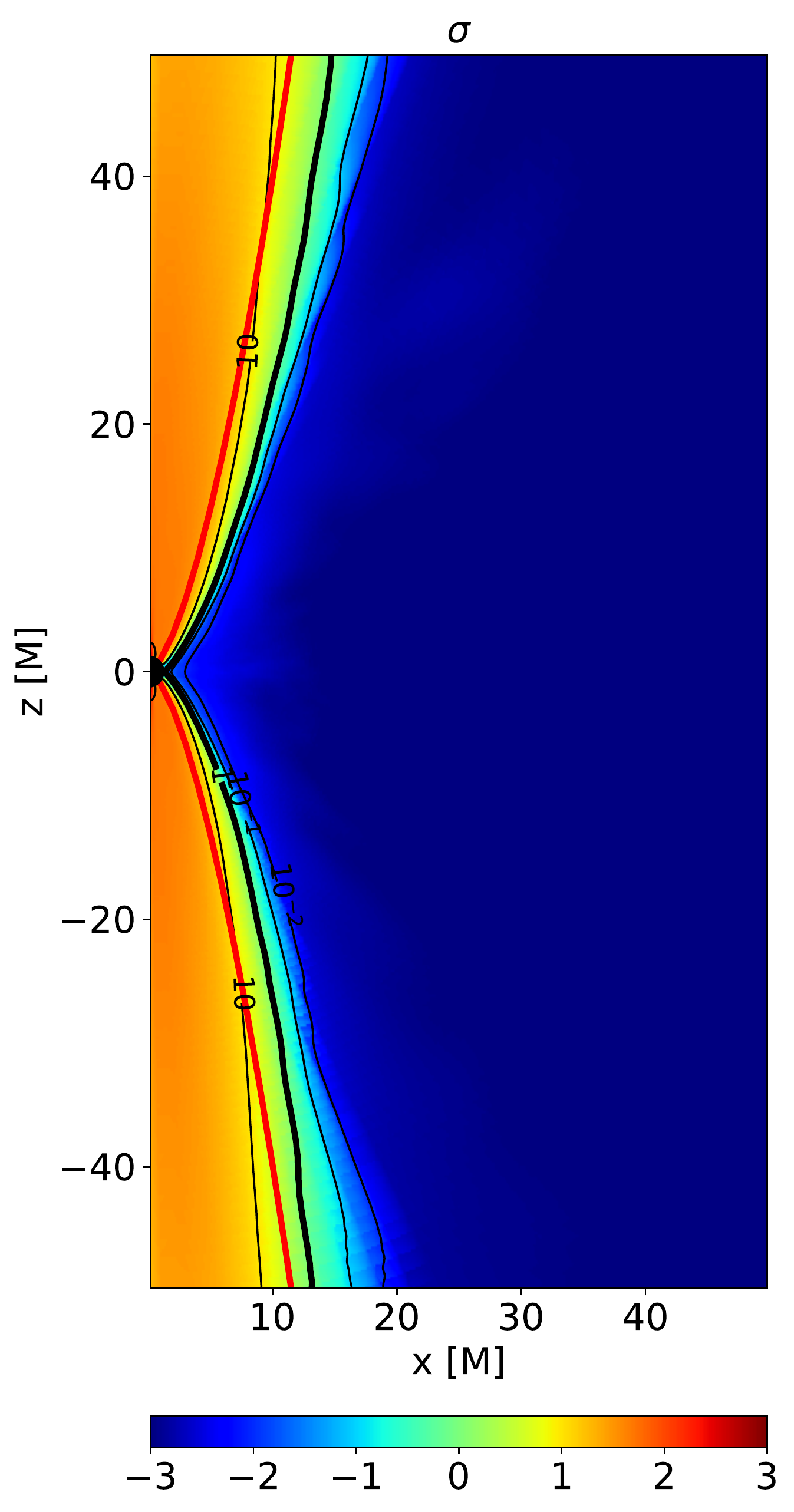}  
    \end{tabular}
    \caption{Time and azimuthal average of the density (left), $\beta$ (middle) and $\sigma$ (right) for R1 (top) and R2 (bottom). The black thick line on the middle and right plots show $\beta = 1$ and $\sigma = 1$. The red line on the right plots represents $z = x^{1.6}$, which describes the surface $\sigma = 1$ in force-free models.
    }
    \label{fig:time_azimuth_integrated}
\end{figure}

\begin{figure}[h!]
    \centering
    \begin{tabular}{ccc}
    \includegraphics[width=0.25\textwidth]{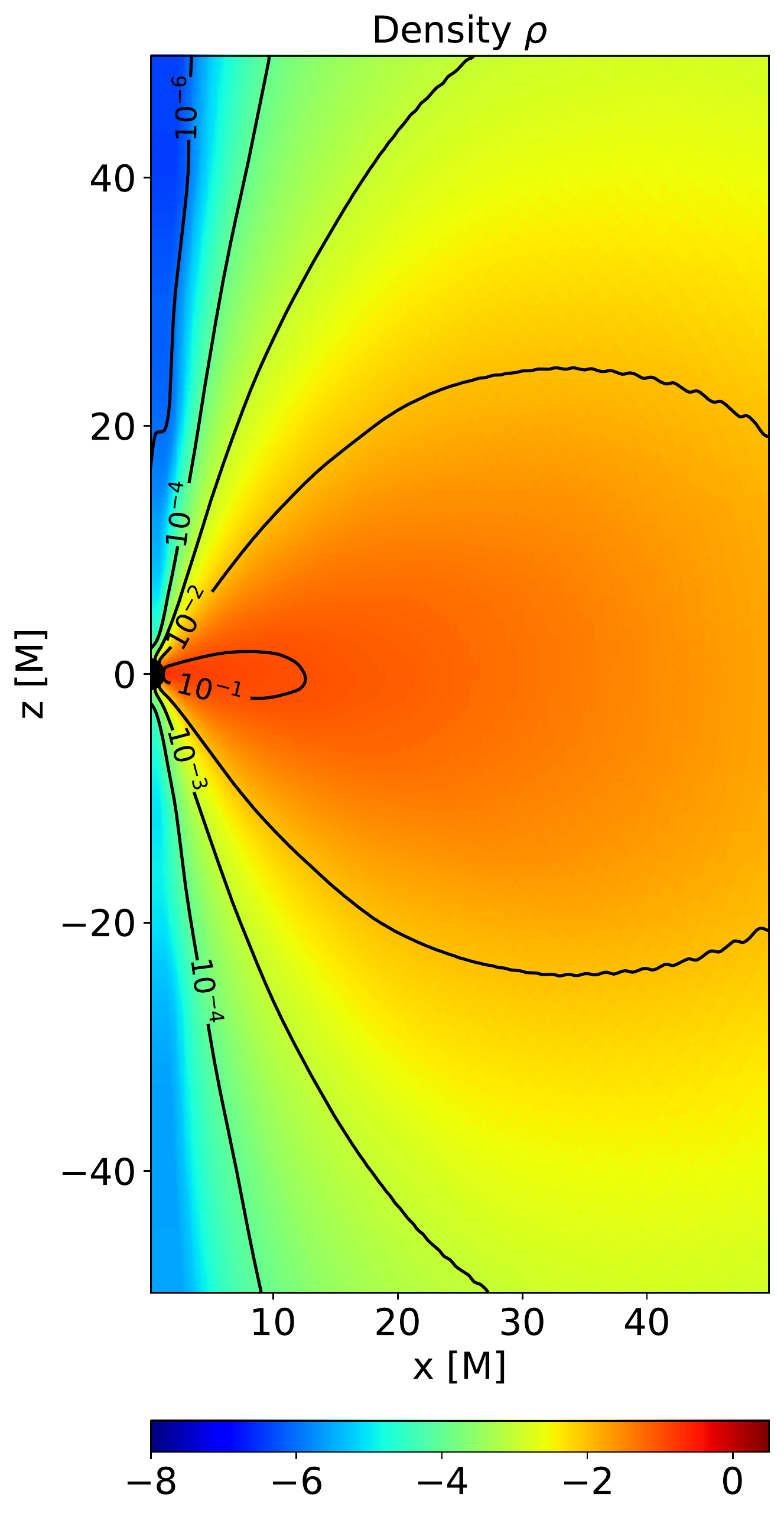}  &
    \includegraphics[width=0.25\textwidth]{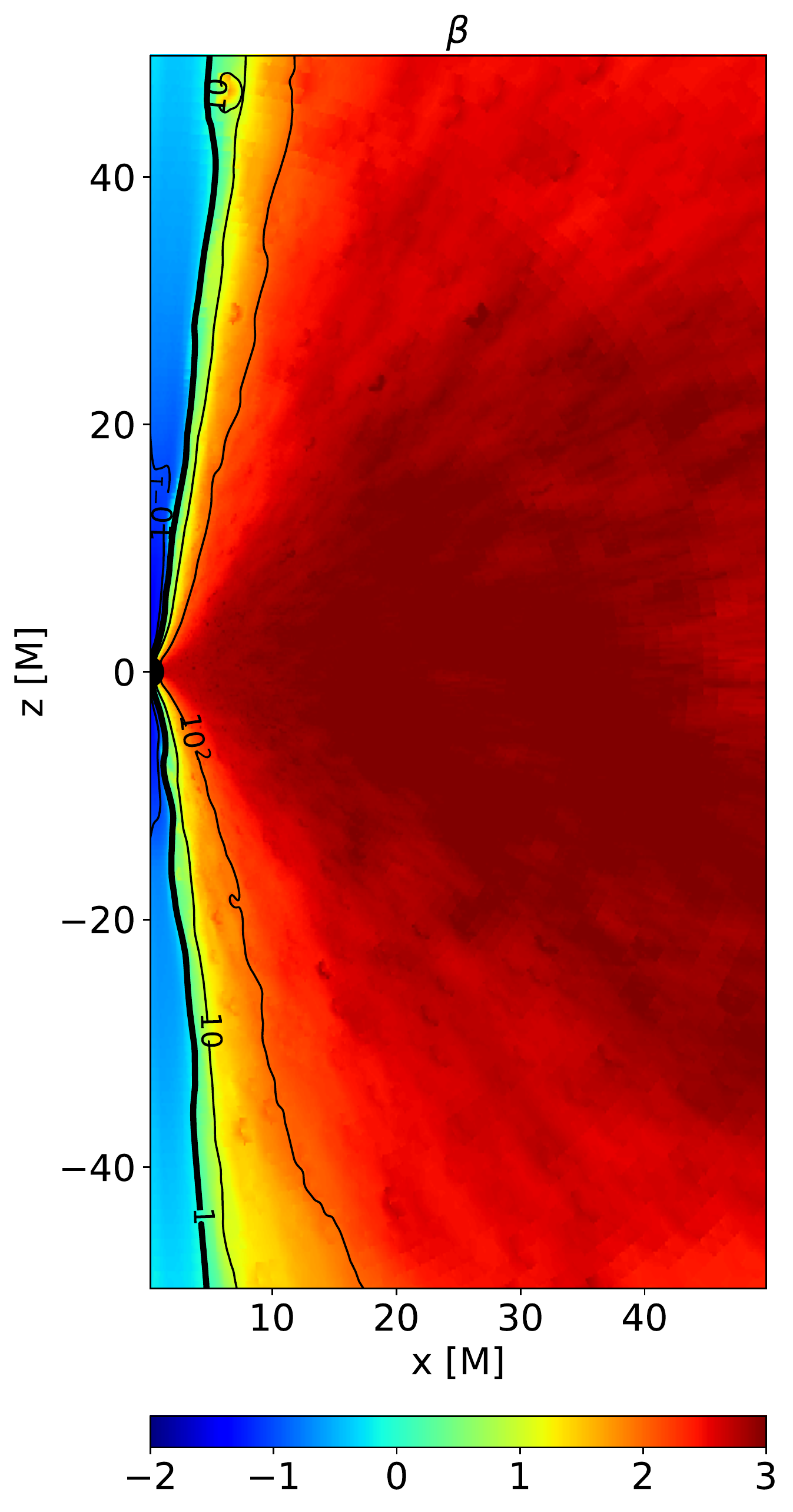} &
    \includegraphics[width=0.25\textwidth]{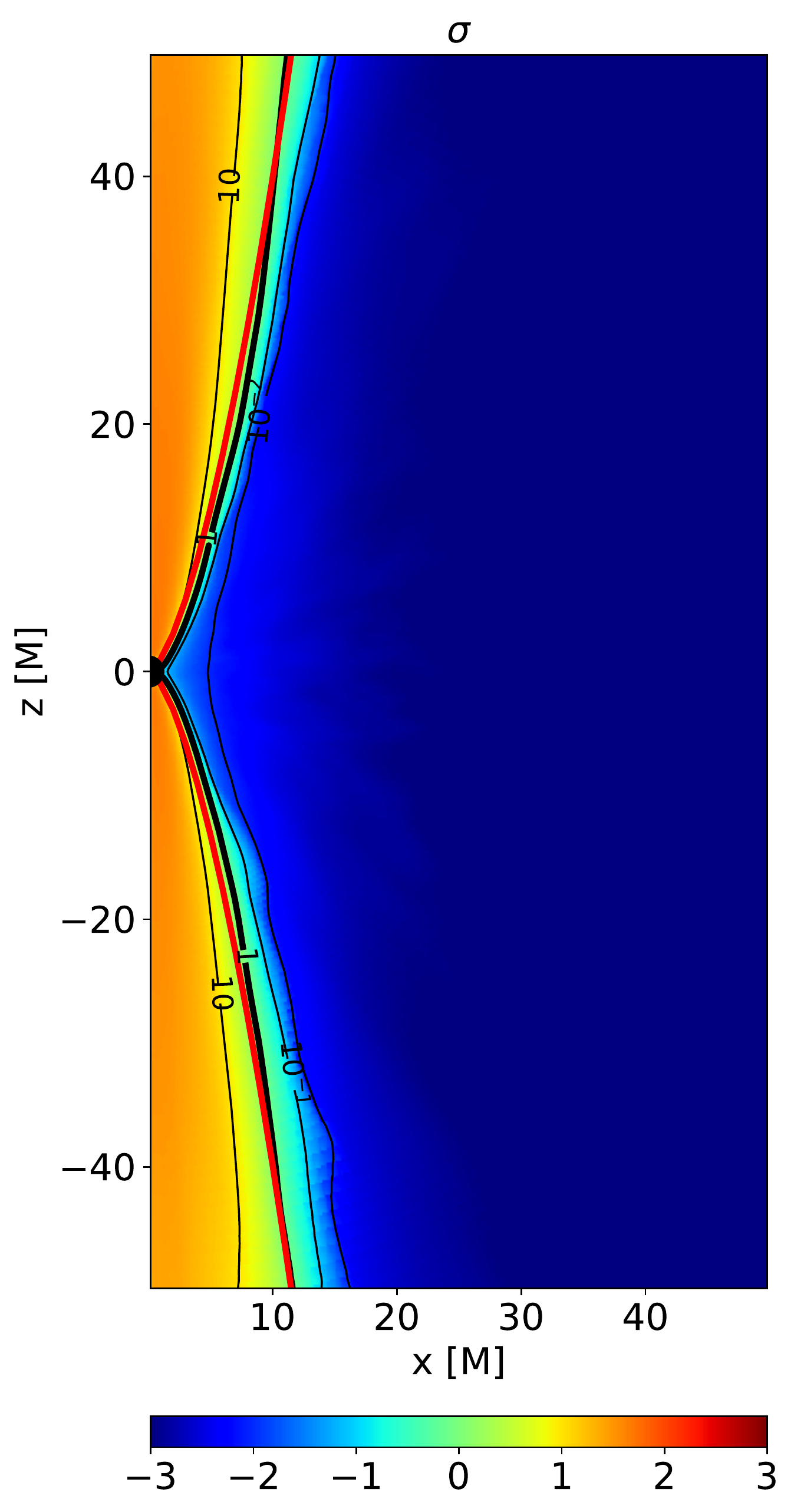} \\ 
    \includegraphics[width=0.25\textwidth]{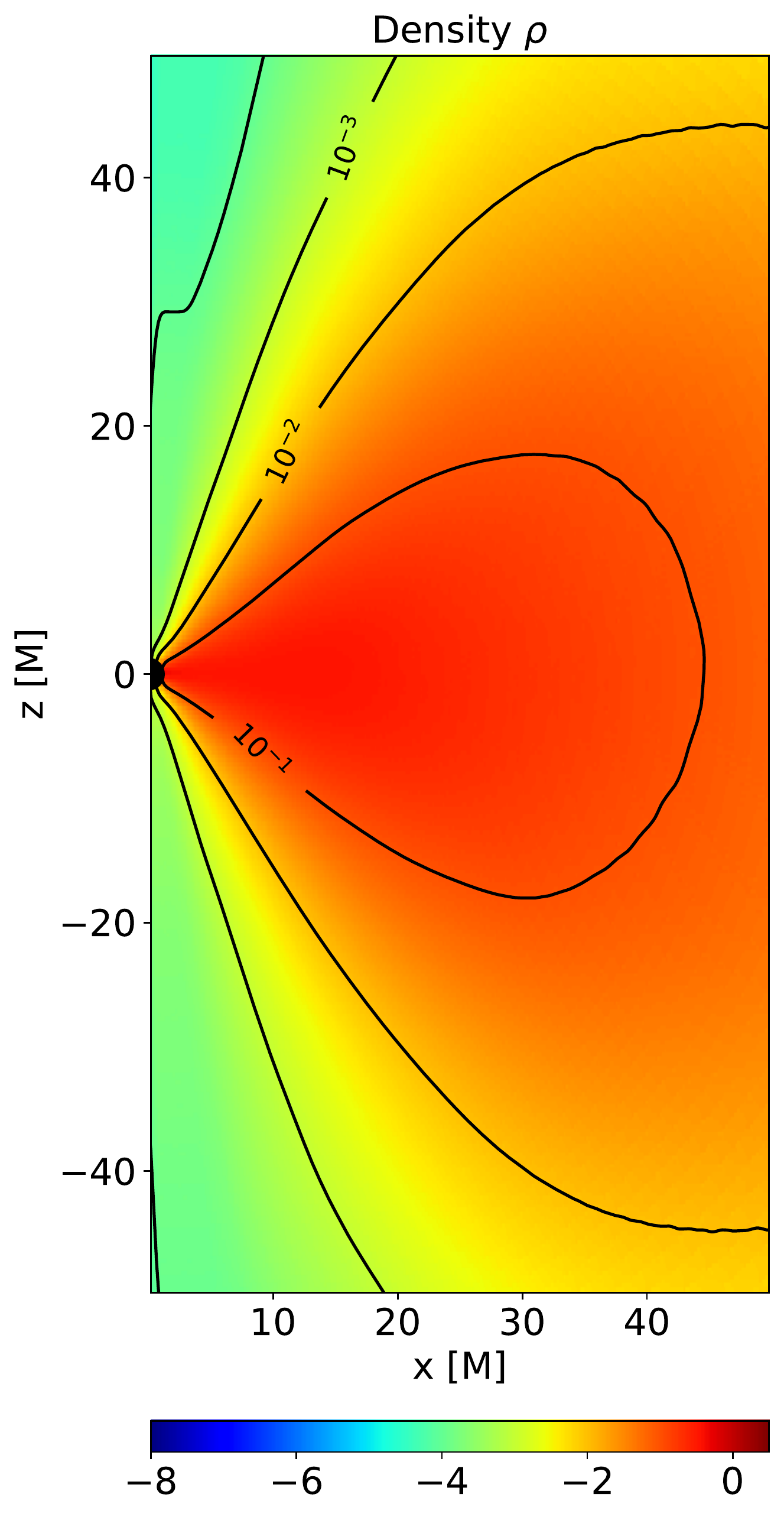}  &
    \includegraphics[width=0.25\textwidth]{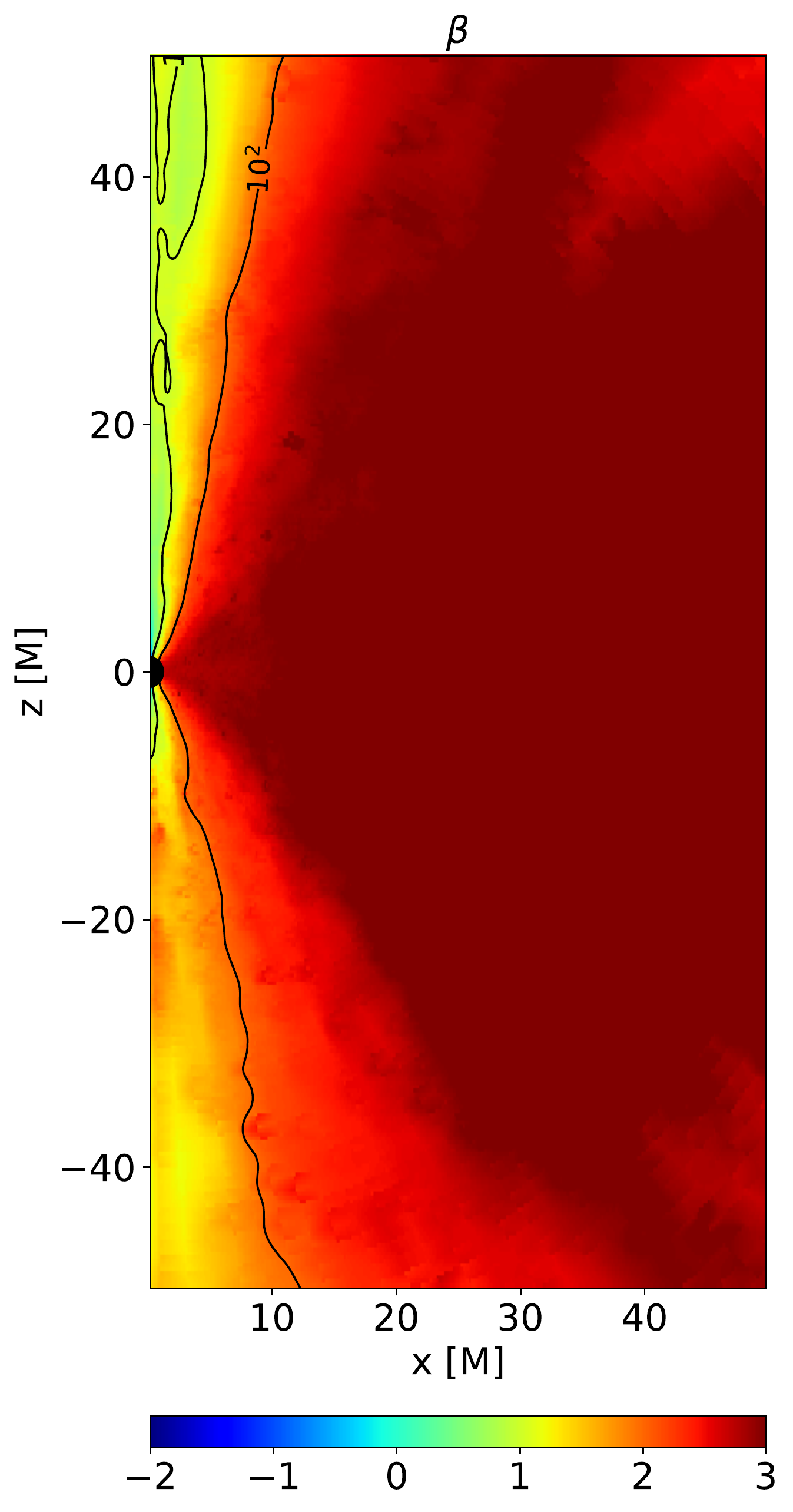} &
    \includegraphics[width=0.25\textwidth]{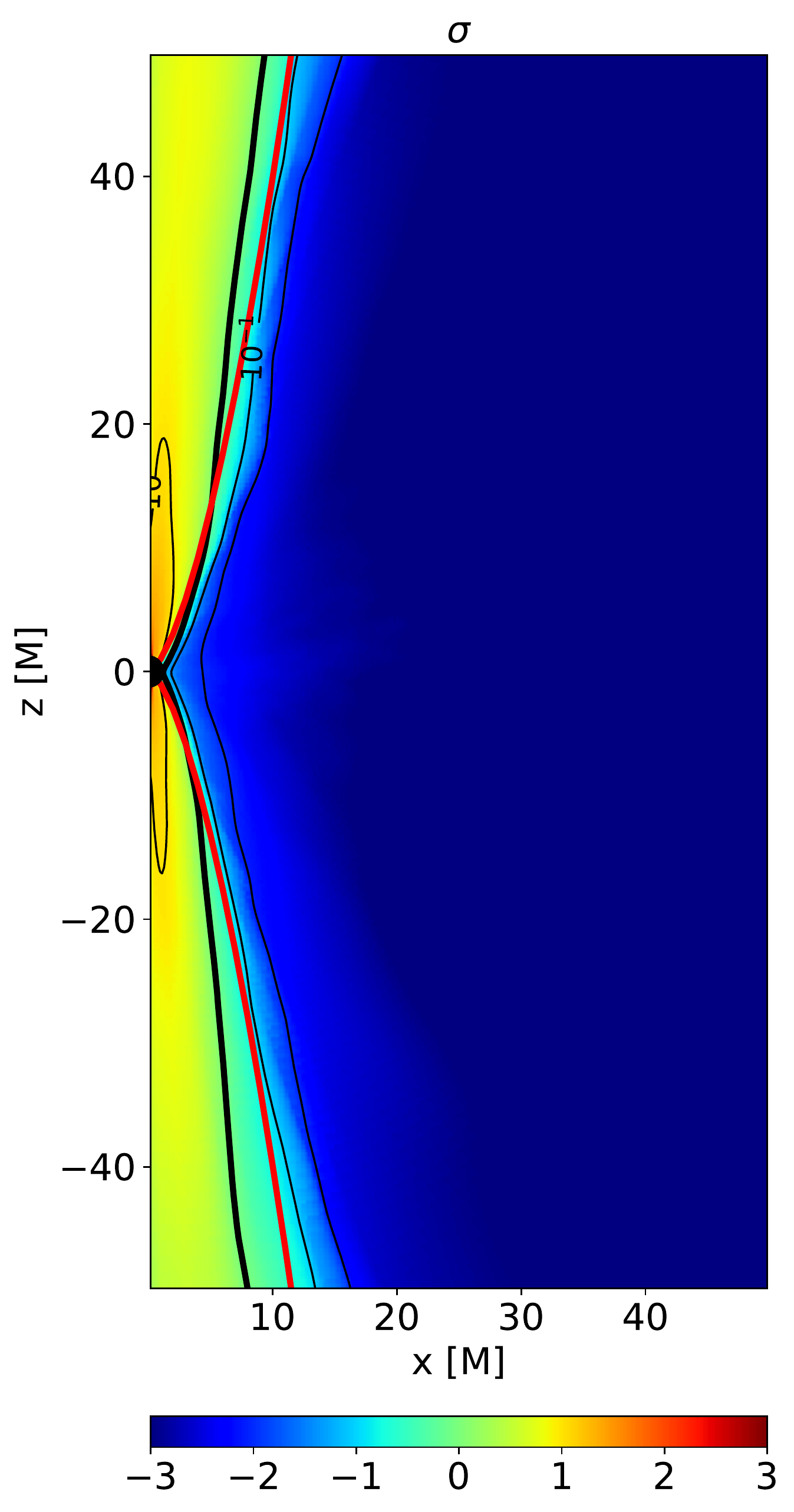}  
    \end{tabular}
    \caption{Same as Figure \ref{fig:time_azimuth_integrated}  for R3 (top) and R4 (bottom). Both have adiabatic index $\hat \gamma = 5/3$ and differ by $r_{\max}$ the value of $\beta_0$. }
    \label{fig:time_azimuth_integrated_R3_R4}
\end{figure}

\begin{figure}
    \centering
    \begin{tabular}{ccc}
    \includegraphics[width=0.25\textwidth]{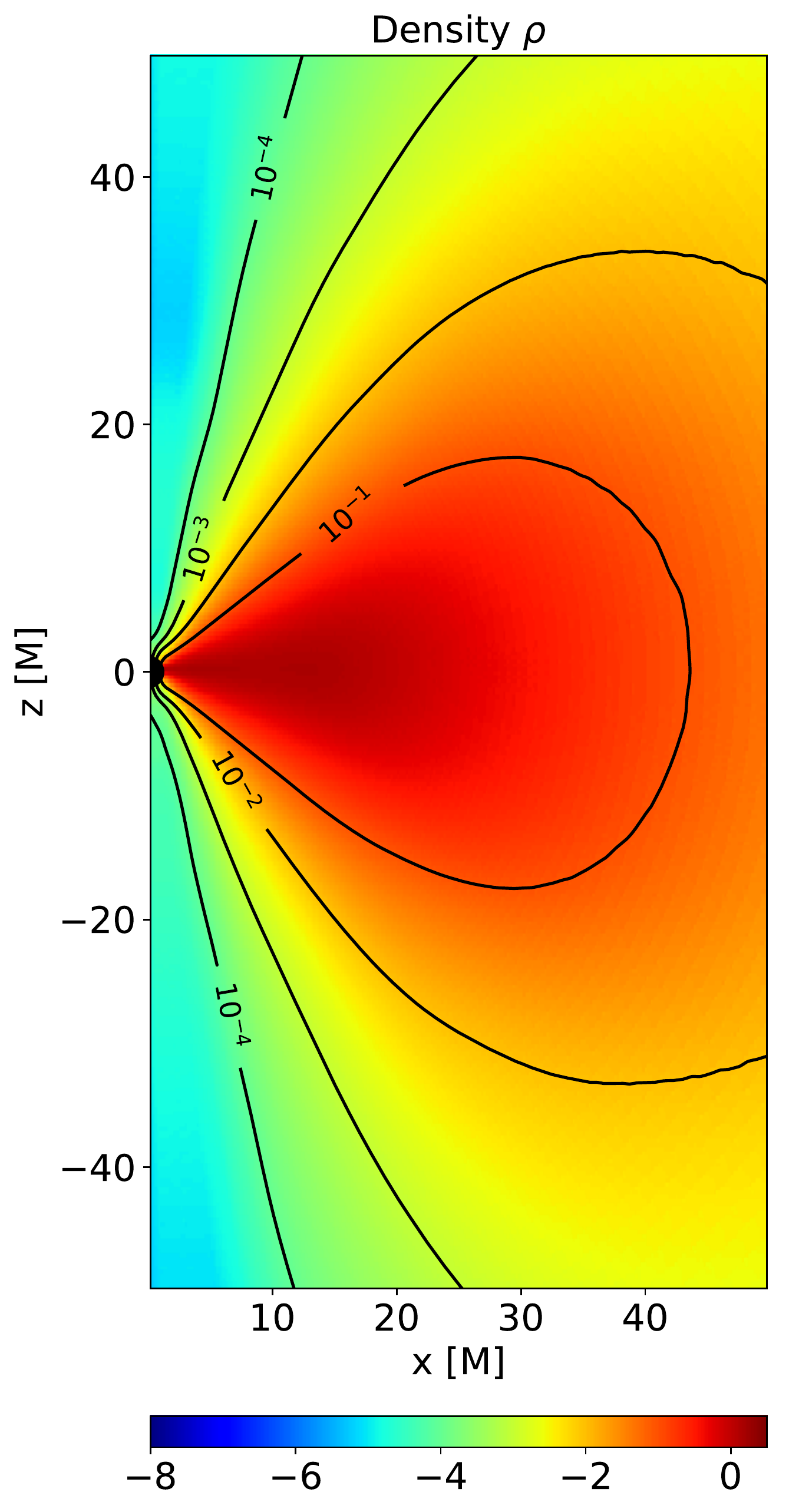}  &
    \includegraphics[width=0.25\textwidth]{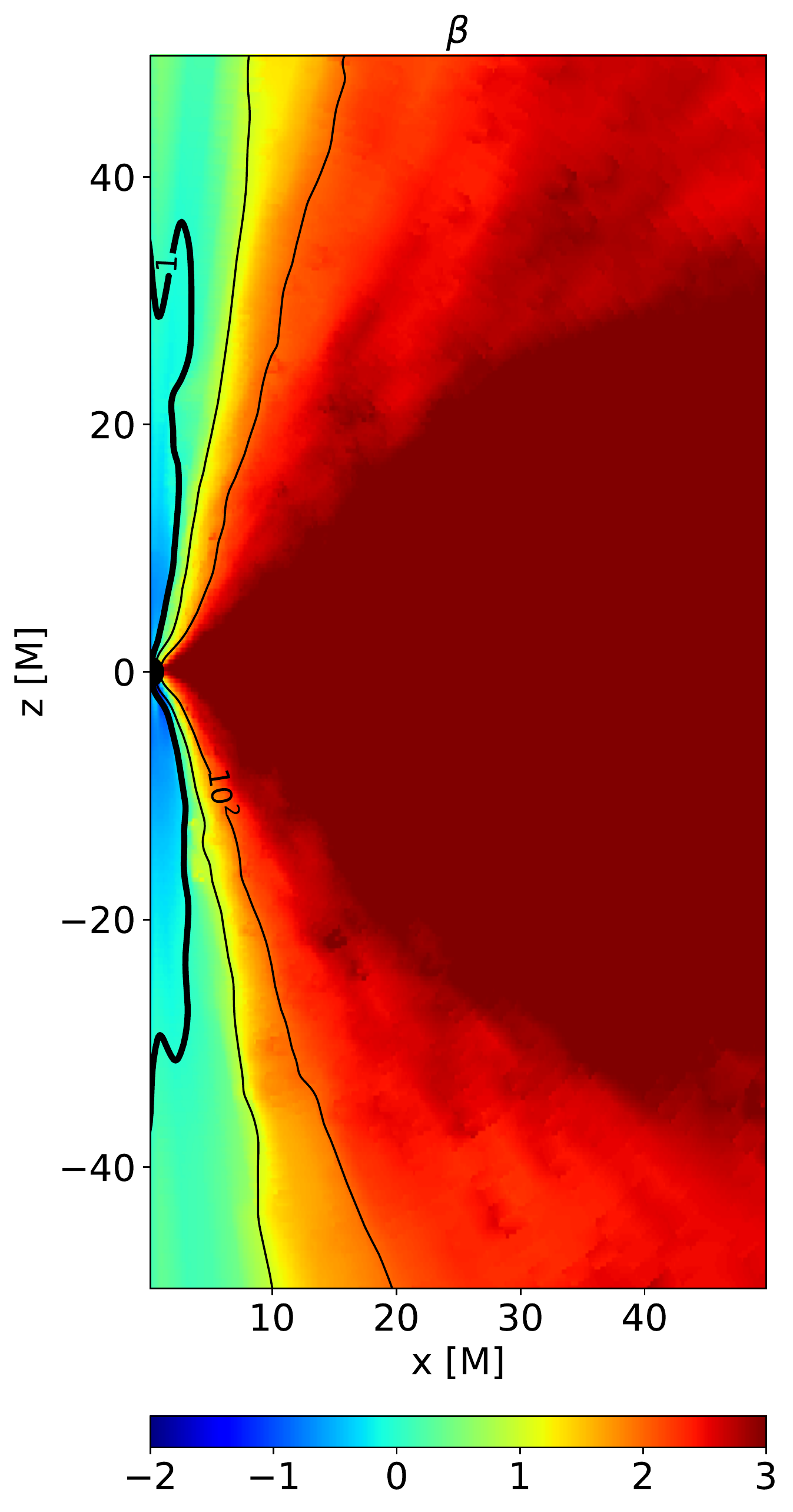} &
    \includegraphics[width=0.25\textwidth]{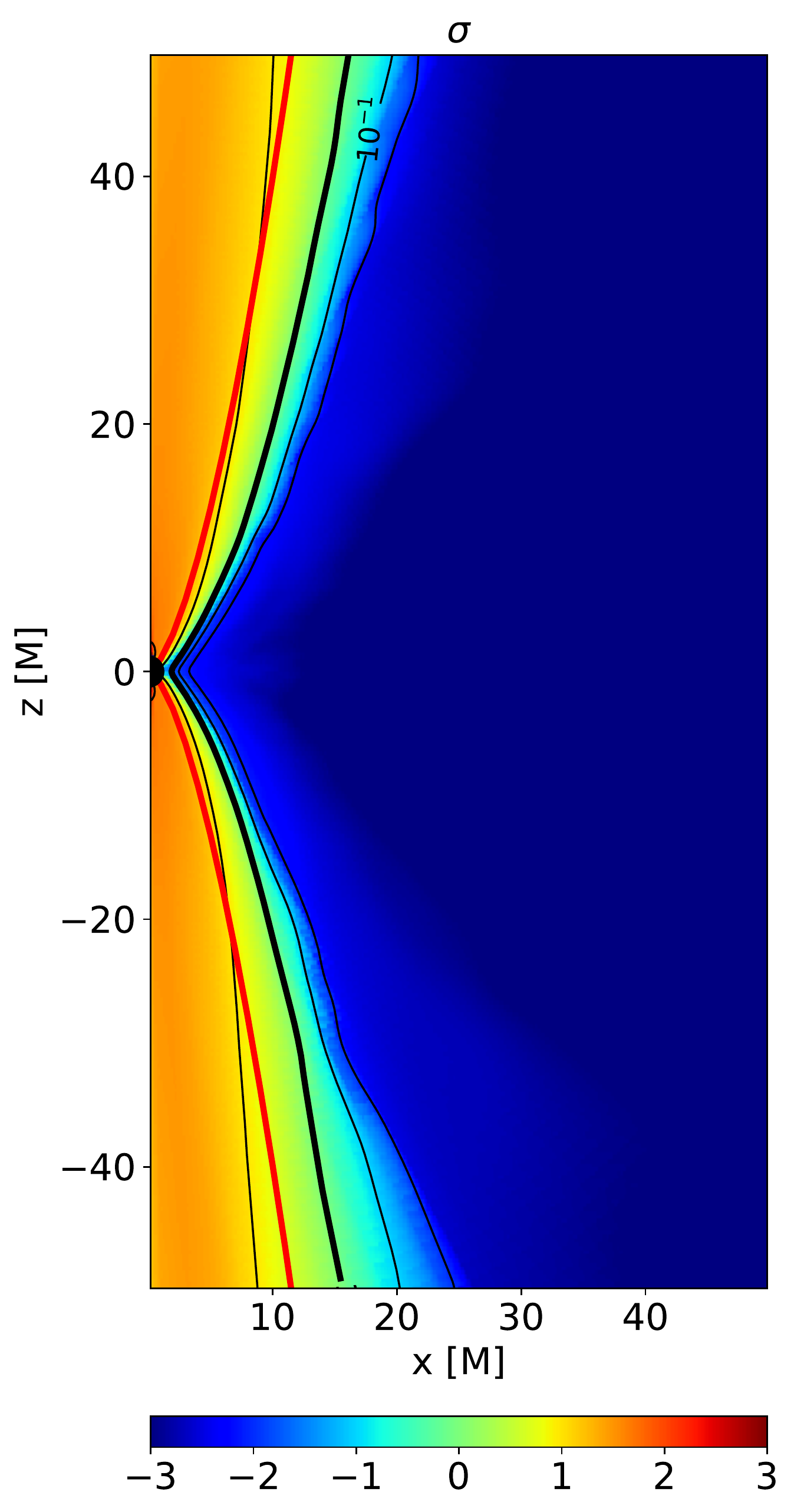} \\ 
    \includegraphics[width=0.25\textwidth]{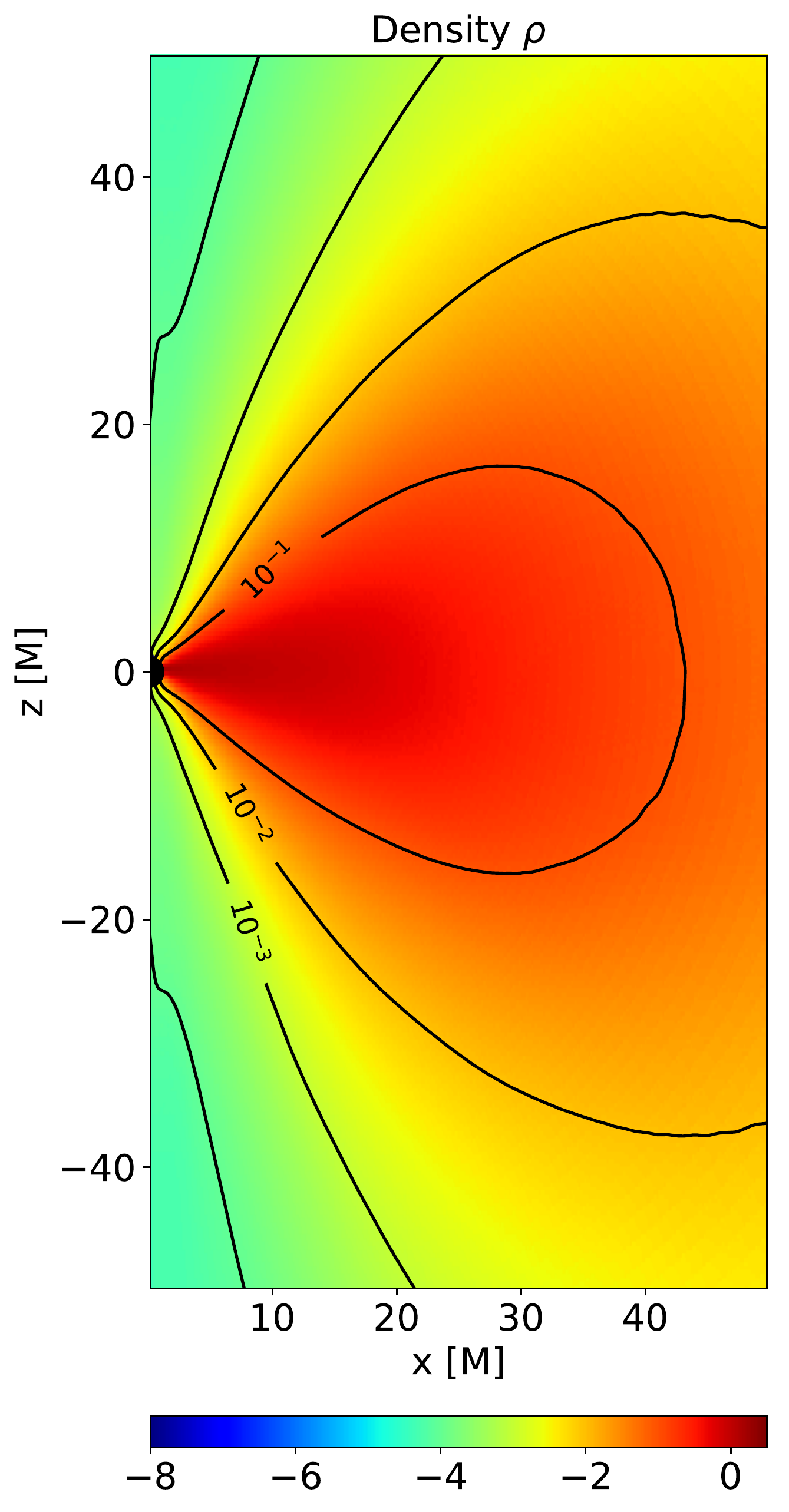}  &
    \includegraphics[width=0.25\textwidth]{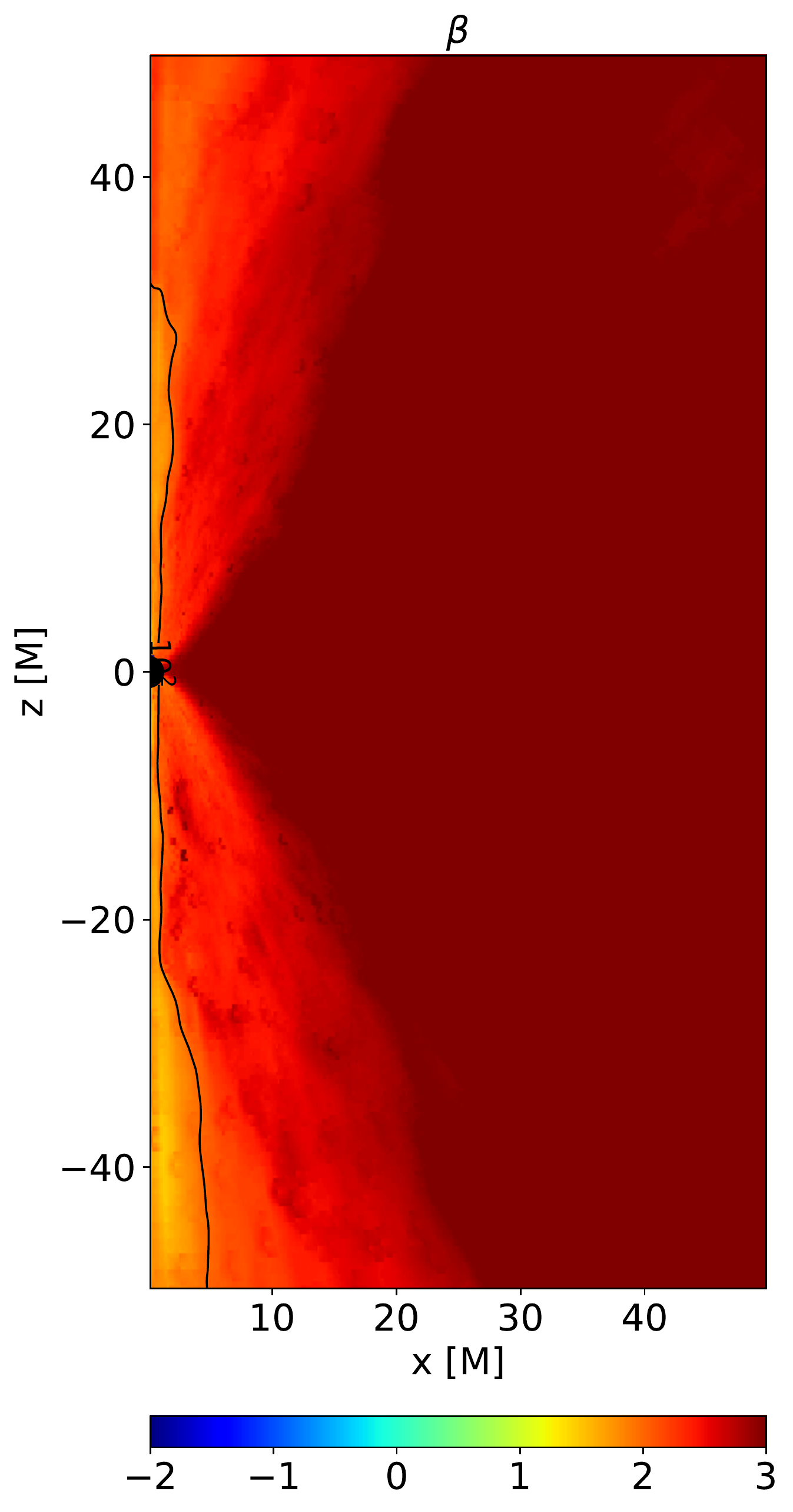} &
    \includegraphics[width=0.25\textwidth]{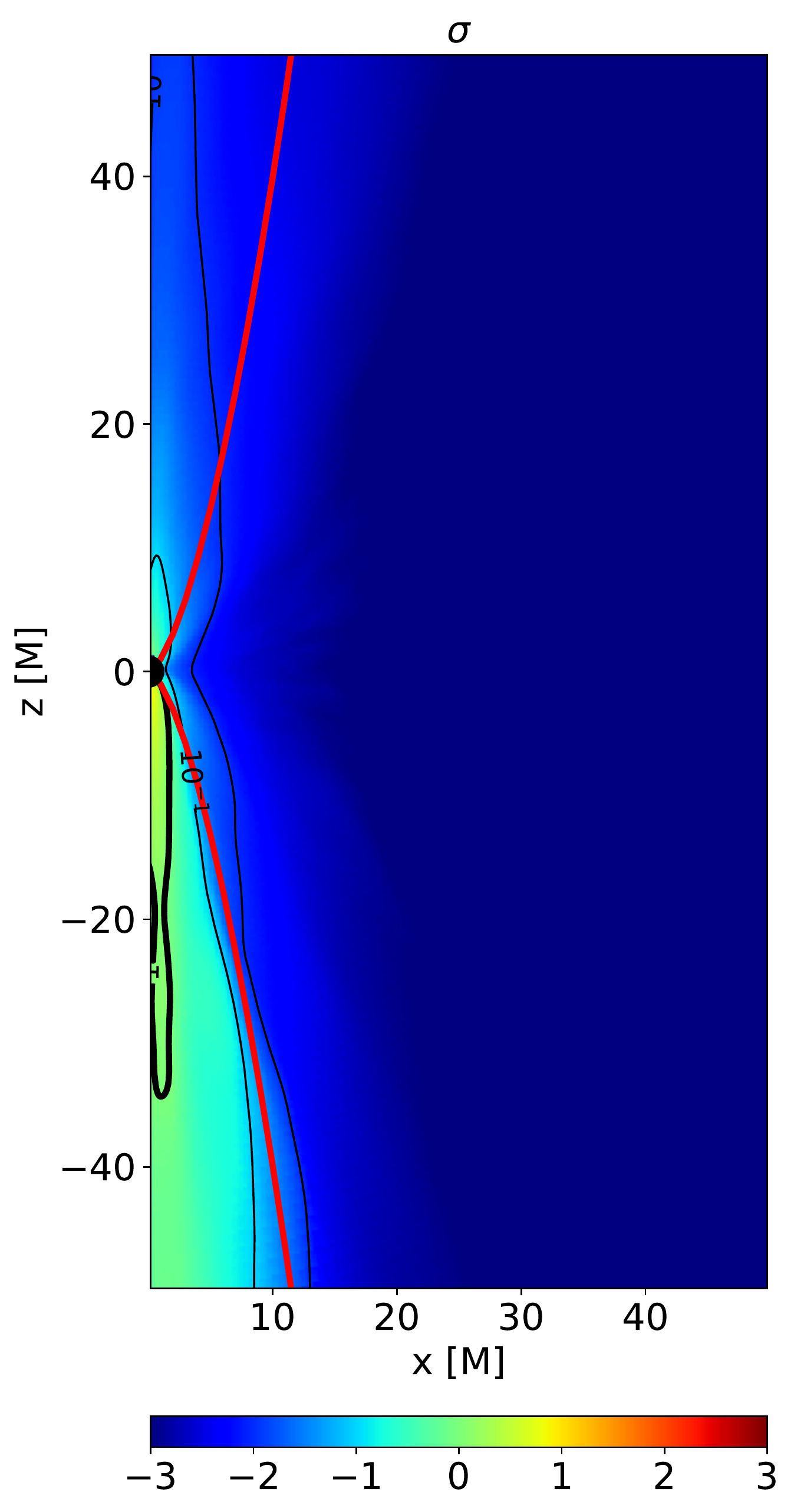}  
    \end{tabular}
    \caption{Same as Figure \ref{fig:time_azimuth_integrated}  for R5 (top) and R6 (bottom). Both runs assume initial disk configuration as expected for adiabatic index $\hat \gamma = 5/3$, yet evolve using $\hat \gamma = 4/3$. They do not reach a steady state at the end of the simulation time.}
    \label{fig:time_azimuth_integrated_R5_R6}
\end{figure}

\begin{figure}
    \centering
    \begin{tabular}{ccc}
    \includegraphics[width=0.25\textwidth]{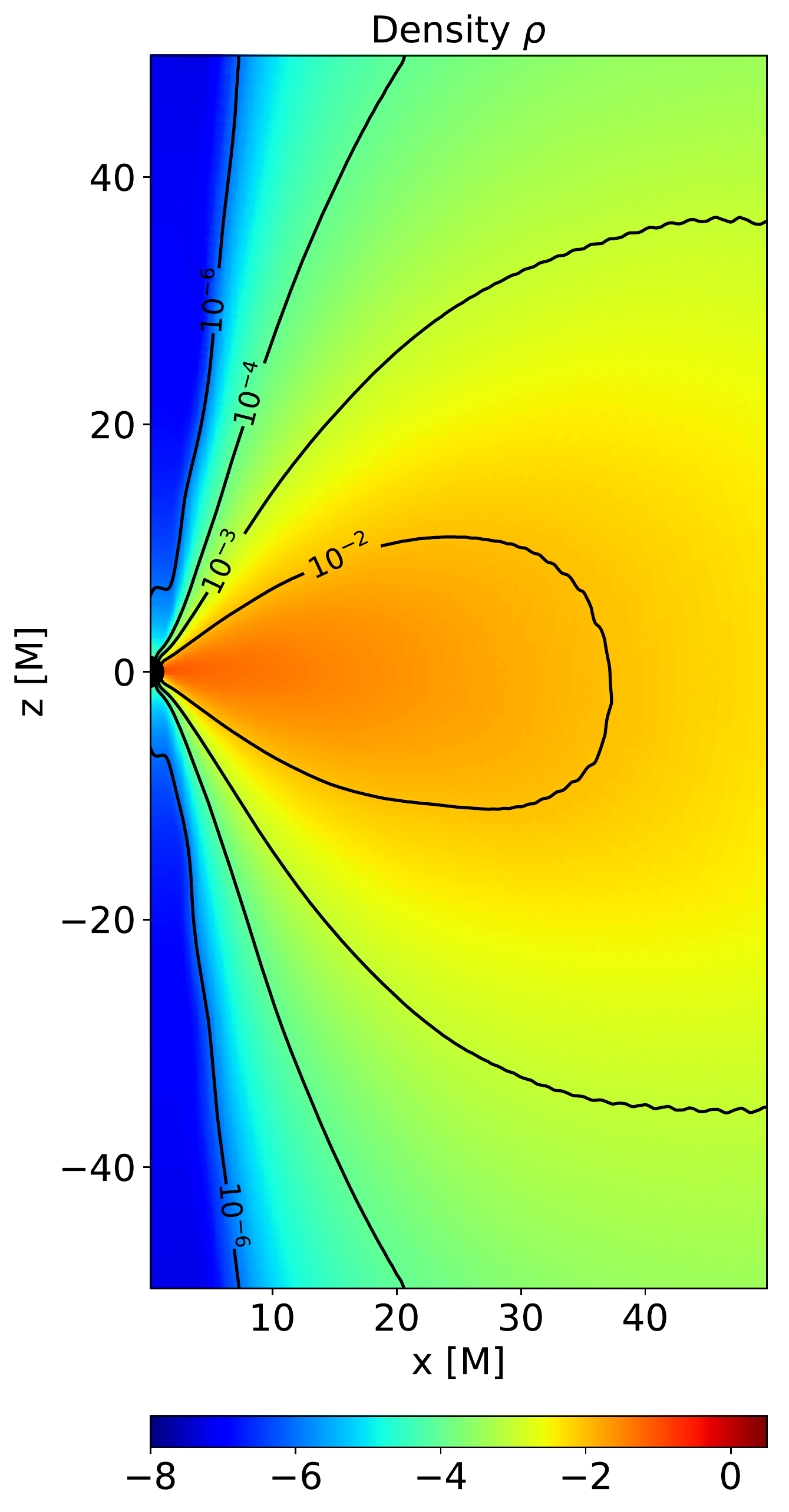}  &
    \includegraphics[width=0.25\textwidth]{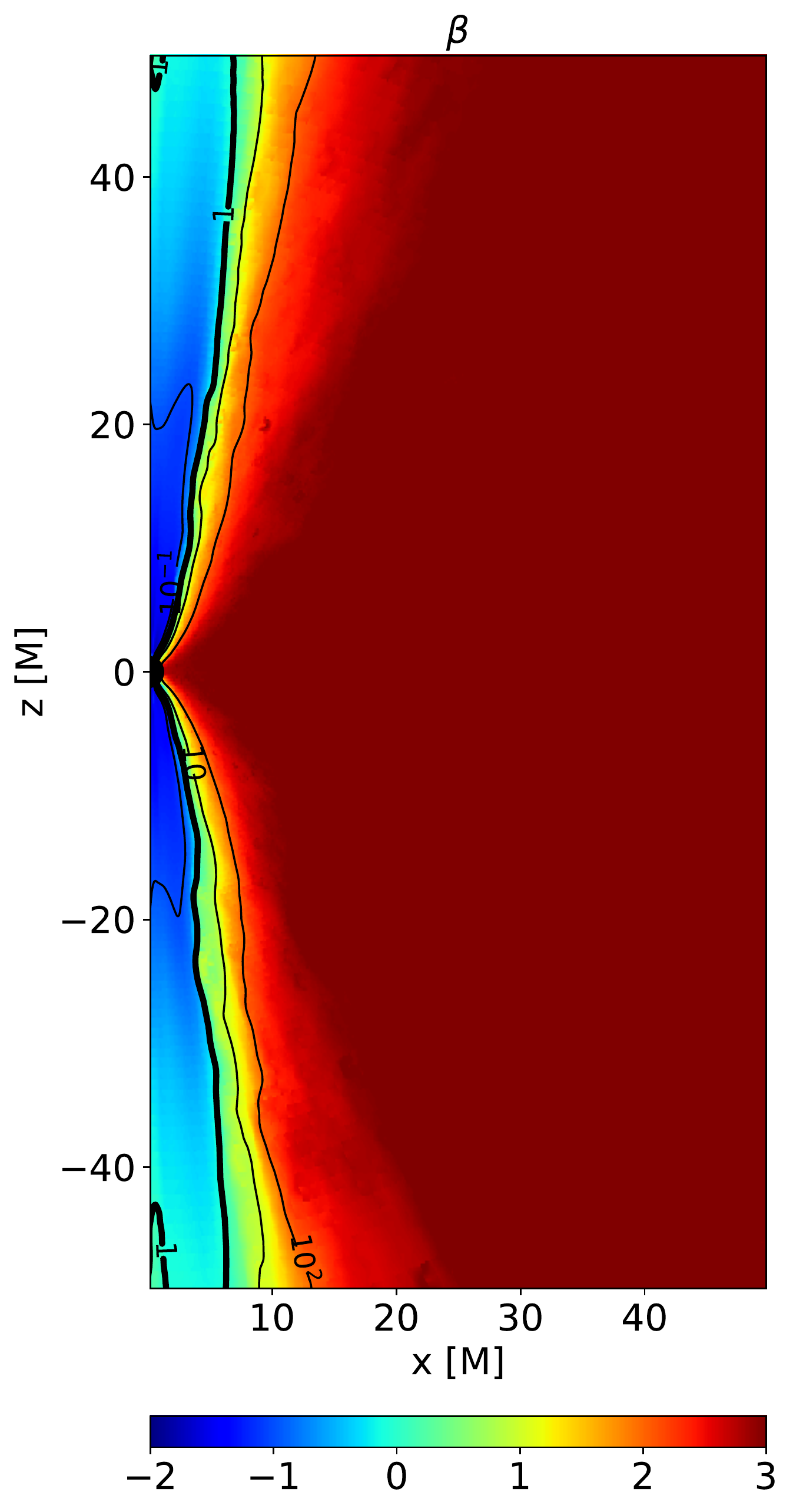} &
    \includegraphics[width=0.25\textwidth]{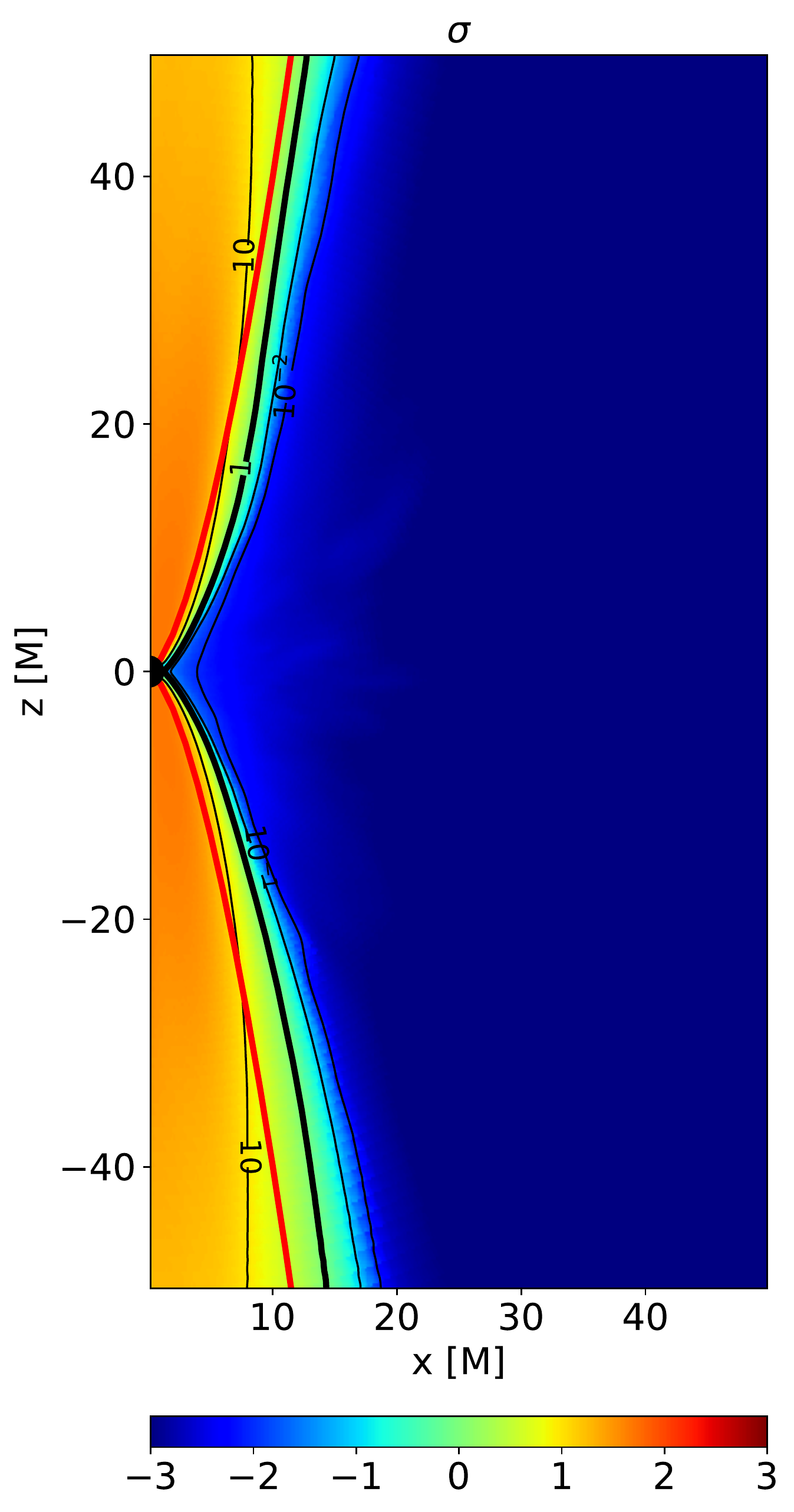} \\ 
    \includegraphics[width=0.25\textwidth]{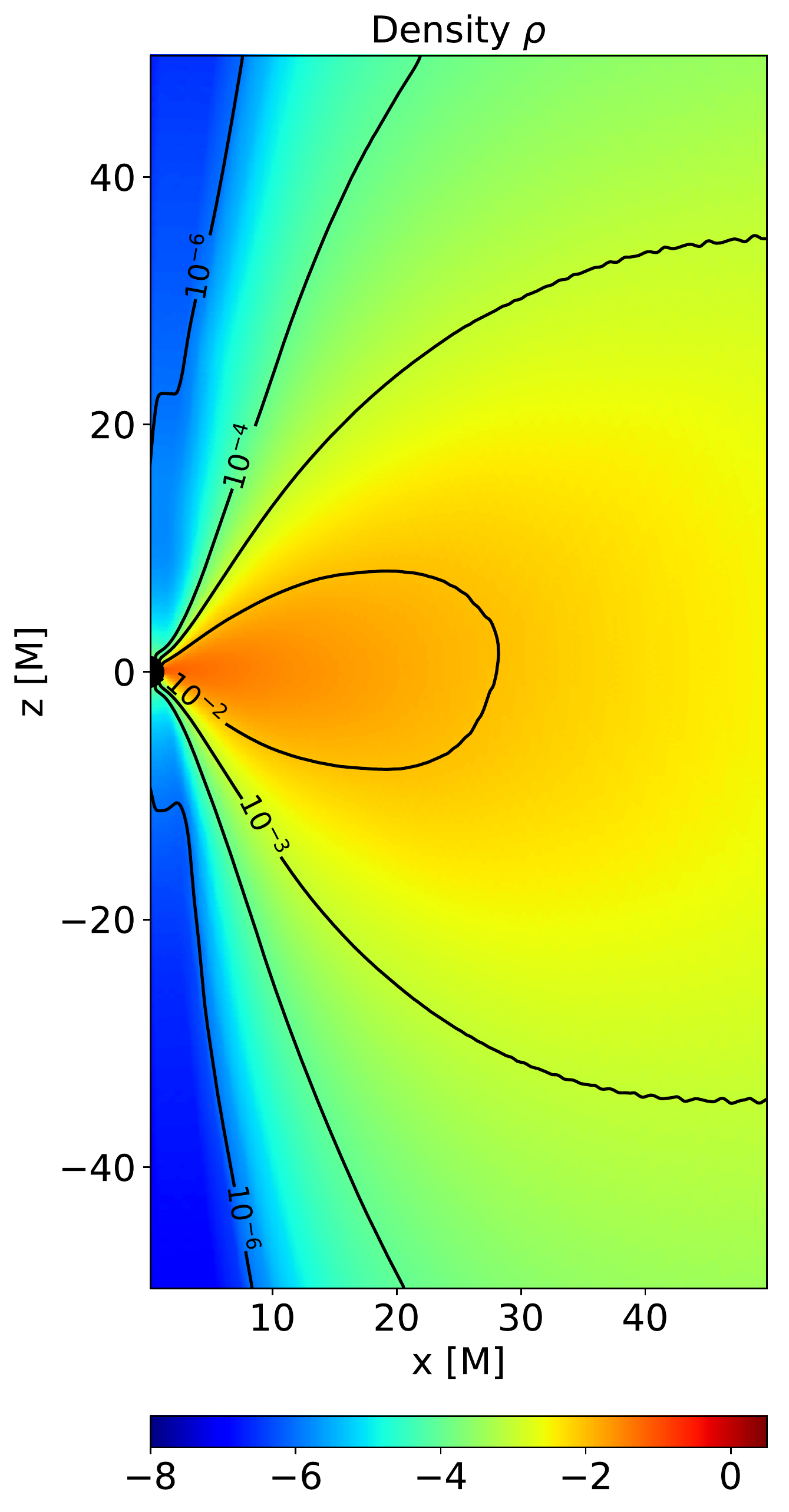}  &
    \includegraphics[width=0.25\textwidth]{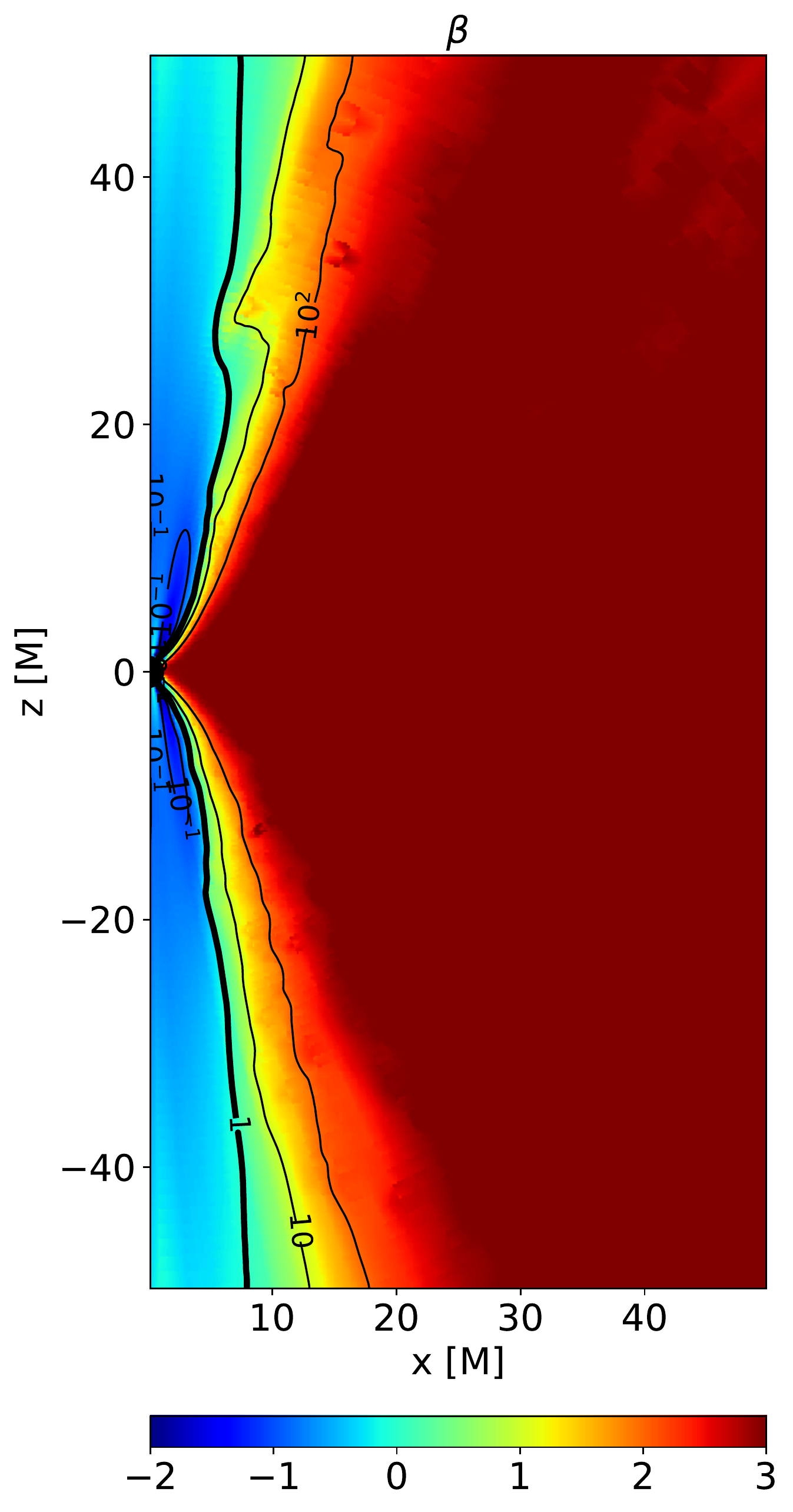} &
    \includegraphics[width=0.25\textwidth]{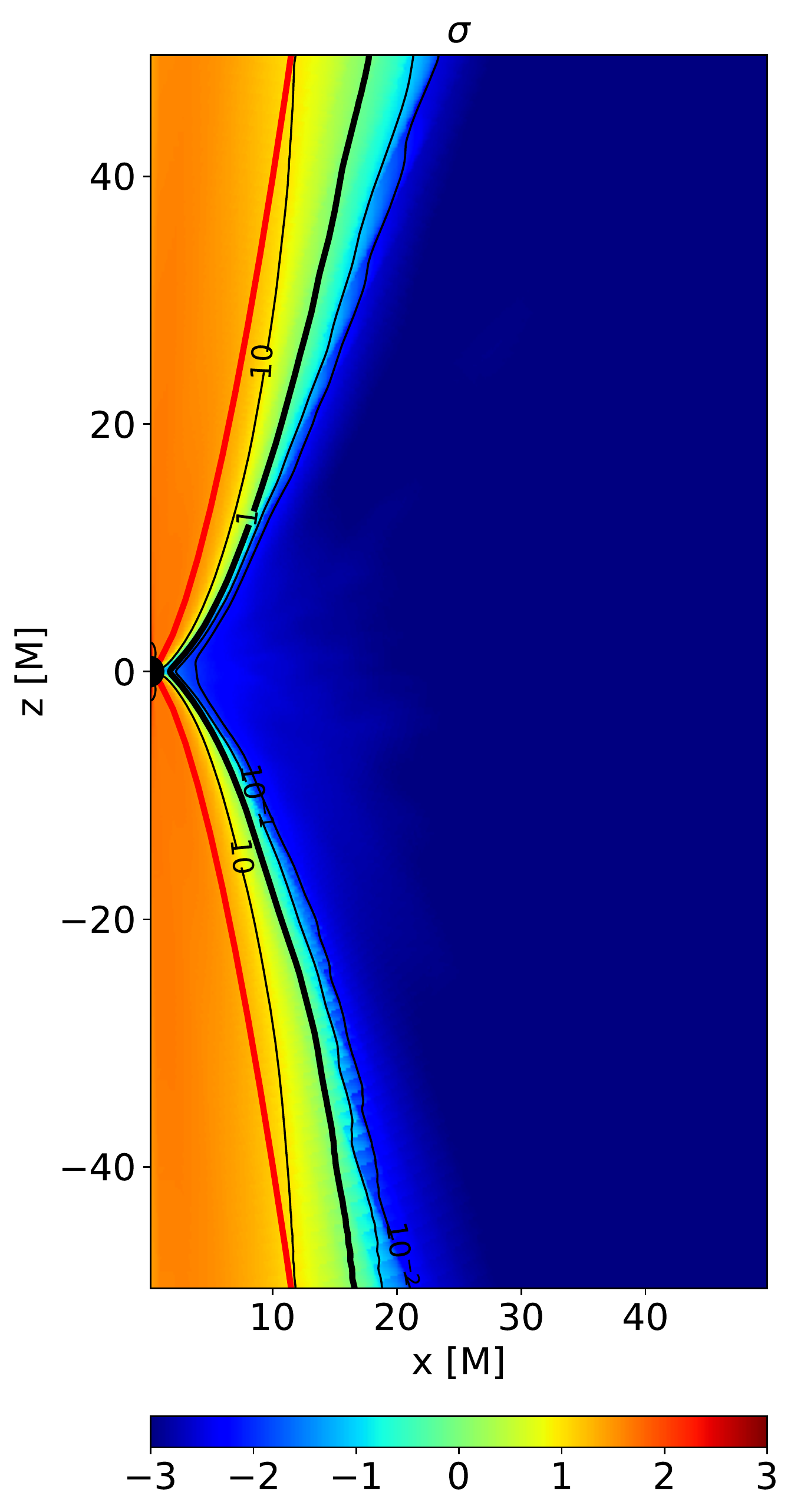}     \end{tabular}
    \caption{Same as Figure \ref{fig:time_azimuth_integrated}  for R7 (top) and R8 (bottom). They differ by the initial value of $\beta_0$.}
    \label{fig:time_azimuth_integrated_R7_R8}
\end{figure}

\begin{figure}
    \centering
    \begin{tabular}{ccc}
    \includegraphics[width=0.25\textwidth]{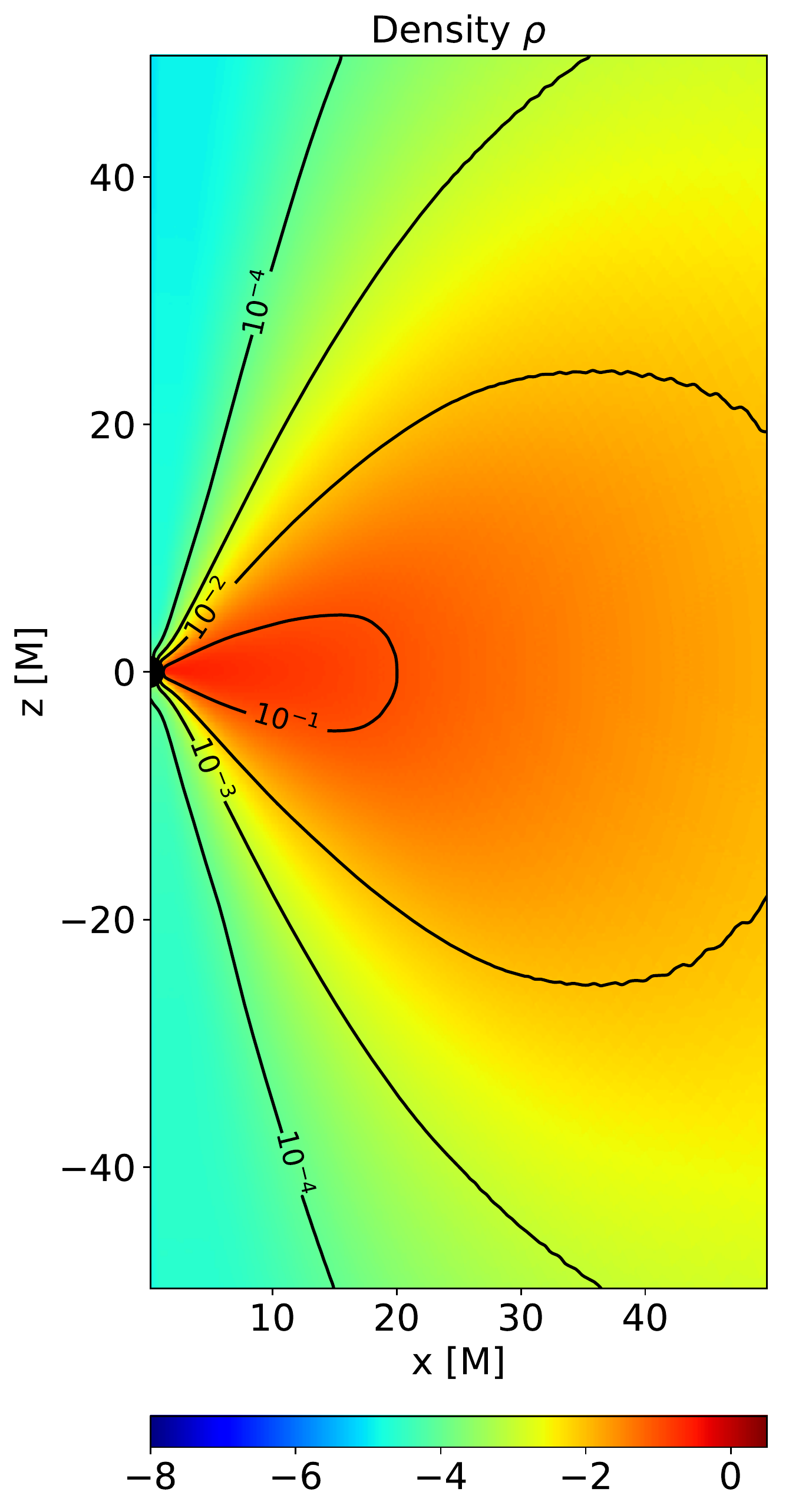}  &
    \includegraphics[width=0.25\textwidth]{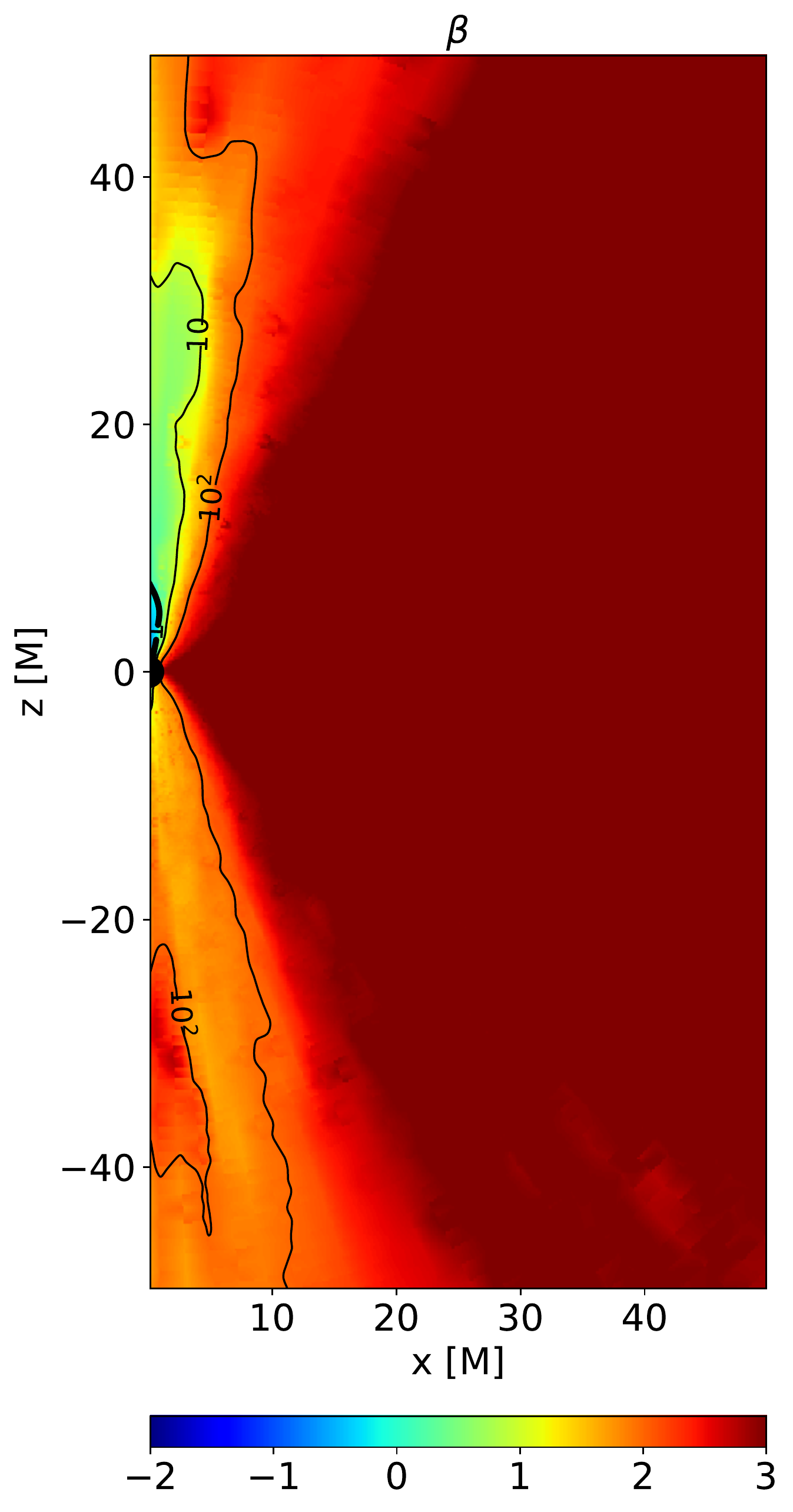} &
    \includegraphics[width=0.25\textwidth]{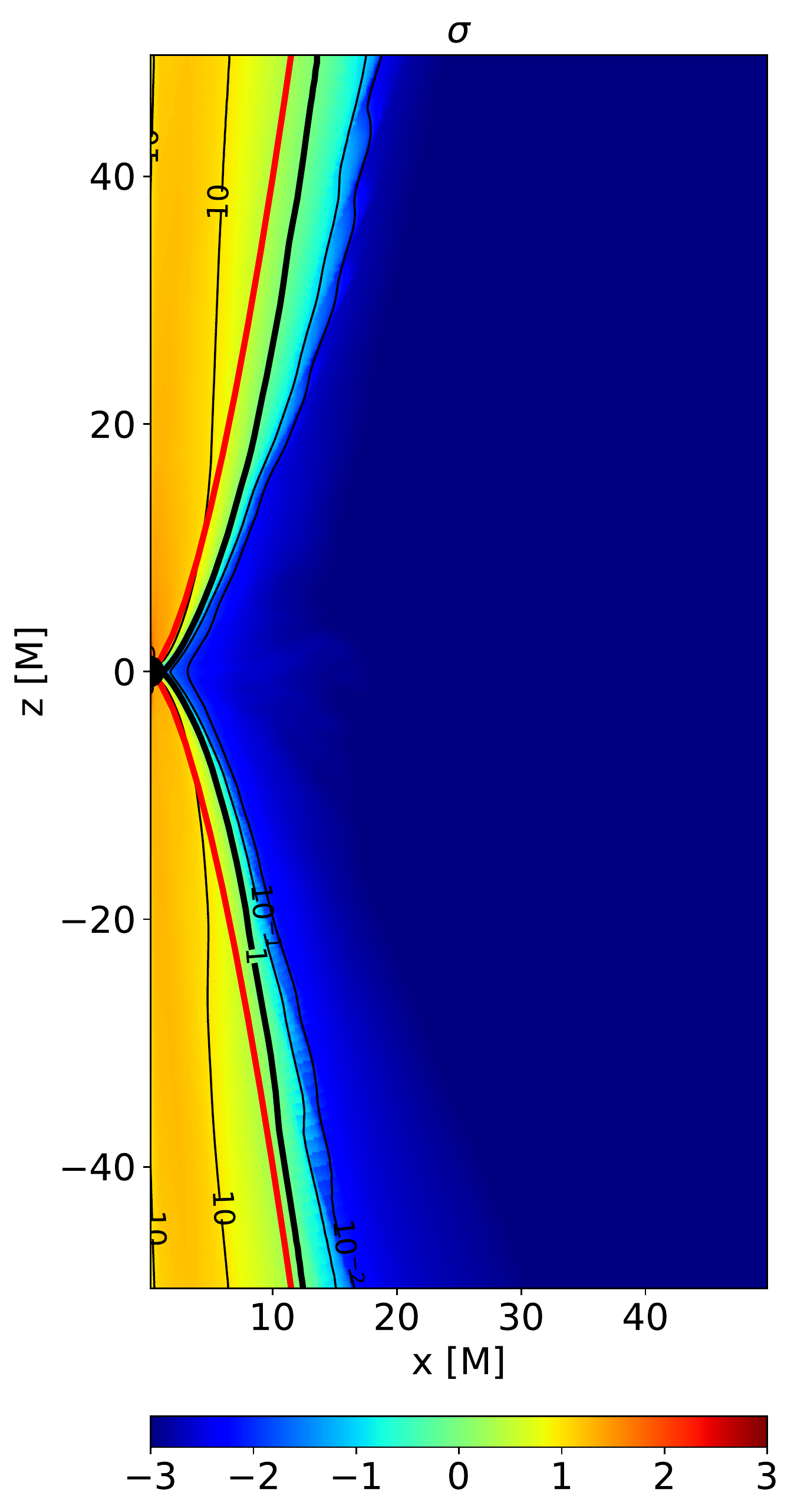}
    \end{tabular}
    \caption{Same as Figure \ref{fig:time_azimuth_integrated} for R9). This is a large disk ($R_{\max} = 13$M) with relativistic adiabatic index $\hat \gamma = 4/3$. The filled region near the polar axis indicates that a relativistic jet is not formed.}
    \label{fig:time_azimuth_integrated_R9}
\end{figure}

We now turn to study the quasi steady state structure of our disk-jet system
in order to constrain the shapes of both the disk and the jet. The temporal and azymuthal average of the density $\rho$,  the
plasma $\beta$ and $\sigma$ parameters averaged over the period $ 5\times 10^3 < t < 10^4 $M
are displayed in Figures \ref{fig:time_azimuth_integrated} --- \ref{fig:time_azimuth_integrated_R9} 
(Figure \ref{fig:time_azimuth_integrated} for R1 and R2,
Figure \ref{fig:time_azimuth_integrated_R3_R4} for R3 and R4, Figure \ref{fig:time_azimuth_integrated_R5_R6} for R5 and R6, Figure
\ref{fig:time_azimuth_integrated_R7_R8} for R7 and R8 and Figure 
\ref{fig:time_azimuth_integrated_R9} for R9).  As seen in those figures, in simulations R1, R2, R3,
R7 and R8 the quasi steady-state equilibrium is composed of three regions. (i) A highly
magnetized region characterised by $\rho \ll 1$, $\beta \ll 1$ and
$\sigma \gg 1$- this is the jet. (ii) A weakly magnetized disk ($\rho \sim 1$, $\beta \gg 1$
and $\sigma \ll 1$). And (iii) an intermediary narrow region where $\rho \ll 1$,
$\beta \sim 1$ and $\sigma \sim 1$ (around the thick black line in Figures \ref{fig:time_azimuth_integrated} ---  \ref{fig:time_azimuth_integrated_R9}). Although formally delimited by the
Bernoulli parameter $u^t > 1$
\citep[which is close to the $\beta = 1$ surface, see \textit{e.g.}][]{MG04,PCN19},
this intermediary region is called the jet sheath \citep{DMA12,MF13,MFN16,DOP19}. We note
that \citet[][]{MG04} also introduced the concept of corona by defining a more
extended region around the jet sheath defined by $ 1 < \beta < 3$
\citep[see also][]{DHK03}.

Runs R4 and R9, which have the largest initial disks and therefore the largest initial mass, 
do not show the formation of a jet; at best R9 has an intermittent one. This
is clearly visible in Figures \ref{fig:time_azimuth_integrated_R3_R4}
and \ref{fig:time_azimuth_integrated_R9}. The polar region is filled by low
to average density plasma with the $\beta$ parameter in the order of the
unity. 
As shown in Table \ref{tab:diagnostic_horizon}, R4 and R9 are
characterized by a small MAD parameter $\langle \Phi_B \rangle_t < 1.5$,
showing that the magnetic flux threading the black-hole is too weak to support
the launch of a jet in those simulations. Yet, R4 and R9 are the runs with the
largest mass accretion rate (discarding out-of equilibrium runs R5 and R6). 

Some differences in the disk and jet structures between the different runs can be seen from the figures. We describe in the sub-sections below these differences and their origin.

\subsection{Adiabatic index: R2 vs R3 and R4 vs R9}

\label{sec:adiabatix_index}

In the list of our simulations, there are two sets which vary only by the
value of their adiabatic index (relativistic-$\hat \gamma = 4/3$ vs. 
non-relativistic, $\hat \gamma =5/3$). The first set of simulations comprises
R2 and R3 and the second one is R4 and R9. The two sets differ by a
different value of the radius of maximum pressure, $r_{\rm max} = 12$M and
$r_{\rm max} = 13$M, as well as by the initial plasma $\beta$ parameter with
$\beta_0 = 100$ and $\beta_0 = 44$, the smaller initial disks (R2, R3) assume the larger
$\beta_0$. Both sets of simulations share the same characteristics: 1) both
R2 and R3 have a jet, while both R4 and R9 do not; 2) the MAD parameter is
consistent being around 2 for R2 ad R3 and being smaller - $\Phi_B \sim 1.4$ for R4
and R9. A noticeable difference is that the jet in R3 is narrower
than that of R2, as is clearly seen in Figures \ref{fig:time_azimuth_integrated} and
\ref{fig:time_azimuth_integrated_R3_R4}.

\begin{figure}
    \centering
    \includegraphics[width = 0.75 \textwidth]{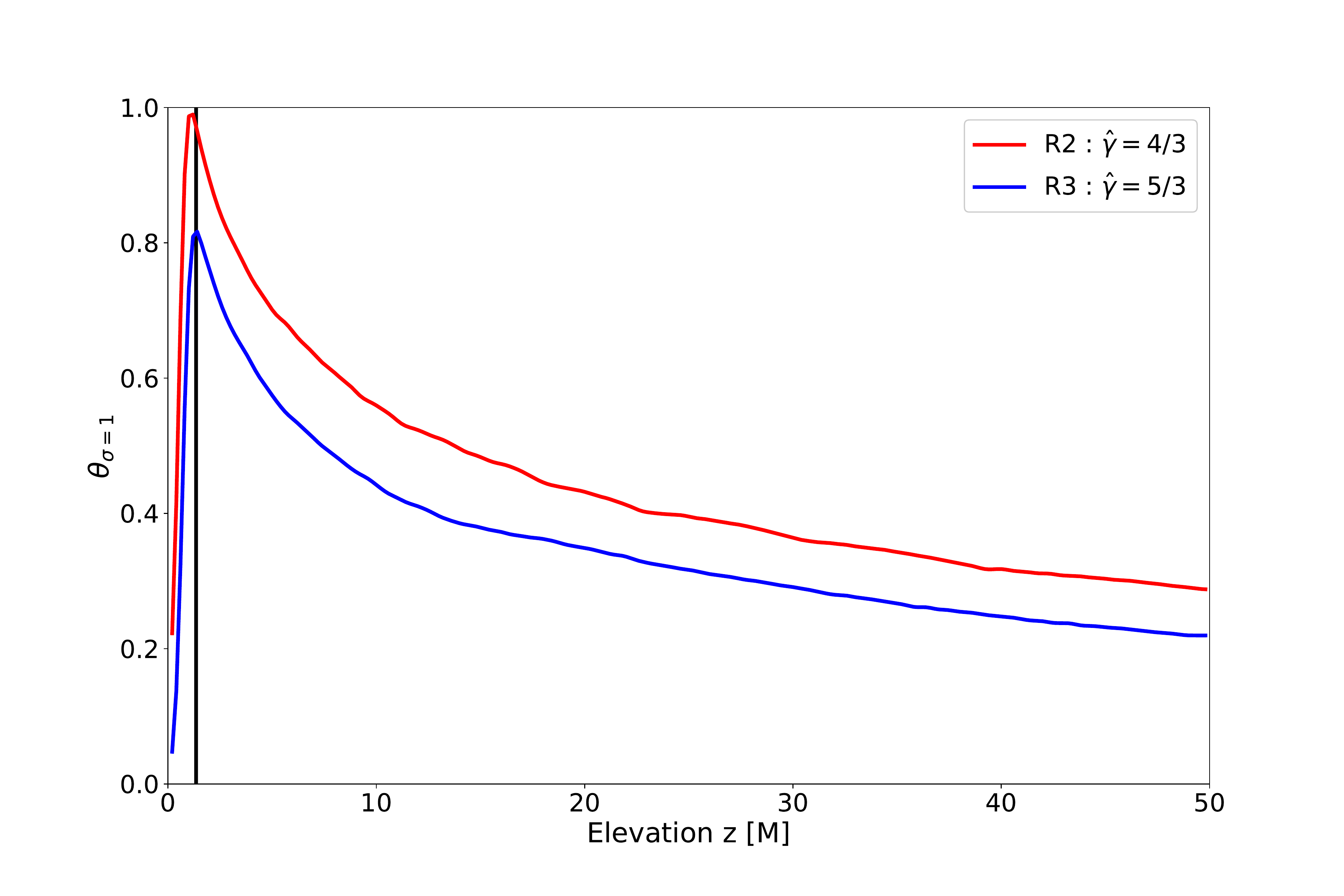}
    \caption{Time and azymuthal angle ($\phi$) average location in the $z-\theta$ plane of the $\sigma = 1$ surface (plotted by the thick black line in Figures \ref{fig:time_azimuth_integrated} and \ref{fig:time_azimuth_integrated_R3_R4})
    for runs R2 and R3. Both runs have a similar initial setup, differing only by the value of the adiabatic index ($\hat \gamma = 4/3$ in R2, $\hat \gamma = 5/3$ in R3). Larger adiabatic index results in higher gas pressure inside the disk, making the resulting jet narrower. The
    vertical black line marks the outer horizon radius.}
    \label{fig:sigma=1_r2_R3}
\end{figure}

This difference is further illustrated in Figure \ref{fig:sigma=1_r2_R3}, where we compare the $\sigma = 1$ surface of runs R2 and R3, both showing jets. In the figure, we show the time average angle at which
$\sigma = 1$ as a function of the elevation $z = r
\cos(\theta)$. It is clear that R3 has a wider jet than R2, at all altitudes. We explain this result by the fact that a larger adiabatic index results in a higher gas pressure inside the disk, making the resulting jet narrower.
Yet we note that we cannot completely exclude the possibility that at least part of this effect is due to the fact that the higher adiabatic index implies that the initial disk in R3 is thicker, and is $\sim 2.5\times$ more massive than that of R2. This extra mass by itself may affect the resulting jet structure. We believe that this is the main source of the difference in the disk structures seen in the more massive disks comparison (R4 vs. R9), where lower adiabatic index implies less dense inner disk (compare Figures \ref{fig:time_azimuth_integrated_R3_R4} and \ref{fig:time_azimuth_integrated_R9}).

Except for the jet opening angle, we do not observe additional substantial differences that can be attributed directly to the adiabatic index. 
This is consistent with conclusions reached in several past studies
\citep[e.g.,][]{MG04, MM07}. Furthermore, it is currently not clear what observational
data could be used to constrain the adiabatic index of the flow. For
example, \citet{BMD20} studied the variability of the mass accretion rate
in very long numerical simulations of accretion disks with relativistic
and non-relativistic adiabatic index. They did not find major qualitative
differences between their simulations and their post-processed results,
specifically for the variability in the mass accretion rate. 




\subsection{Effect of initial magnetic field parameter, $\beta_0$ : R1, R2, R7 and R8}

\label{sec:magnetic_field}


We conducted four simulations aimed at exploring the effect of the initial value of the parameter $\beta_0$ (maximum value of the initial magnetic field pressure inside the disk, normalized to maximal gas pressure) on the disk and jet structures. These are R1, R2 ($\beta_0 = 100$), R7 ($\beta_0 = 44$) and R8 ($\beta_0 = 20$).

We first validated that the resolution we use is sufficient to ensure full development of the MRI. For that, the first two runs we conducted, 
R1 and R2 only differ by
their azimuthal resolution, with R2 having a two times higher resolution than
R1. No major differences between the results of R1 and R2 are found. We therefore conclude that
the resolution of R2 ($N_r \times N_\theta \times N_\phi = 256 \times 128 \times 128$) is sufficient to support the conclusion 
presented herein. We note however that the viscous spreading is sensitive to
the angular resolution in the $\theta$ direction, and to the reconstruction method (PPM vs PLM)
\citep{KNC11, PCN19}. On the other hand, for resolutions higher than  $192^3$ 
it was shown by \citet{PCN19} that the resolution-dependence is weak. 
Although this is an important limitation of our numerical models
that must be kept in mind, we find that our results are in excellent agreement with the
numerical results presented in \citet{PCN19}, and are therefore reliable. 

\begin{figure}
    \centering
    \includegraphics[width=0.75\textwidth]{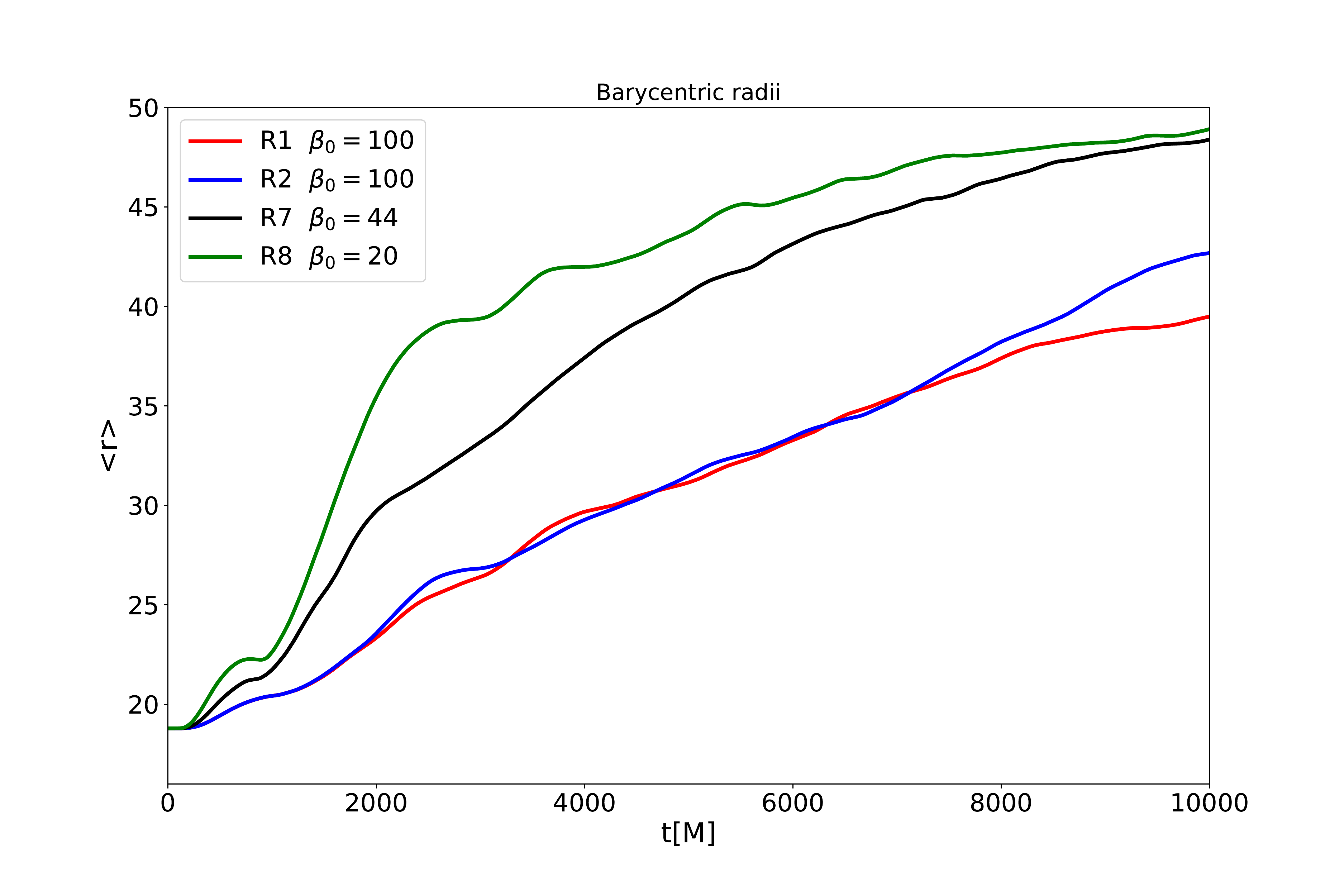}
    \caption{Time evolution of the barycentric radius for the four runs with different initial magnetization R1, R2 ($\beta_0=100$), R7 ($\beta_0=44$) and R8 ($\beta_0 = 20$). 
    Maximum integration radius $r_{\max} = 80$ is taken in all calculations (see Equation  \ref{eq:barycentric_radius} and the following discussion). The increase in the barycentric radius is due to viscous spreading. Clearly, the larger the initial magnetic field (the smaller $\beta_0$) the faster the disk expands, implying that the initial viscosity is directly related to the initial magnetization, $\beta_0$.
    } 
    \label{fig:barycentric_radius_beta}
\end{figure}

The effect of the initial magnetic field normalisation $\beta_0$ is better
seen and understood from Figure \ref{fig:barycentric_radius_beta}, which shows
the temporal evolution of the barycentric radius, given by Equation \eqref{eq:barycentric_radius}, for the 4 runs. It is clear that the larger the initial 
$\beta_0$, the faster is the spread of the disk. This is to be expected as the 
Maxwell viscous coefficient scales $ \propto b^r b^\phi$, namely with the magnetic field components \citep{KHH05,PSK13}. Although the MRI 
turbulence should develop independently of the initial condition and drive
the viscous spreading, the initial system evolution is driven by the initial
value of $\beta_0$, with a larger $\beta_0$ leading to a slower initial disk spreading.

\begin{figure}
    \centering
    \begin{tabular}{cc}
    \includegraphics[width = 0.50\textwidth]{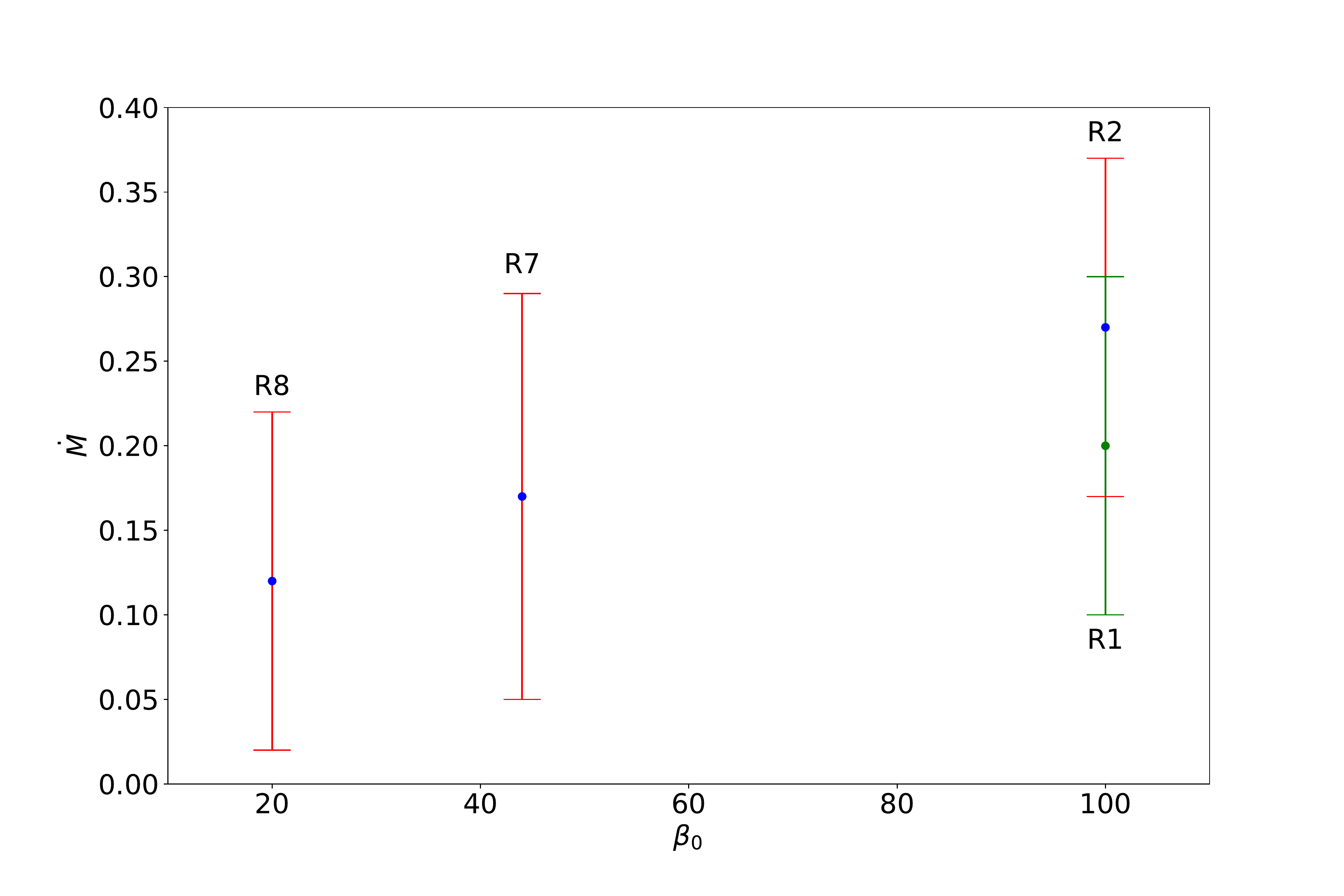} &
    \includegraphics[width = 0.50\textwidth]{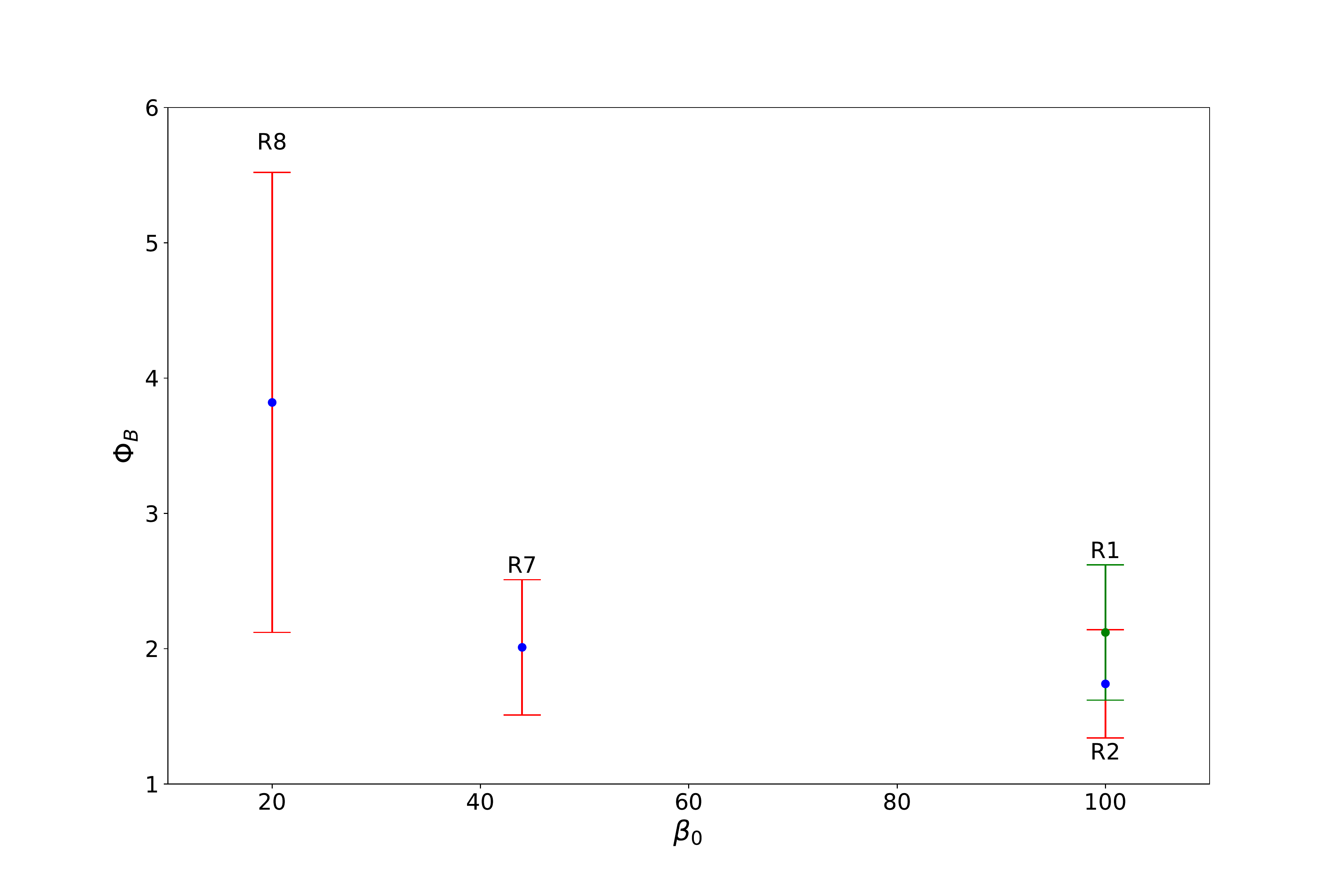}
    \end{tabular}
    \caption{Left - Time average mass accretion rate $\dot M$ for run R1, R2, R7 and R8 which only differ by their value of $\beta_0$. Error bars represent $2\sigma$ variance. Right - Same but for the MAD parameter $\Phi_B$.
    }
    \label{fig:compare_beta_0}
\end{figure}

In Figure \ref{fig:compare_beta_0}, we show $\dot M$ and $\Phi_B$ as a function of $\beta_0$. 
The mass accretion rate is consistent with being the same for all 4 simulations, with a trend towards increase of the accretion rate with $\beta_0$. This is in agreement with the results
of \citet{BHK08}, who found a negligible influence of the magnetic field on the
accretion flow. However, we do find a large spreading for the
average MAD parameter at the horizon $\langle \Phi_B \rangle_t$:
it is a factor $\sim 2$ larger for R8 ($\beta_0 =20$) than for R1, R2 ($\beta_0=100$) while the MAD parameter for R7 ($\beta_0=44$) is in between those two values. All these four 
simulations are able to launch jets. 

\begin{figure}
    \centering
    \includegraphics[width=0.75\textwidth]{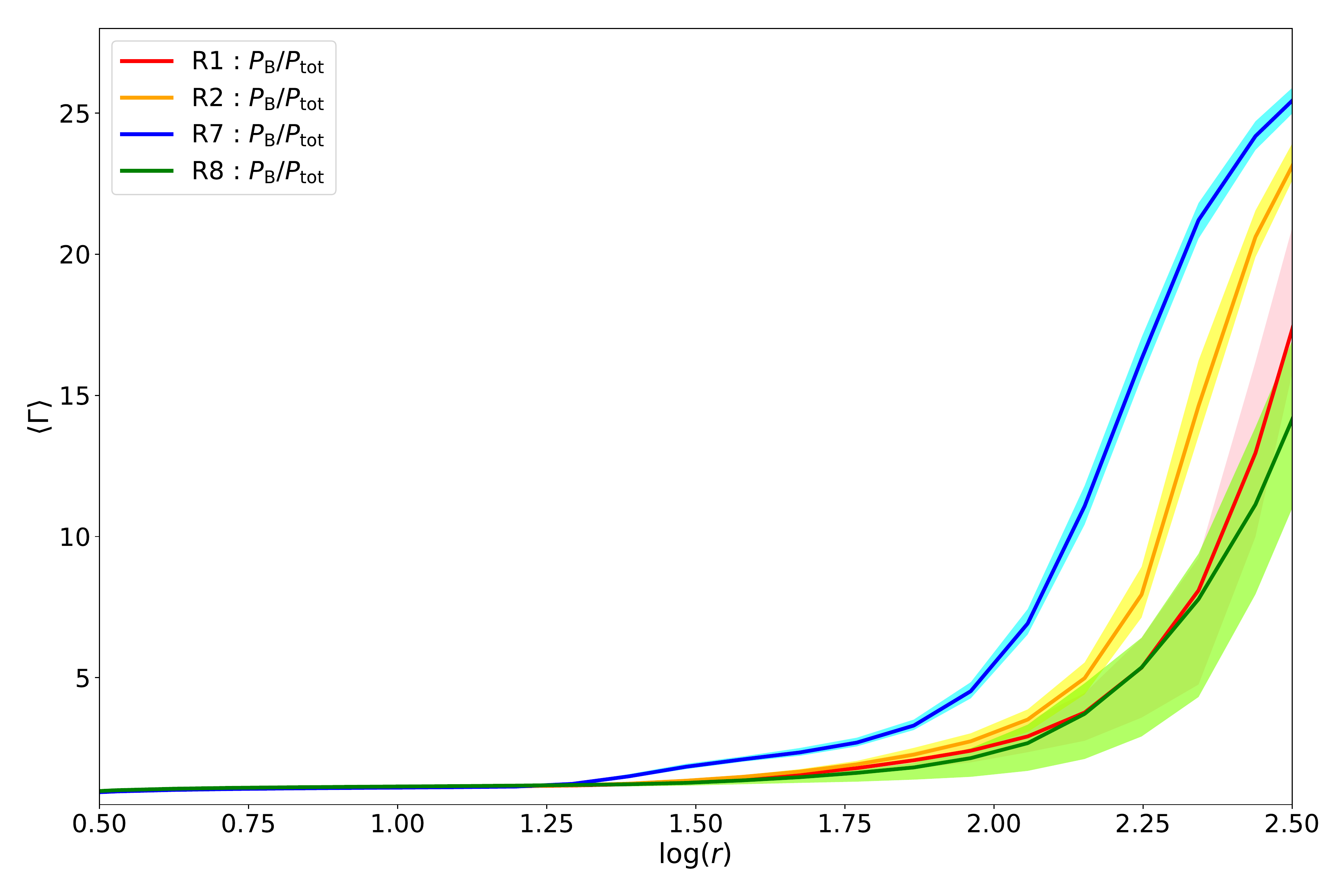} 
    \caption{Time- and angle-average Lorentz factor of the jet (defined by $\sigma>1$) for the same simulations, as a function of radius.}
    \label{fig:jet_lorentz_factor_different_beta}
\end{figure}

For all four simulations, we show in Figure \ref{fig:jet_lorentz_factor_different_beta}
the jet Lorentz factor (Equation \eqref{eq:Lorentz_factor}), where the
integration is restricted to regions in which $\sigma > 1$. In all
cases, the Lorentz factor remains small, around a few, below radius
$r \lsim 30$M. Above this radius, it starts to increase faster to reach an asymptotic value 
$\Gamma \sim 20$. This evolution with radius is in qualitative agreement
with the results presented in \citet[][see their Figure 7]{FWR12}, although the
final values are larger than the value of $\sim 7- 10$ found by \citet{McK06}, \citet{FWR12} and \citet{CLT19}. 
We note however that the initial torus used in the two
aforementioned papers are substantially different than the one used in our
simulations.

We tentatively explain this difference in Lorentz factor at $r \sim 200$M by
the fact that our numerical grid is not suitably chosen to study jets at large
distances from the black-hole. Indeed, as jets from such types of simulations
are expected to be closer to parabolic than to a radial geometry
\citep[e.g.][]{TMN08,NAH18}, numerical codes need a special type of grid that
warps towards the pole to resolve the jet all the way to the outer domain
boundary \citep[e.g.,][]{MTB12, RTQ17}. The simulations presented here do not
use such a grid. The approximate force free solution for the surface
$\sigma =1$ is well approximated by $z \propto R^\nu$, with $1.3 \leq \nu \leq 2$
\citep{TMN08}. Here, $R = r \sin \theta$ is the cylindrical radius. Comparing this jet
evolution to the angular grid we use here, we find that at radius
$r = 200M$ the jet is resolved by only 3 (7) grid cells in the $\theta$
direction for $N_\theta = 128$ ($N_\theta = 256$) respectively. This number
of cells between the jet boundary and the pole is clearly insufficient to
properly capture the jet dynamics at height $z \gsim 200$M.


\subsection{Effect of $r_{\rm max}$ : R7 vs R9 }
\label{sec:6.5}

We next examined the effect of the disk size, as is prescribed by the radius of maximum pressure, $r_{\max}$. 
Simulations R7 and R9 only differ by their initial value of $r_{\rm max}$, namely $r_{\max} (R7) = 12$~M, and $r_{\max} (R9) = 13$~M. Both have a relativistic adiabatic index
and an initial $\beta$ parameter $\beta_0 = 44$. Increasing the radius of
maximum pressure sharply changes the disk morphology. Firstly, the disk of
R9, having a larger $r_{\rm max}$, is more extended than that of R7. Secondly,
as a result of the normalisation to the unity of the maximum disk density,
the disk mass is larger by a factor $\sim 3$ (see
Table \ref{tab:tab_run_caracteristics}). 

The jet in R9 seems to be intermittent, with the polar region getting filled
by material from the disk. It is interesting to note that this difference is
not due to the initial normalisation of the magnetic field, $\beta_0$, which is the same in both runs. Indeed,
R3 and R4 differ by both $r_{\rm max}$ and $\beta_0$, which is scaled such that 
the ratio of mass to magnetic field energy is constant. The same difference,
i.e. the absence of a jet, is observed for that set of simulations.
The comparison of R7 and R9 thus shows that the difference in $\beta_0$ does not produce the major effect on the ability to launch a jet in the SANE state. Rather, the mass distribution seems to be the dominant factor. Alternatively, the initial magnetic field topology may also have an effect on the ability of the system to launch a jet, but in this current work we limit our simulations to a single initial magnetic field configuration, as is given in Equation \ref{eq:potetial_vector}.

\subsection{Out of equilibrium initial condition}
\label{sec:out_of_equilibrium_results}

Runs R5 and R6 are different from the others as their initial setup is out
of equilibrium, with a disk which is not supported by its initial pressure.
To obtain this effect, we numerically initialized the disks using a non-relativistic 
adiabatic index $\hat \gamma = 5/3$ (thus, the initial disk is similar to R4), while in calculating their evolution, we assumed $\hat
\gamma = 4/3$. These two runs are identical except for the grid size, which is larger in R6 by a factor of 2 in the $\theta$ direction. 

\begin{figure}
    \centering
    \includegraphics[width = 0.75\textwidth]{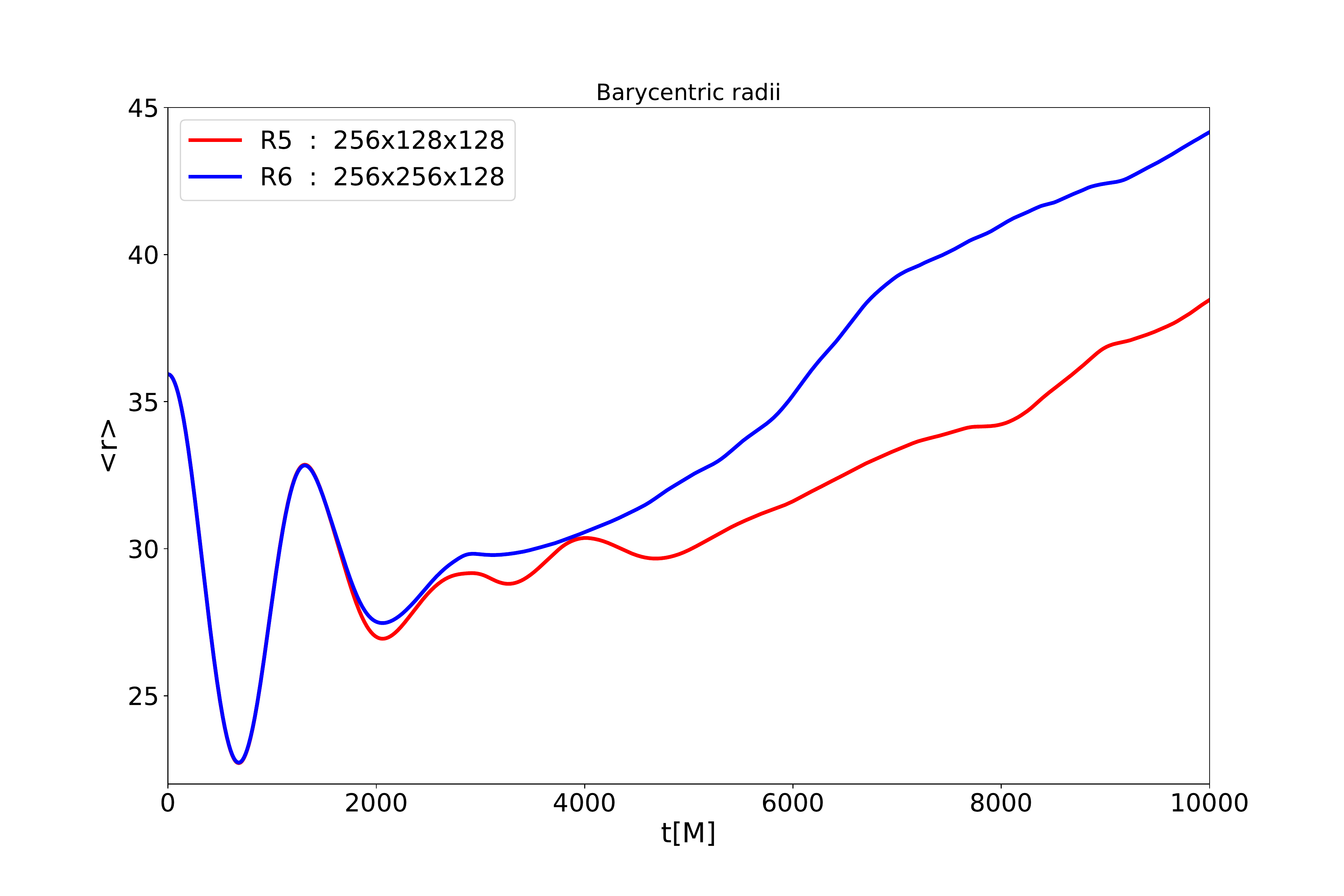}
    \caption{Barycentric radius for the two non-equilibrium simulations. For both simulations, the disk initially contracts, bounces two times and then start expanding similarly to the other disks, which starts from an equilibrium state. }
    \label{fig:unsteady_barycenter}
\end{figure}

\begin{figure}
    \centering
    \includegraphics[width = 0.75\textwidth]{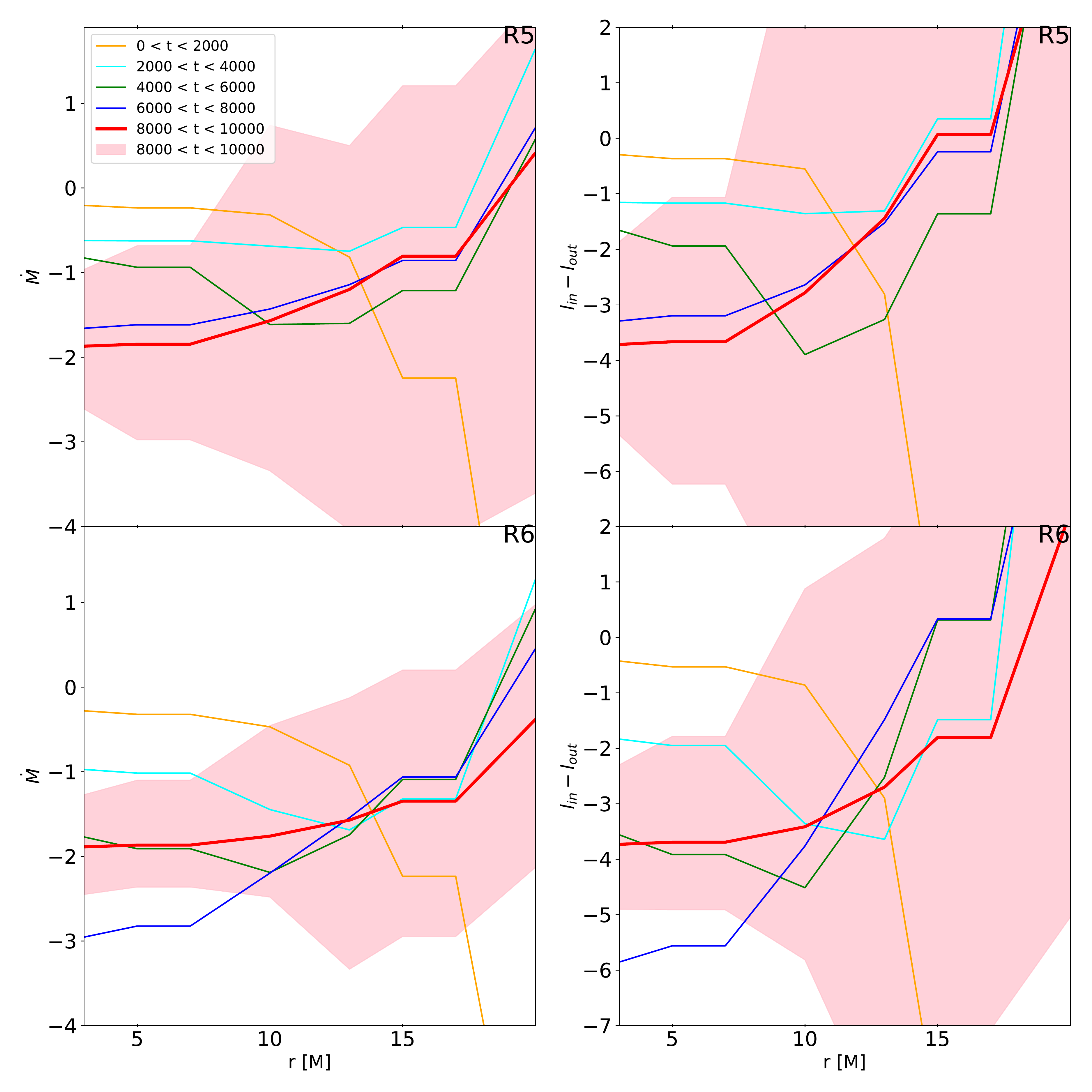}
    \caption{Same as Figure \ref{fig:MdotLinlout_r} for R5 and R6. Left - Time-averaged mass accretion rate as a function of radius $r$ for different time intervals. Right - Time averaged angular momentum flux as a function of radius. The pink region corresponds to the 2-$\sigma$ variance during the last time interval, $8000 < t < 10^4$M. From this figures, it is clear that at the end of our run, quasi steady state is not achieved. }
    \label{fig:mdot_lin_lout_r5_r6}
\end{figure}

Initially, the disks contract rather than expand due to the viscous stresses.
This is clearly visible in Figure \ref{fig:unsteady_barycenter} which shows
the time evolution of the barycentric radius for both runs. After the initial
contraction (at $t\lsim 1000$~M), the disks bounce a couple of times, after
which the disks start to expand similarly to that of the other simulation.
This is also visible in Figure \ref{fig:mdot_lin_lout_r5_r6} in which the
variations of $\dot M$ and $l_{in} - l_{out}$ between $8\times 10^3 < t
< 10^4$ are the largest while the average are not yet independent on the
radius at any distance from the black-hole. This means that at the end of
the simulation, R5 and R6 did not yet reach (quasi) steady-state. 

During the initial transition period, the initial MAD parameter reaches
a relatively large value of $\sim 10$. This results in an initially strong
jet, lasting during the transition period of  $t \sim 3\times 10^3$M.
However, after this period ends, the MAD parameter sharply drops to an
average value of $ \Phi_{\rm B}(R6) = 0.96$, which is the smallest average
value of all the runs we have. Despite the large mass accretion rate at late
times (after the transition period), the jet is not sustained, but rather
disappears. This is clearly shown in Figure \ref{fig:time_azimuth_integrated_R5_R6}. 
This result therefore demonstrates the ability of transient systems to
produce transient jets.

A more extending magnetized region could have contributed to sustain
the magnetic flux threading the horizon and potentially sustain a jet
in these configurations as well. This underlines the sensitivity to the
initial condition of a jet in the SANE regime. 



\section{Conclusions}

\label{sec:discussion}

Using CUDA-C and OpenMP, we wrote a GPU-accelerated code which solves the GR-MHD
equations in the context of accretion disk and jet system around a rotating
black-hole. The code is currently designed to run on a single multi-GPU
workstation. This is currently sufficient to perform
simulations of a system composed of a thick accretion disk around a black hole, and the
jets resulting from the accretion, with a resolution sufficient for MRI to
develop according to the quality factor $Q^{(i)}$ given by Equations
\ref{eq:volume_average_q}. In the future, we are planning an extension with MPI
to enable the code to run on a multi-node architecture. 

This code adds to two existing GR-MHD codes aimed at studying accretion disks that were designed to run on GPUs \citep{CFG17, LHT18}. In this paper, we detailed some of the strategies we used to increase the compute efficiency of our code. Specifically, the computation along direction $\phi$ is critical to reduce memory loads of the metric terms and improve the computational speed.

We have used our new numerical code to perform several simulations. We
present in this paper the results of 9 such simulations of ADAF disks accreting
in the SANE regime. In this first paper we focus on this regime, as it enables a relatively simple comparison of our results with that of previous simulations \citep{PCN19} thereby providing confidence in the correctness of our numerical scheme. We then change the initial conditions to study a yet unexplored region. Our disk have different initial parameters, such as
$r_{\rm max}$, $\beta_0$ and initial adiabatic indexes. From those simulations,
we reach several conclusions.

First, we simulated four disks with different initial magnetic field normalisation
in order to study the impact of the initial setup (section 
\ref{sec:magnetic_field}). We found that the mass accretion
rate is comparable in all cases, with a slight increase for small $\beta_0$ (so a
large initial magnetization). The MAD parameter however is different for all
four simulations, being larger for smaller $\beta_0$. We thus conclude that while in the SANE state, the mass accretion rate only weakly depends on the initial magnetization, $\beta_0$.

Second, interpreting our
results for disks with different masses, but with
a constant ratio of initial mass to magnetic energy (section \ref{sec:6.5}), we find that the accretion
rate and the presence of a jets mostly depend on the initial mass distribution
rather than on the initial magnetic field, provided that $\beta_0$ is large. When the magnetic flux threading the horizon drops, the jet disappears. This result therefore demonstrates that the existence of the jet is not a linear function of the initial magnetic field strength.

Third, the effect of the adiabatic index on the
morphology of the accretion system and on the accretion rate is investigated in section \ref{sec:adiabatix_index}.
For the structure of the disk, only small differences were found, in agreements with previous
results \citep{MG04, MM07}. However, we find that the jet is narrower with
non-relativistic adiabatic index $\hat \gamma = 5/3$ than it is for relativistic
adiabatic index, $\hat \gamma = 4/3$, as seen from Figure \ref{fig:sigma=1_r2_R3}. We thus conclude that the structure of the disk/jet is a weak function of the adiabatic index of the gas, although a relativistic gas tends to result in a wider jet. 

We have also presented the results of two simulations with out-of-equilibrium
disks. Those two simulations only differ by the angular resolution so
their reliability can be checked. We found that (i) those two simulations
have the smallest MAD parameter and that they are unable to form a jet, and
(ii) that after a transition period a disk similar to the one obtained for
the other simulations is obtained, showing that the disk structure seems to
be robust to change in the initial configuration but that the existence of the
jet in the SANE regime is not.

Overall, our simulations show how sensitive the disk and the accretion
properties are to the initial conditions of the simulation in the SANE regime.
We found that the disk structure is robust while the presence of a jet is strongly
dependent on the initial mass distribution, with large disks only forming transient
jets at the onset of the simulations. For small initial $r_{\rm max}$, the disk
and jet structures are independent on the magnetization, while the adiabatic index
of the gas only changes the opening angle of the jet.


To conclude, in the context of the results from the EHT collaboration and
the imaging of the very center of an accretion system \citep{EHT19, EHT21},
detailed predictions can now be directly tested. However, producing those
predictions requires (i) state of the art GRMHD codes, (ii) radiative transfer
codes and (iii) access to HPC facilities, most of which are now equipped with
GPU accelerators, to obtain meaningful results. In this paper, we presented a
first step towards the creation of a new GRMHD numerical code with the capability
of being accelerated by GPUs.



\begin{acknowledgments}
We thank Oliver Porth for helpful discussions. DB and AP acknowledge support from the European Research Council via the ERC consolidating grant $\sharp$773062 (acronym O.M.J.). B.-B.Z. acknowledges support by the National Key Research and Development Programs of China (2018YFA0404204), the National Natural Science Foundation of China (Grant Nos. 11833003, U2038105, 12121003, 11922301, 12041306, 12103089), the science research grants from the China Manned Space Project with NO.CMS-CSST-2021-B11, the Natural Science Foundation of Jiangsu Province (Grant No. BK20211000), and the Program for Innovative Talents, Jiangsu. This work is performed on a HPC server equiped with 8 Nvidia DGX-V100 GPU modules at Nanjing University. We acknowledge the IT support from the computer lab of School of Astronomy and Space Science at Nanjing University. Guoqiang Zhang also acknowledges support by the China Scholarship Council for 1 year study in Bar-Ilan University.
\end{acknowledgments}

\bibliography{biblio}

\appendix

\section{Impact of resolution}

\label{sec:resolution_effects}.

\begin{figure}[t]
\centering
\begin{tabular}{cc}
\includegraphics[width = 0.44 \textwidth]{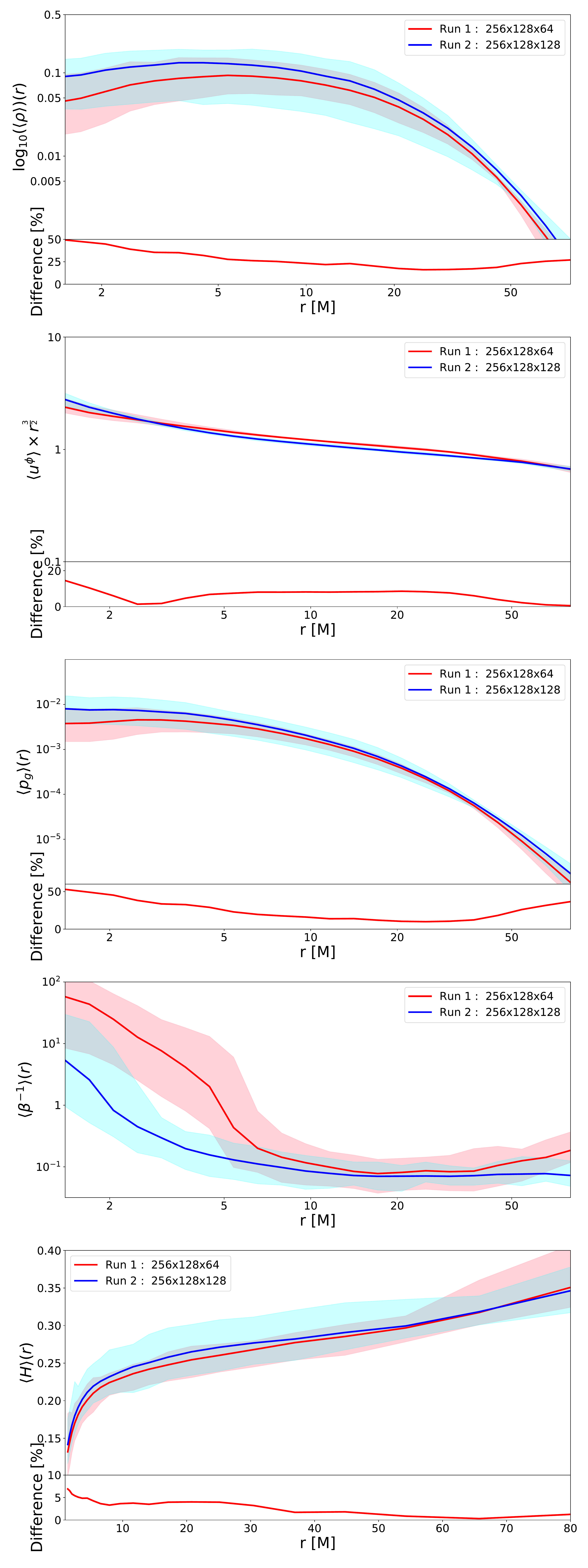} &  
\includegraphics[width = 0.44 \textwidth]{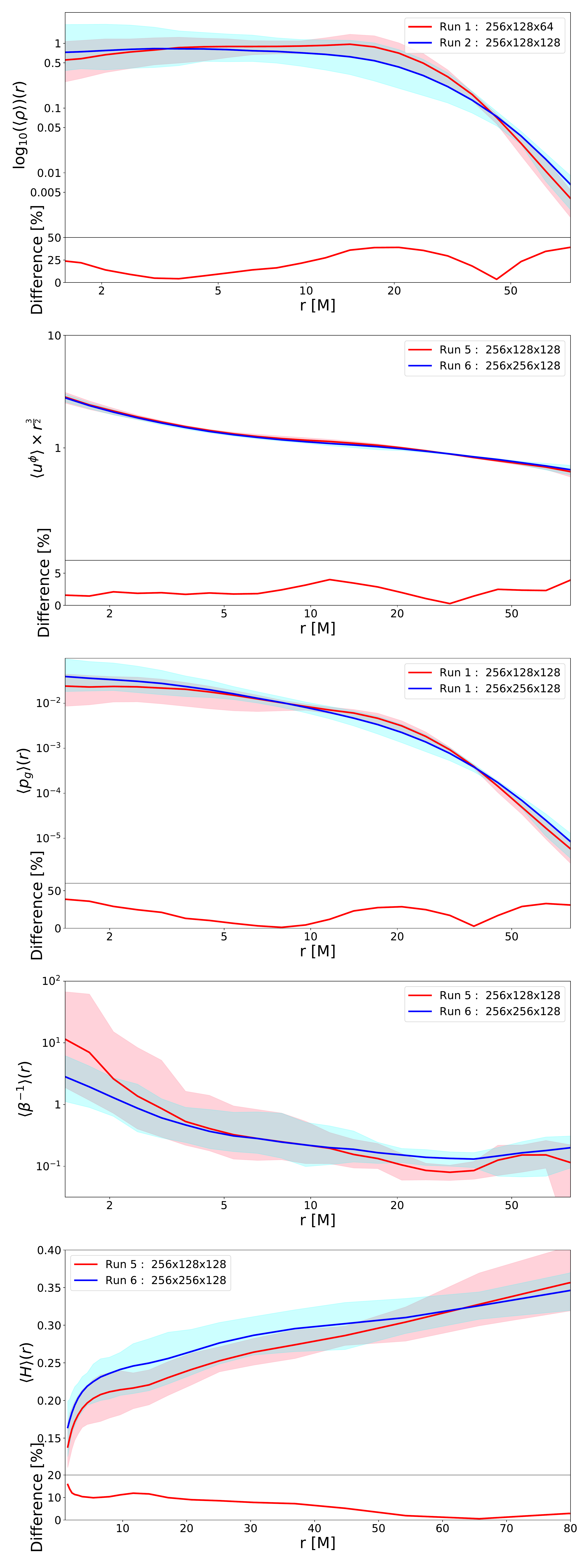}
\end{tabular}
\caption{Comparing the effect of varying the resolution. Left - comparison between R1 and R2. Right - comparison between R5 and R6. From top to bottom : density, poloidal velocity $u_\phi$, gas pressure $p_g$, plasma $\beta$ and disk width $H$. The thick lines are averaged over time such that $ 5\times 10^3 < t < 10^4$ and the  shaded regions correspond to the maximal variation amplitude for data exported every $\Delta t = 5$M.}
\label{fig:resolution}
\end{figure}

We have two set of similar simulations which only differ by their $\theta$ or $\phi$ resolution : 
\begin{itemize}
    \item R1 and R2 only differ by the increase in resolution of $N_\phi$ which is doubled for R2.
    \item R5 and R6 which only differ by the increase in resolution of $N_\theta$ which is doubled for R6.
\end{itemize}
Figure \ref{fig:resolution} compares the difference between $\langle \rho \rangle$, $\langle u^\phi \rangle$, $\langle p_g \rangle$, $\langle \beta^{-1} \rangle$ and $\langle H \rangle$ as a function of radius. Only $\beta^{-1}$ is sharply different between R1 and R2, while all quantities are in good agreement for R5 and R6. Moreover, we checked that the profile for R2 is compatible with the results from \citet{PCN19}.  We therefore conclude that a resolution of 256x128x128 is the minimal resolution required to sufficiently resolve our accretion disks.


\end{document}